\documentclass[a4paper,11pt]{article}
\pdfoutput=1 

\usepackage{jheppub} 
\usepackage{bm,amssymb,slashed,graphicx,multirow,soul,mathtools,xspace,array}  
\usepackage{float} 
\allowdisplaybreaks
\usepackage{bbold}
\usepackage{subfigure}

\newcommand{\C}{{\cal C}}

\newcommand{\mev}{{\rm MeV}}

\newcommand{\todo}[1]{{\color{red} \ifmmode\else[todo]\fi #1}}
\usepackage[usenames,dvipsnames]{xcolor}
     \definecolor{hgreen}{rgb}{0,.3,0}
      \definecolor{darkgreen}{rgb}{0.3,.8,0.2}
     \definecolor{hred}{rgb}{.3,0,0}
     \definecolor{hblue}{rgb}{0,0,.3}
     \definecolor{LightGray}{gray}{0.95}
     
\usepackage{listings}

\newcommand{\lp}{\left(}
\newcommand{\rp}{\right)}

\newcommand{\g}{\gamma}

\newcommand{\MeV}{\text{MeV}}
\newcommand{\Br}{\text{Br}}

\newcommand{\Neff}{$N_{\rm eff}$}

\newcommand{\beq}{\begin{equation} }
\newcommand{\eeq}{\end{equation}} 
\newcommand{\bi}{\begin{itemize} }
\newcommand{\ei}{\end{itemize} }

\definecolor{Red}{rgb}{1.,0.,0.}
\definecolor{Grn}{rgb}{0.,0.75,0.}
\definecolor{Blu}{rgb}{0.,0.,1.}

\let\Re\relax
\DeclareMathOperator{\Re}{Re}

\DeclareMathOperator{\Tr}{Tr}

\allowdisplaybreaks  

\title{\boldmath Enhanced neutrino polarizability}

\author[1]{S. Bansal,}
\author[2]{G. Paz,}
\author[2,3]{A.A. Petrov,}
\author[4]{M. Tammaro,}
\author[1]{J. Zupan,}

\affiliation[1]{Department of Physics, University of Cincinnati, Cincinnati, Ohio 45221,USA}
\affiliation[2]{Department of Physics and Astronomy,
Wayne State University, Detroit, Michigan 48201, USA}
\affiliation[3]{Department of Physics and Astronomy,
University of South Carolina, Columbia, South Carolina 29205, USA}
\affiliation[4]{Jozef Stefan Institute, Jamova 39, Ljubljana, Slovenia}

\emailAdd{saurabhbansal20@gmail.com}
\emailAdd{gilpaz@wayne.edu}
\emailAdd{apetrov@sc.edu}
\emailAdd{michele.tammaro@ijs.si}
\emailAdd{zupanje@ucmail.uc.edu}

\abstract{We point out that neutrinos can have enhanced couplings to photons, if light (pseudo)scalar mediators are present, resulting in a potentially measurable neutrino polarizability. We show that the expected suppression from small neutrino masses can be compensated by the light mediator mass, generating  dimension 7 Rayleigh operators at low scales. We explore the rich phenomenology of such models, computing in details the constraints on the viable parameter space, spanned by the couplings of the mediator to neutrinos and photons. Finally, we build several explicit models that lead to an enhanced neutrino polarizability by modifying the inverse see-saw majoron, i.e., the pseudo-Nambu-Goldstone boson of the $U(1)_L$ global lepton number responsible for generating small neutrino masses.
}

\begin{document} 

\maketitle

\flushbottom

\section{Introduction}
\label{sec:intro}

Electromagnetic interactions of neutrinos serve as a 
primary venue for discovering new physics interactions.
It can be viewed as a qualitatively different pathway to 
uncover physics beyond the standard model compared to the observation of neutrino masses two decades ago \cite{Super-Kamiokande:1998kpq}. For instance, if neutrinos are of the Majorana type, their masses do point to a new physical scale, $\Lambda$, since in this case the neutrino masses are generated through a non-renormalizable dimension-5 Weinberg operator, ${\cal L}\supset  y'_{ij} \big(\bar L_i^c H^{c \dagger} H^\dagger L_{j}\big)/\Lambda$. However, it is equally possible that neutrinos are of the Dirac type, in which case the neutrino masses are due to the renormalizable Yukawa interactions, ${\cal L}\supset  y_{ij} \big(\bar \nu_{R i}  H^\dagger L_{j}\big)$.
To be certain that the neutrino masses imply the existence of a new physics scale, $\Delta L=2$  neutrinoless double $\beta$ decay needs to be discovered first, see, e.g., \cite{Gonzalez-Garcia:2002bkq,Gonzalez-Garcia:2007dlo,DellOro:2016tmg}.

In contrast, if neutrinos are found to couple directly to photons in the current or immediately planned experiments, this would unambiguously point to the existence of a new physical scale. The operators of the lowest dimension, invariant under $SU(2)_L\times U(1)_Y$, that couple neutrinos to photons $F_{\mu\nu}$ are the dipole operators, which for Dirac neutrinos are of dimension 6, $\big(\bar \nu_{R i}  \sigma^{\mu\nu} H^\dagger L_{j}\big)B_{\mu\nu}/\Lambda^2$, and the dimension 8 Rayleigh operators such as $\big(\bar \nu_{R i} H  L_{j}\big)B_{\mu\nu}B^{\mu\nu}/\Lambda^4$ (similar operators can be written for the weak isospin fields $W_{\mu\nu}^a$ by direct substitutions of the weak hypercharge fields $B_{\mu\nu}$). After the Higgs obtains a vev, $H=(0, v)/\sqrt2$,  these operators lead to neutrino dipole moments, $\bar \nu_{ R i} \sigma_{\mu\nu} \nu_{Lj} F^{\mu\nu}$, and neutrino polarizability\footnote{In the manuscript we use interchangeably neutrino polarizability and neutrino Rayleigh operators.},  $\bar\nu_{R i} \nu_{L j} F^{\mu\nu} F_{\mu\nu}$, respectively. The Dirac neutrino mass term, $m_\nu \bar \nu_L \nu_R$, as well as the neutrino dipole moments and the neutrino polarizability operators, are all chirality flipping. The new physics that generates at some loop-level the neutrino dipole moments and/or the neutrino polarizability is, therefore, expected to generate at the same loop-level also the contributions to the neutrino masses. Unless there are large cancellations between tree level and radiatively generated contributions to the neutrino masses, the dipole moments and polarizability thus need to be tiny, effectively proportional to the tiny neutrino masses, $m_\nu$, and out of reach of the experiments. In this manuscript, we show that this is not necessarily the case for Rayleigh operators, for which the $m_\nu$ suppression can be parametrically compensated if the couplings to photons arise from tree-level exchanges of light new physics. 

Similar naive dimensional analysis arguments apply to Majorana neutrinos, though with several important differences. First,  if neutrinos are  Majorana, the same operators: the neutrino mass term, the dipole, and the Rayleigh operators, require an extra Higgs insertion compared to Dirac neutrinos.  That is, for Majorana neutrinos the mass term is of dimension 5, the dipole operators are of dimension 7, $\big(\bar L_{ i}^c H  \sigma^{\mu\nu} H^\dagger L_{j}\big)B_{\mu\nu}$, while Rayleigh operators are of  dimension 9, $ \big(\bar L_i^c H^c{}^\dagger   H^\dagger L_{j}\big)B_{\mu\nu}B^{\mu\nu}$ (and similarly for $W^a_{\mu\nu}$). More importantly, these operators violate the lepton number by $\Delta L=2$. This breaking is expected to be small, explaining why the neutrino masses are small and implying that the neutrino magnetic moment and neutrino polarizability will be small. 

There are, however, exceptions to this general rule. First of all, for Majorana neutrinos, the tensor and scalar neutrino currents have definite symmetry under the interchange of the neutrinos (unlike in the case of Dirac neutrinos). 
Since $\bar \nu_{i L}^c \sigma_{\mu\nu} \nu_{j L}=-\bar \nu_{j L}^c \sigma_{\mu\nu} \nu_{i L}$ is odd, while $\bar \nu_{i L}^c\nu_{ j L}= \bar \nu_{j L}^c  \nu_{i L}$ is even under the interchange of the two neutrinos, any new physics that is odd under the same flavor exchange will only contribute to the neutrino magnetic moments and not to the neutrino masses \cite{Voloshin:1987qy}. This has been used in Refs. \cite{Babu:1989wn,Babu:1990wv,Babu:2020ivd,Babu:2021jnu} to build explicit models of enhanced neutrino magnetic moments. 

No such symmetry distinguishes the neutrino mass operator from the Rayleigh operators since the neutrino currents in both are exactly the same. Neutrino polarizability is thus inevitably suppressed by the same small $\Delta L=2$ breaking spurion as neutrino masses. That is, neutrino polarizability is model-independently proportional to tiny neutrino masses. However, it can still be parametrically enhanced if generated by a tree-level exchange of a light scalar or pseudo-scalar mediator. A prototypical example is a pseudo-Nambu-Goldstone boson (pNGB) due to spontaneous breaking of the lepton number -- the majoron, which couples derivatively to the $\Delta L=2$ current, $i\bar \nu_L^c \nu_L \partial_\mu \phi/f_\phi\to -i m_\nu \nu_L^c \nu_L  \phi/f_\phi$. Generically,  majoron also couples to photons through a higher dimension operator, $\phi FF/\Lambda_\gamma$. For the minimal majoron, this operator is additionally suppressed by the majoron mass squared, $m_\phi^2$, while this suppression is absent in non-minimal models. At energies below  $m_\phi$ this then leads to the neutrino polarizability of the form $\nu\nu FF \times (m_\nu/f_\phi) \times 1/(m_\phi^2 \Lambda_\gamma)$.  The small majoron mass compensates for the $m_\nu$ suppression, leading to parametrically enhanced neutrino polarizability within reach of astrophysical and terrestrial experiments. In this manuscript, we perform the first phenomenological analysis of the existing constraints and possible future probes of neutrino polarizability over a wide range of mediator masses, from eV, i.e., comparable to the neutrino masses, up to the GeV scale. 
 
 The paper is organized as follows. In Section~\ref{sec:nu:photons}, we introduce the neutrino dipole, anapole, and polarizability operators within an EFT framework. The enhanced neutrino polarizability via a light mediator exchange is detailed in Sec.~\ref{subsec:toymodel}. In Section~\ref{sec:cosmology}, we explore the consequences of this interaction for cosmological observables such as Cosmic
 Microwave Background (CMB) and Big Bang Nucleosynthesis (BBN). In Section~\ref{sec:StarCooling}, we analyze bounds from anomalous star cooling rates due to the production of light $\phi$ particles. At higher energy scales, the Rayleigh operator can be probed with neutrino scatterings in terrestrial experiments, including the production of $\phi$ particles in colliders; these are discussed in Section~\ref{sec:NeutrinoScatteringExperiments}. In Section~\ref{sec:model}, we discuss UV complete models that lead to enhanced neutrino polarizability, focusing on spontaneously broken $U(1)_L$.
Our conclusions are summarized in Section~\ref{sec:Conclusions}. Appendix~\ref{sec:app:notations} contains our notation and conventions, while appendix~\ref{sec:app:cooling:rates} contains further details on the calculation of production rates of light (pseudo)scalars in stellar cores. Appendix \ref{sec:RareDecays}  contains further details on constraints from invisible decays of heavy (pseudo)scalars. 

\section{Neutrino couplings to photons}
\label{sec:nu:photons}

Neutrino couplings to photons arise from higher dimensional operators. Using the notation of Ref.~\cite{Altmannshofer:2018xyo} and restricting the discussion to low energies, well below the electroweak symmetry breaking scale, the relevant operators are given by\footnote{The dimension six anapole moment operator induces a contact interaction and can be replaced through the use of the equation of motion by the four fermion operators, a choice made in the construction of the complete basis in Ref. \cite{Altmannshofer:2018xyo}. See Section \ref{sec:neutrino:anapole} for further details.} (see also Appendix \ref{sec:app:notations}), 
\beq
\label{eq:L:EFT}
\begin{split}
{\cal L}_{\rm EFT} &\supset  \sum_{i>j}\frac{\C_{1,ij}^{(5)}}{\Lambda} \frac{e}{8\pi^2}  (\bar \nu_i  \sigma^{\mu \nu} P_L \nu_j) F_{\mu\nu} 
+\frac{1}{2} \sum_{i,j} \biggr[ \frac{\C_{1,ij}^{(7)}}{\Lambda^3}\frac{\alpha}{12\pi} (\bar \nu_i P_L  \nu_j) F_{\mu\nu} F^{\mu\nu}
\\
&\qquad 
 + \frac{\C_{2,ij}^{(7)}}{\Lambda^3}\frac{\alpha}{8\pi} (\bar \nu_i P_L  \nu_j) F_{\mu\nu}\tilde  F^{\mu\nu} \biggr] +{\rm h.c.}+\cdots,
\end{split}
\eeq
with ellipses denoting higher dimension terms.
The indices $i,j=e,\mu,\tau$ represent the SM neutrino flavors, while $F_{\mu\nu}$ is the electromagnetic field strength tensor, with $\tilde F_{\mu\nu}=\frac{1}{2}\epsilon_{\mu\nu\rho\sigma} F^{\rho\sigma}$ its dual. Here, and in the rest of the paper, the neutrinos, $\nu_i$, are assumed to be Majorana fermions.
Throughout the manuscript, we also use the four-component notation with the conventions from Ref. \cite{Dreiner:2008tw}, so that $\nu=\nu^c$.  

The dimension 5 operators in \eqref{eq:L:EFT} encode the neutrino dipole moments.
For Majorana neutrinos the flavor conserving dipole moments vanish  because the dipole is antisymmetric in flavor indices, $ (\bar \nu_i  \sigma^{\mu \nu} P_L \nu_j)=- (\bar \nu_j  \sigma^{\mu \nu} P_L \nu_i)$. 
The dimension-7 Rayleigh operators, on the other hand, are symmetric in flavor indices, $\C_{1,ij}^{(7)}=\C_{1,ji}^{(7)}$, and thus mediate also flavor diagonal transitions. The definitions of the Wilson coefficients in \eqref{eq:L:EFT} include the loop factor, anticipating that in many models the operators would be generated at one loop, while $\Lambda$ is the mass scale associated with the masses of particles running in the loop (see also the discussion below and in Section \ref{sec:model}).

Below we will also use a short hand notation, where $\Lambda$ is absorbed in the definitions of the Wilson coefficients that now become dimensionful, 
\beq
\label{eq:Chat}
\hat \C_{1(2),ij}^{(7)} \equiv \C_{1(2), ij}^{(7)}/\Lambda^3.
\eeq
Quite often we will also assume that the neutrino polarizability is flavor diagonal, so that (no summation implied)
\beq
\label{eq:C12:flav:diag}
\hat \C_{1(2),ij}^{(7)} = \hat \C_{1(2),i}^{(7)} \delta_{ij},
\eeq
and similarly for dimensionless Wilson coefficients, $\C_{1(2),ij}^{(7)}$. For flavor universal case we will denote
\beq
\label{eq:C12:flav:uni}
\hat \C_{1(2),ij}^{(7)} = \hat \C_{1(2)}^{(7)} \delta_{ij},
\eeq
Finally, we also define 
\beq
\label{eq:C12:Re}
\hat \C_{1(2)}^{\rm Re} = \sum_{ij} 2 \Re\Big[\hat \C_{1(2),ij}^{(7)}\Big].
\eeq

In the remainder of this section we discuss in more detail the neutrino dipole moments (Sec.~\ref{sec:neutrino:dipole}), neutrino anapole moments (Sec.~\ref{sec:neutrino:anapole}), and neutrino polarizability (Sec.~\ref{subsec:toymodel}), including possible enhancements. 

\subsection{Neutrino dipole moments}
\label{sec:neutrino:dipole}
\begin{figure}
 	\centering
 	\includegraphics[width=0.4\linewidth]{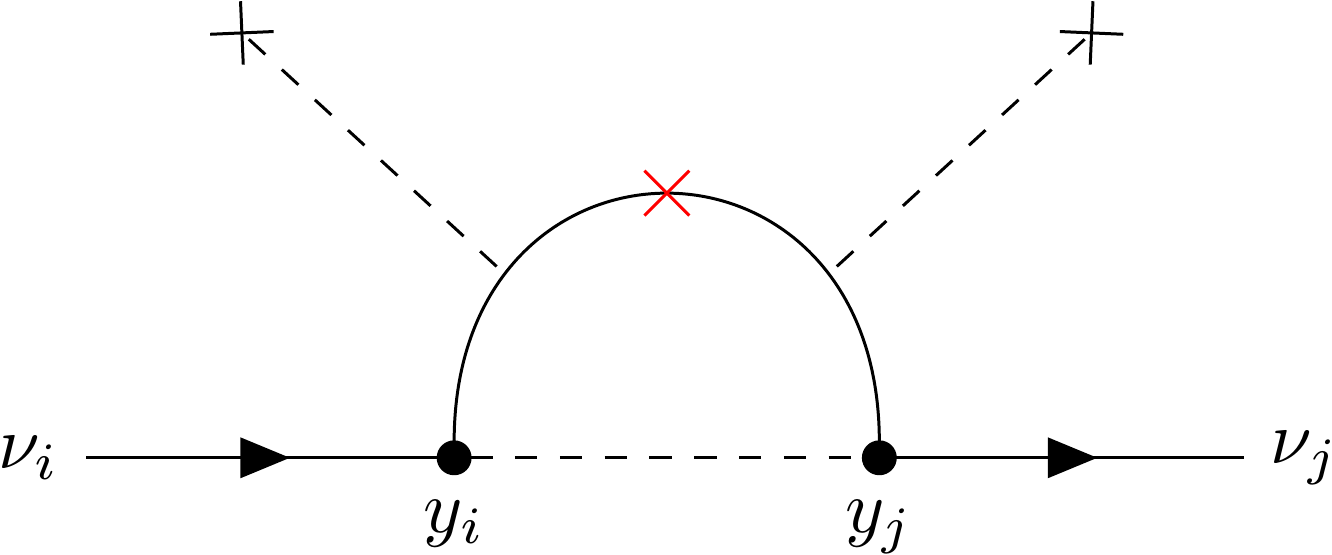}\hspace{0.5cm}
 	 	\includegraphics[width=0.4\linewidth]{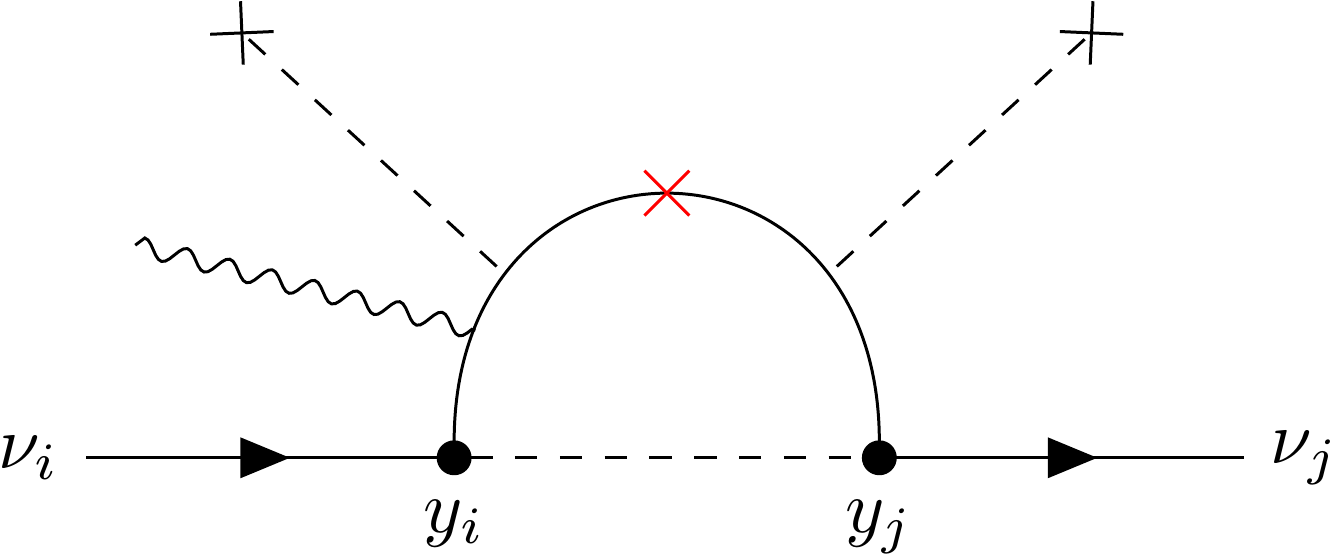}\vspace{0.5cm}
 	 	 	\includegraphics[width=0.4\linewidth]{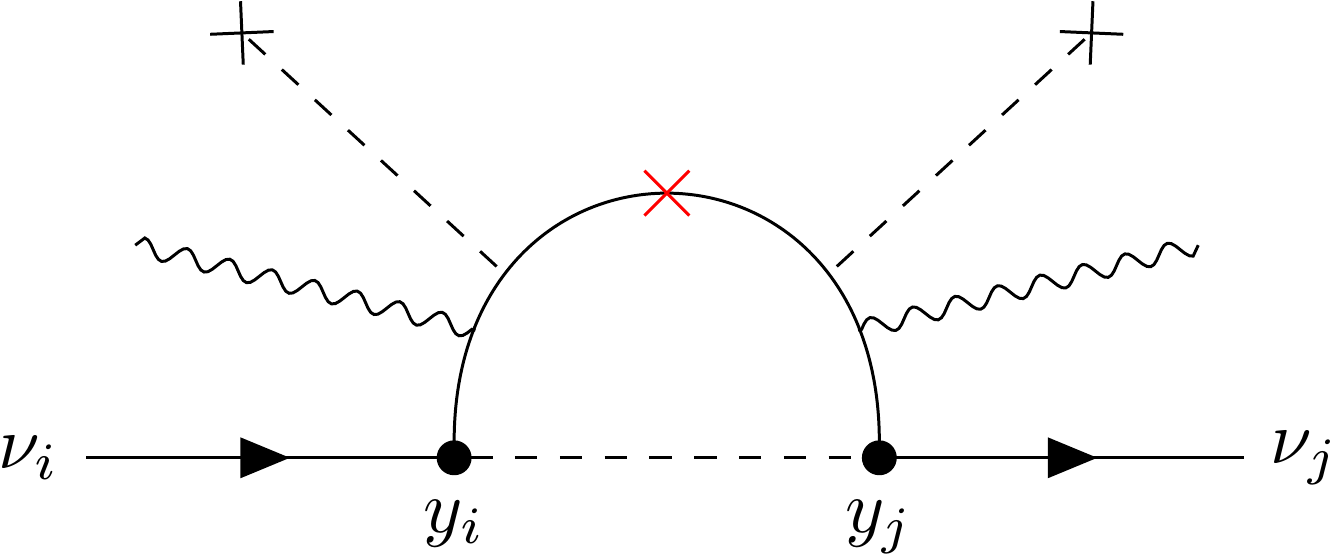}
 	\caption{Sample one loop contributions to the neutrino mass matrix (top left), dipole moments (top right) and polarizability (bottom). The Higgs vev insertions are denoted with black cross, while the $\Delta L=2$ mass insertion by a red cross. }
 	\label{fig:neutrino:riadiative}
 \end{figure}

The neutrino  dipole moments are tightly constrained from the searches for  solar neutrino scatterings on electrons by Borexino, which gives at 90\%CL $\mu_\nu^{\rm eff}<2.8 \cdot 10^{-11} \mu_B$ \cite{Borexino:2017fbd}, 
where the $\mu_\nu^{\rm eff}$ is a linear combination of magnetic moments that depends on flavor composition of neutrino flux on Earth, for details see Refs. \cite{Borexino:2017fbd,XENON:2020rca}, and also Appendix \ref{sec:app:notations}. Interperting both measurements as bounds and taking $\C_{1,ij}^{(5)}=1$, this translates to $\Lambda\gtrsim 10^6$ GeV~\cite{Altmannshofer:2018xyo}. While the bound on $\Lambda$ is impressive, it is useful to compare it with the typical sizes of neutrino masses,
\beq\label{eq:mass:nu}
{\cal L}\supset - \frac{1}{2}(m_\nu)_{ij} \bar \nu_i P_L \nu_j +{\rm h.c.}.
\eeq
For concreteness let us assume that the neutrino dipole moments are generated at one loop, so that parametrically
\beq
\label{eq:mnu:scaling}
(\lambda_\nu)_{ij} \mu_B \equiv  \frac{\C_{1,ij}^{(5)}}{\Lambda} \frac{e}{4\pi^2} \sim e \frac{y_i y_j}{16\pi^2} \frac{v^2}{M^3}=2.8 \times 10^{-11}\,\mu_B \frac{y_i y_j}{\big(M/2.4 {\rm~TeV}\big)^3} , 
\eeq
where $M$ is the typical mass of new physics particles in the loop, $y_i$ their couplings to neutrinos, and we included two insertions of the Higgs electroweak vev, $v=246$ GeV, as required to project out only the neutrino part of the electroweak leptonic doublet. Sample diagrams for the one loop radiative corrections are shown in Fig.~\ref{fig:neutrino:riadiative} (top right). Generically, the same loop, but without attached photon, Fig.~\ref{fig:neutrino:riadiative} (top left), will also contribute to the neutrino masses
\begin{align}
\label{mnu:scaling}
m_\nu&\sim \frac{y_i y_j}{16\pi^2} \frac{v^2}{M}=0.05 {\rm~eV}\, \frac{y_i y_j}{\big(M/7.7 \cdot 10^9 {\rm~TeV}\big)}.
\end{align}
In both \eqref{eq:mnu:scaling} and \eqref{mnu:scaling} we assumed that the $\Delta L=2$ mass insertion, denoted with red cross in Fig.~\ref{fig:neutrino:riadiative},  is of the same size as the typical mass $M$ of the new particles.
Comparing \eqref{eq:mnu:scaling} with the experimental bound, $\mu_\nu^{\rm eff}<2.8 \cdot 10^{-11} \mu_B$  \cite{Borexino:2017fbd}, shows that the neutrino magnetic moments can be large enough to be observed in the near future only if the related radiative corrections to the neutrino masses are suppressed below the generic expectations given by Eq.~\eqref{mnu:scaling}. 

Such a suppression of the neutrino masses is possible due to the {\it Voloshin mechanism} \cite{Voloshin:1987qy}, i.e., exploiting the fact that the operator $\bar \nu_i^c  \sigma^{\mu \nu} P_L \nu_j$ is antisymmetric, while $\bar \nu_i^c  \nu_j$ is symmetric under the exchange of flavor indices. An explicit realization are models with approximate  horizontal $SU(2)_H$ symmetry, in which $(\nu_e, \nu_\mu)$ form a doublet of $SU(2)_H$, while $\nu_\tau$ is a singlet \cite{Babu:2020ivd}. The $SU(2)_H$ allows a nonzero magnetic dipole term, $ \bar \nu_i  \sigma^{\mu \nu} P_L \nu_j\epsilon^{ij}$, which is a singlet under the horizontal symmetry. The neutrino mass terms vanish in the limit of unbroken $SU(2)_H$, $ \bar \nu_i   \nu_j\epsilon^{ij}=0$, due to the symmetric nature of the mass term. The neutrino masses are thus proportional to the charged lepton masses that break the $SU(2)_H$, giving rise to small enough neutrino masses without tuning, while  neutrino magnetic moments can be observably large \cite{Babu:1989wn,Babu:2020ivd,Babu:1990wv,Leurer:1989hx}. The Voloshin mechanism can be applied also to the transition dipole moments to  sterile neutrinos \cite{Brdar:2020quo}.

\subsection{Neutrino anapole moments}
\label{sec:neutrino:anapole}
The anapole or toroidal moment of the neutrino is represented by a dimension 6 operator 
\beq
\
{\cal L}_{\rm EFT} = \sum_{i,j}\frac{\C_{F,ij}^{(6)}}{\Lambda^2} \bar \nu_i \gamma^\mu \gamma_5 \nu_j \partial^\nu F_{\mu\nu},
\eeq
which has no classical analogue in the mulitipole expansion. The operator breaks both charge, $C$, and parity, $P$, but conserves the time-reversal symmetry $T$. This is immediately apparent in the non-relativistic limit, where the interaction Hamiltonian is ${\cal H}_{a} \sim \vec \sigma\cdot \vec J_{\rm em}$. The anapole moment was first proposed by Zeldovich in 1958 \cite{Zeldovich:1958} and can be viewed as the direct interaction between neutrino and the electromagnetic current 
\beq
 \partial^\nu F_{\mu\nu}= J_\mu^{\rm em} =\sum_f e Q_f \bar f \gamma_\mu f.
\eeq
Here, the sum runs over the SM fermions with charges $Q_f$ and mass smaller then the scale $\mu\lesssim 2$ GeV, at which we define the EFT. The anapole moment operator does not lead to an emission of a propagating photon, but rather to a short range interaction described by dimension 6 four fermion operators
\beq
\label{eq:anapole:4fermion}
{\cal L}_{\rm EFT} = \sum_{i,j,f} \frac{e Q_f \C_{F,ij}^{(6)}}{\Lambda^2} \, \bar \nu_i \gamma^\mu \gamma_5 \nu_j \, \bar f \gamma_\mu f.
\eeq
That is, the anapole operator can be replaced by the sum over four-fermion operators. 
We refer the interested reader to Ref. \cite{Altmannshofer:2018xyo} for the discussion of the phenomenology of non-standard neutrino interactions due to such point-like four-fermion interactions. 

The anapole moment of the neutrino is related to the neutrino charge radius \cite{Degrassi:1989ip}.  Defining the effective electromagnetic form factor of the neutrino by the relation $\langle \nu_i|J_\mu^{\rm em}|\nu_j\rangle =F_{1,ij}(q^2) \bar u_i\gamma_\mu P_L u_j+\ldots$, where we do not display the $F_2$ term, the neutrino charge is $F_1(0)=0$, while its effective mean-square charge radius is
\beq
\langle r^2\rangle_{ij} =6\frac{\partial F_{1,ij}(q^2)}{\partial q^2}\Big|_{q^2=0}.
\eeq
Evaluating the single photon exchange contribution to the scattering of charged SM fermions on neutrinos, the $q^2$ factor in the $F_1'(0) q^2$ term cancels the $1/q^2$ pole, and results in  a contact contribution of the form \eqref{eq:anapole:4fermion}. The neutrino charge radius is therefore directly proportional to the neutrino anapole moment
\beq
\langle r^2\rangle_{ij}=6 \frac{\C_{F,ij}^{(6)}}{\Lambda^2}.
\eeq

\subsection{Light scalar mediator model for enhanced neutrino polarizability}
\label{subsec:toymodel}
In generic new physics models the neutrino polarizability will be highly suppressed. For instance, if the dimension 7 Rayleigh operators in \eqref{eq:L:EFT} result from heavy particles running in a loop, Fig.~\ref{fig:neutrino:riadiative} (bottom),  and if we assume that the neutrino masses are dominated by a similar loop without photons attached, Fig.~\ref{fig:neutrino:riadiative} (top left), the NDA expectation is 
\beq
\label{eq:polarizability:NDA}
\frac{\C_{1(2),ij}^{(7)}}{\Lambda^3}\sim  y_i y_j\frac{v^2}{M^5}\sim \frac{m_\nu}{M^4},
\eeq
where in the last estimate we used the relation \eqref{mnu:scaling}. The searches for new charged particles at the LEP and LHC requires $M\gtrsim {\mathcal O}(100~{\rm GeV})$.
This gives an NDA estimate for the neutrino polarizability that is orders of magnitudes below the present and future experimental sensitivities, see Table \ref{table:bounds_summary_EFT}.

 \begin{figure}
	\centering
	\includegraphics[width=0.36\linewidth]{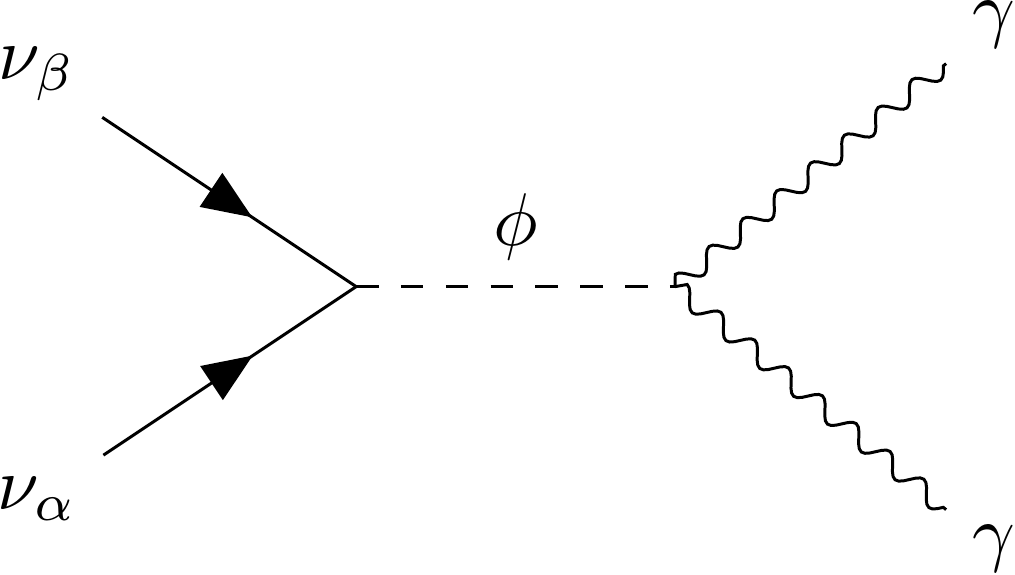}
	\caption{The tree-level $\phi$ exchange that leads to neutrino polarizability once $\phi$ is integrated out. 
	}
	\label{fig:phi_exchange}
\end{figure}

The crucial assumption in the NDA estimate \eqref{eq:polarizability:NDA} was that all the relevant new physics is heavy. If this is not the case, the
effective scale $\Lambda$ can be significantly lower \cite{Paz:2020pbc}. A simple example is a model with a light scalar mediator, $\phi$, that couples to neutrinos, and, through dimension 5 operators, also to photons, 
\beq\label{eq:lagr:phi}
{\cal L}_{\rm int}\supset 
- \frac{\alpha}{12\pi}  \frac{c_\gamma}{f_\phi} \phi F_{\mu\nu} F^{\mu\nu}
 - \frac{\alpha}{8\pi} \frac{c_\gamma'}{f_\phi} \phi  F_{\mu\nu}\tilde  F^{\mu\nu} + \frac{1}{2}c_\nu^{ij} \lp\bar\nu_iP_L\nu_j\rp\phi +{\rm h.c.}.
\eeq
Here, $c_{\gamma}, c_\gamma'$ and $c_\nu^{ij}$ are dimensionless couplings, while $f_\phi$ is the UV scale (for a pNGB $\phi$ the $f_\phi$  is related to the scale of spontaneous symmetry breaking, see Section \ref{sec:model}).  In general UV models we expect, $c_{\gamma,\gamma'}\sim {\mathcal O}(1)$ and $c_\nu\propto m_\nu$, the latter suppressed either by $f_\phi$ or some other dimensionful scale. 

If the scalar mediator is heavy enough to be integrated out, the tree level exchange of $\phi$ generates the Rayleigh operators in \eqref{eq:L:EFT}, cf. Fig.~\ref{fig:phi_exchange},
\beq\label{eq:matching}
\frac{\C_{1,ij}^{(7)}}{\Lambda^3}=c_\nu^{ij} c_\gamma \frac{1}{m_\phi^2 f_\phi}, \qquad \frac{\C_{2,ij}^{(7)}}{\Lambda^3}=c_\nu^{ij} c_\gamma' \frac{1}{m_\phi^2 f_\phi}.
\eeq
For light mediator the effective scale $\Lambda$ is thus parametrically smaller then the UV scale $f_\phi$. For instance, taking $c_\nu^{ij}=10^{-3}$ and setting the other dimensionless couplings to 1,  we have $\Lambda=\{1\,{\rm GeV}, 10\,{\rm MeV}, 100\,{\rm keV}\}$ for $m_\phi=\{1\,{\rm MeV}, 1\,{\rm keV}, 1\,{\rm eV}\}$ and $f_\phi=1$\,TeV.
Whether or not the EFT description can be used for a particular process depends on the typical energy and momentum exchange. These range from eV to GeV for the observables we take into account in the phenomenological analysis, as listed in Table \ref{table:bounds_summary_EFT}.

\begin{table}
\centering
\begin{tabular}{c c c c} 
 \hline\hline
 Process & ${\cal C}_2^{(7)}/\Lambda^3$ (GeV$^{-3}$) & ~EFT thr. (GeV)~ &  Sec.  \\ 
 \hline
 BBN & $-$ & $\sim 10^{-3}$  & \ref{sec:BBN}  
 \\ 
 $\nu$ decay & $1.2\times10^{11}$ & $\sim10^{-10}$ &  \ref{sec:NeutrinoDecay} 
 \\ 
  $\nu$ self-interaction & $-$ & $\sim10^{-6}$ &  \ref{sec:planck}  
 \\ 
 HB star & $1.9\times10^{6}$ & $\sim10^{-5}$ &  \ref{sec:HBstars} 
   \\ 
 SN1987a & $-$ & $\sim10^{-1}$ &  \ref{sec:SN} 
 \\ 
 Borexino & $1.5\times10^{3}$ & $\sim10^{-4}$ &  \ref{sec:Borexino}
  \\ 
 Xenon-nT & $0.5\times10^{3}$ & $\sim10^{-4}$ &  \ref{sec:XenonnT} 
 \\ 
 MiniBoone & $4\times10^{-3}$ & $\sim 1$ &  \ref{sec:MiniBoone} 
 \\ 
 BaBar & $0.2$ & $\sim 10$ &  \ref{sec:ColliderConstraints} 
 \\ 
 $\pi^0\to\g\g\to\nu\nu$ & $4.7\times10^{3}$ & $\sim0.1$ &  \ref{sec:ColliderConstraints} 
 \\ 
  $B^0\to\g\g\to\nu\nu$ & $3.7\times10^{4}$ & $\sim5$ &  \ref{sec:ColliderConstraints} 
  \\
  \hline
  $h\to\g\g\to\nu\nu$ & $1.2$ & $\sim10^2$ &  \ref{sec:ColliderConstraints} 
 \\ 
 \hline\hline
\end{tabular}
\caption{Summary of bounds (2nd column) on the neutrino polarizability Wilson coefficient ${\cal C}_2^{(7)}/\Lambda^3$, assuming flavor universality, Eq. \eqref{eq:C12:flav:diag}. The bounds from processes list in the 1st column were obtained under the assumption that the EFT framework \eqref{eq:L:EFT}  applies, i.e., that the mediators are heavier than the process dependent typical energy and momentum exchanges (3rd column). Further details can be found in sections listed in the fourth column. For  the $h\to\g\g\to\nu\nu$ decay (last row) the bound is on ${\cal C}_1^{(7)}/\Lambda^3$. 
}\label{table:bounds_summary_EFT}
\end{table}

\begin{figure}
	\centering
	\includegraphics[width=0.48\linewidth]{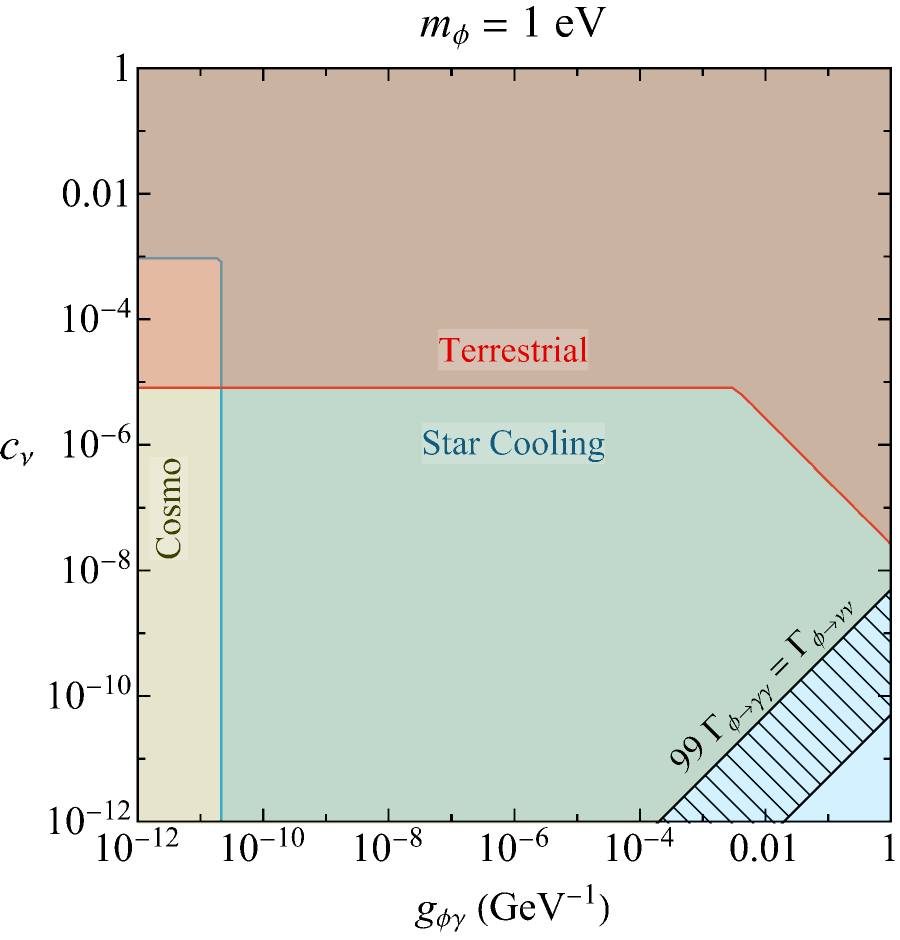}
	\quad
	\includegraphics[width=0.48\linewidth]{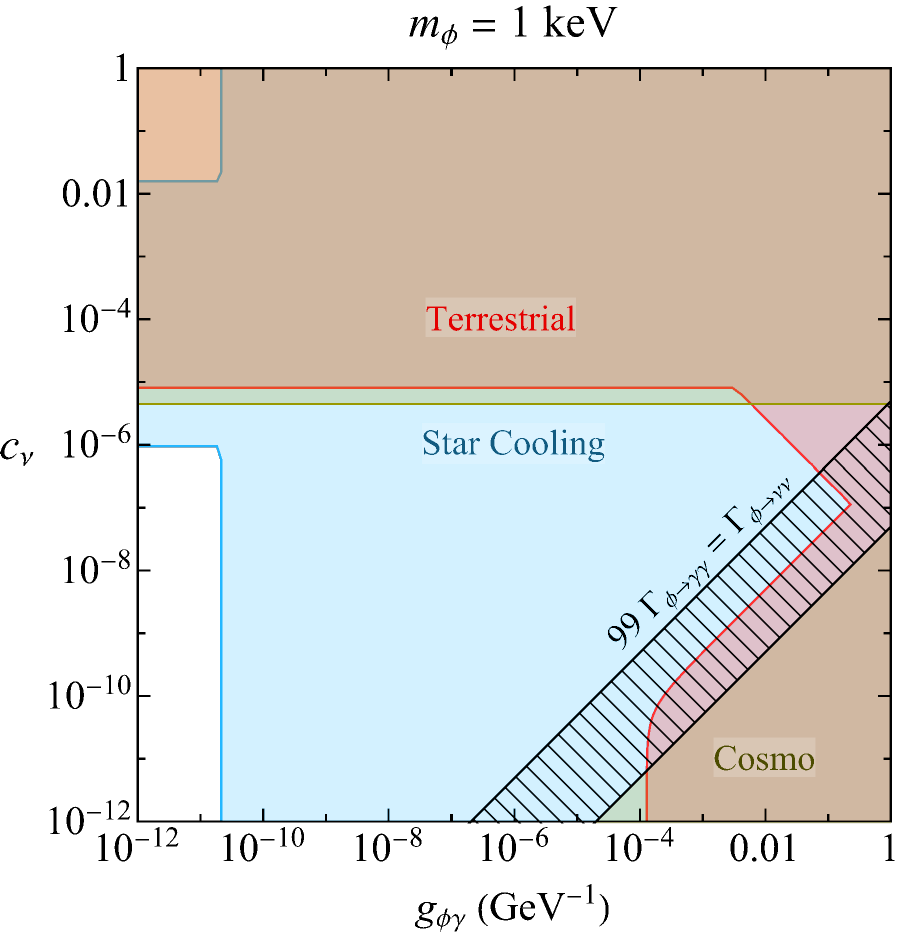}
	
	~

	\includegraphics[width=0.48\linewidth]{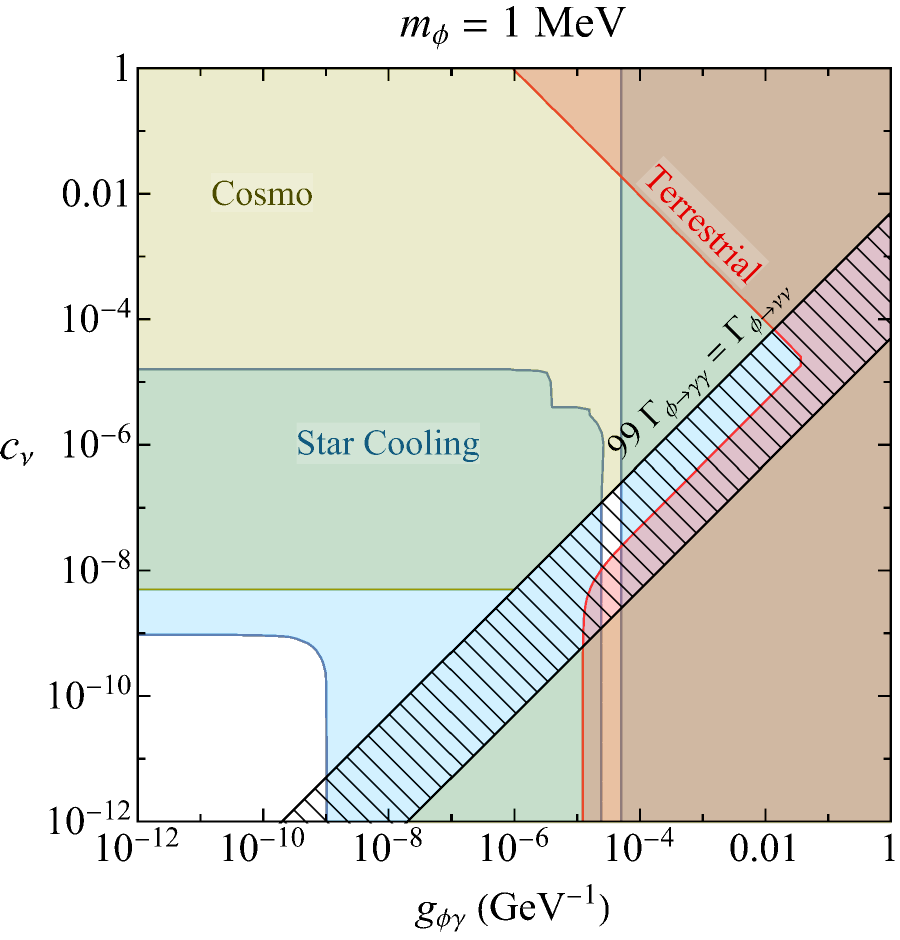}
	\quad
	\includegraphics[width=0.48\linewidth]{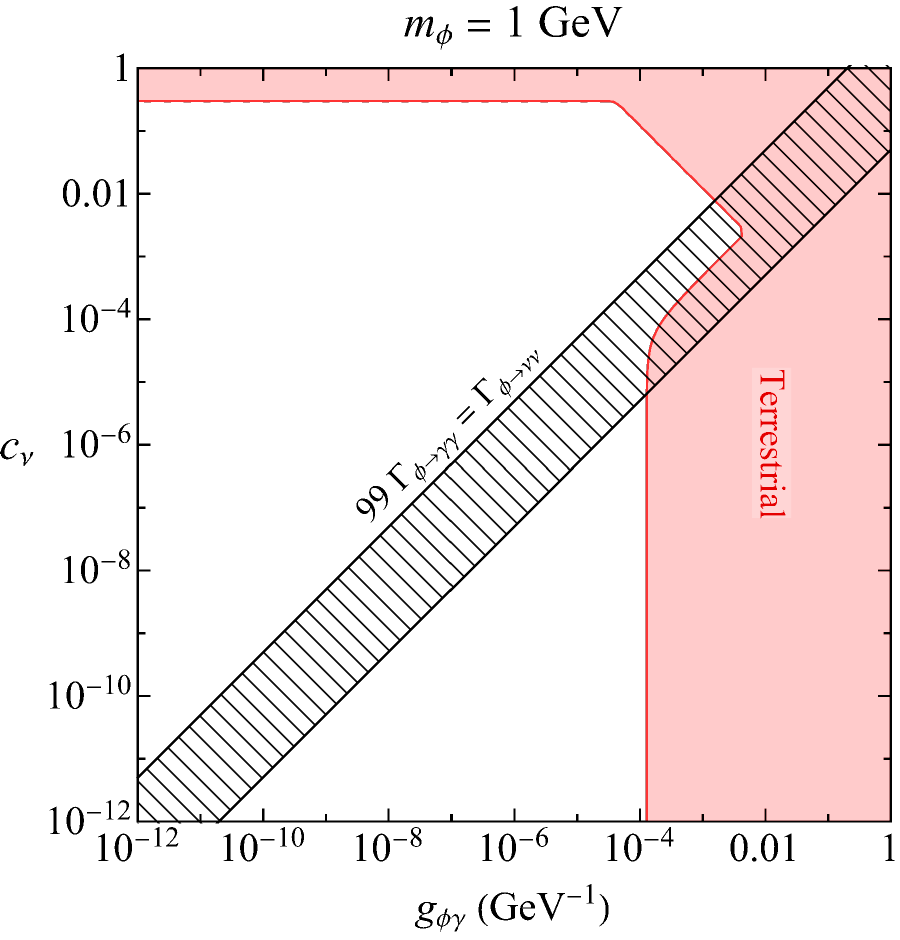}
	\caption{ Summary of the bounds on the pseudoscalar mediator model for enhanced neutrino polarizability, Eq. \eqref{eq:lagr:phi}, for four mass benchmarks, $m_\phi = 1\,$eV, $1\,$keV, $1\,$MeV, $1\,$GeV, as indicated. The green, blue and red regions indicate exclusion from cosmological (Sec.~\ref{sec:cosmology}), stellar cooling (Sec.~\ref{sec:StarCooling}), and terrestrial (Sec.~\ref{sec:NeutrinoScatteringExperiments}) constraints, respectively. 
	The transition region  between neutrino dominated (above) to photon dominated (below) bounds is shown as hatched, with the upper (lower) boundary corresponding to ${\cal B}(\phi \to \gamma\gamma)=1\%$ (${\cal B}(\phi \to \nu\nu)=1\%)$.
	}
	\label{fig:Summary:AllBounds}
\end{figure}

In the remainder of the paper we discuss different probes of neutrino polarizability, summarized in Fig.~\ref{fig:Summary:AllBounds} and in Tables~\ref{table:bounds_summary_EFT}, \ref{table:bounds_summary_PHI:ev,kev}, \ref{table:bounds_summary_PHI:mev,gev}.
We derive bounds both assuming an EFT, Eq.~\eqref{eq:L:EFT}, and assuming the scalar mediator model, Eq.~\eqref{eq:lagr:phi}, for four mass benchmarks, $m_\phi = 1$\,eV, 1\,keV, 1\,MeV, 1\,GeV. For easier comparison with the literature, we take in the numerical analysis the couplings to neutrinos to be flavor universal, 
\beq
\label{eq:cnu}
c_\nu^{ij}= i c_\nu \delta_{ij},
\eeq
 set $c_\g = 0$, and keep $c_\gamma'\ne0$. For such purely pseudoscalar $\phi$, we adopt the commonly used notation  
\beq
\label{eq:gphigamma} 
g_{\phi\gamma} \equiv \frac{\alpha}{2\pi}\frac{c_\gamma'}{f_\phi}.
\eeq

Most of the experimental probes we consider are not sensitive to photon polarization and do not distinguish $c_\g$ from $c_\g'$. In fact, most of the phenomenology is governed by the partial decay widths for $\phi$ to photons or neutrinos,
\begin{align}
\Gamma(\phi\to\gamma\gamma)&=\lp\frac{\alpha}{8\pi}\rp^2\frac{m_\phi^3}{4\pi f_\phi^2} \biggr[\frac{4}{9}(c_\g)^2+(c_\g')^2\biggr] \,,\label{eq:phidecayrate:gg}
\\
\Gamma(\phi\to \nu\nu) &=\frac{m_\phi}{32\pi} \sum_{ij}\big|c_\nu^{ij}\big|^2   \,.\label{eq:phidecayrate:nunu}
\end{align} 
  
Numerically, the ratio of $\phi\to \gamma\gamma$ and $\phi\to \nu\nu$ branching fractions is given by
\beq\label{eq:Brfraction}
\frac{{\cal B}(\phi\to\gamma\gamma)}{{\cal B}(\phi\to \nu\nu)}=6.7\times10^{-19}\times \biggr(\frac{m_\phi}{{\rm MeV}}\biggr)^2 \lp \frac{{\rm TeV}}{f_\phi} \rp^2 \lp\frac{(2c_\g/3)^2 + c_\g^{'2}}{\big|c_\nu^{ij}\big|^2}\rp,
\eeq
For comparable values of $c_{\gamma, \gamma'}$ and $c_\nu$, with $f_\phi$ in the TeV regime therefore $\phi$ predominantly decays to neutrinos. This is, for instance, a typical situation for the enhanced neutrino polarizability model in Sec.~\ref{sec:enhanced:polariz:model}, in which $c_\nu\sim m_\nu/f_\phi'$ with $f_\phi'\ll f_\phi$. 
For the modified majoron model in Sec.~\ref{sec:Majoron:inverse:see-saw}, on the other hand, $c_\nu\sim m_\nu/f_\phi$ so that $c_\nu\sim 10^{-13}$ for $f_\phi\sim 1\,$TeV, and thus the decays to photons dominate. In the phenomenological analysis in Sections~\ref{sec:cosmology} to~\ref{sec:ColliderConstraints} we treat  $c_\gamma'$ and $c_\nu$ as free parameters (assuming flavor universal neutrino couplings), and set $c_\gamma=0$ (except for bounds from Higgs decays, see below). For the parameters used in the numerical analysis, Eqs.~\eqref{eq:cnu}, \eqref{eq:gphigamma}, the ratio of branching fractions is given by 
\beq\label{eq:Brfraction:ggamma}
\frac{{\cal B}(\phi\to\gamma\gamma)}{{\cal B}(\phi\to \nu\nu)}=0.17\times \biggr(\frac{m_\phi}{{\rm MeV}}\biggr)^2 \lp \frac{g_{\phi\gamma}}{10^{-4}\, \text{GeV}^{-1}} \rp^2 \lp\frac{10^{-7}}{ c_\nu}\rp^{2}.
\eeq

\begin{table}[t]
\renewcommand*{\arraystretch}{1.2}
\centering
\begin{tabular}{ c c c c } 
 \hline\hline
 Process & ~ $m_\phi = $ eV ~ & ~ $m_\phi = $ keV ~  & Sec.  \\  
 \hline
 BBN & $c_\nu\lesssim 4\times10^{-5}$ & $c_\nu\lesssim 4.4\times10^{-6}$ & \ref{sec:BBN}  
 
 \\ 
  $\nu$ self-interaction & $c_\nu<2.8\times10^{-7}$ & $c_\nu<2.8\times10^{-4}$ & \ref{sec:planck}  
 \\
 HB star & $g_{\phi\g}\in [3.5\times10^{-3},10^{-11}]$ & $g_{\phi\g}\in [3.3\times10^{-3},10^{-11}]$ & \ref{sec:HBstars} 
  \\ \hline
   \multirow{2}{4em}{SN1987a} & $g_{\phi\g}\in[10^{-2},5\times10^{-6}] $ & $g_{\phi\g}\in[10^{-2},5\times10^{-6}] $ & \multirow{2}{2em}{\ref{sec:SN} }
 \\ 
  & $c_\nu\in[10^{-3}, 1]$ & $c_\nu\in[10^{-6}, 10^{-2}]$ & 
 \\ \hline
 Borexino & $c_\nu g_{\phi\g}<5.3\times10^{-8}$ & $c_\nu g_{\phi\g}<5.3\times10^{-8}$ & \ref{sec:Borexino}
  \\  
 Xenon-nT & $c_\nu g_{\phi\g}<2.5\times10^{-8}$ & $c_\nu g_{\phi\g}<2.5\times10^{-8}$ & \ref{sec:XenonnT} 
 \\  
 MiniBoone & $c_\nu g_{\phi\g}<4\times10^{-6}$ & $c_\nu g_{\phi\g}<4\times10^{-6}$ & \ref{sec:MiniBoone} 
 \\ 
 $M/\tau$ rare dec. & $c_\nu <4\times10^{-3}$ & $c_\nu <4\times10^{-3}$ & \ref{sec:ColliderConstraints} 
 \\ 
  $0\nu2\beta$ & $c_\nu <8\times10^{-6}$ & $c_\nu <8\times10^{-6}$ & \ref{sec:ColliderConstraints} 
 \\ 
   Beam dump & $-$ & $g_{\phi\g} <10^{-2}$ & \ref{sec:ColliderConstraints} 
 \\ 
   $e^+e^-\to3\g$ & $-$ & $-$ & \ref{sec:ColliderConstraints} 
 \\ 
 $\pi^0\to\g\g\to\nu\nu$ & $c_\nu g_{\phi\g}< 2\times10^{-2}$ & $c_\nu g_{\phi\g}< 2\times10^{-2}$ & \ref{sec:ColliderConstraints}
 \\ 
  $B^0\to\g\g\to\nu\nu$ & $c_\nu g_{\phi\g} < 180$ & $c_\nu g_{\phi\g} < 180$ & \ref{sec:ColliderConstraints}
 \\ 
 BaBar & $ g_{\phi\g}<1.5\times10^{-4}$ & $g_{\phi\g}<1.5\times10^{-4}$ &  \ref{sec:ColliderConstraints}
 \\
 \hline 
  $h\to\g\g\to\nu\nu$ & $c_\nu g_{\phi\g} < 2.4$ & $c_\nu g_{\phi\g} < 2.4$ & \ref{sec:ColliderConstraints}
  \\
 \hline\hline
\end{tabular}
\caption{The bounds on flavor universal pseudoscalar mediator couplings  to neutrinos, $c_\nu$, Eq. \eqref{eq:cnu}, and photons, $g_{\phi\g}$ (in GeV$^{-1}$), Eq. \eqref{eq:gphigamma}, or the product of the two, for various processes (1st column), for two mass benchmarks, $m_\phi = 1\,$eV, $1\,$keV (2nd and 3rd columns). Further details are given in sections listed in the last column. The BBN bounds also require $\Gamma_{\phi\to\gamma\gamma} \ll \Gamma_{\phi\to\nu\nu}$, while the quoted SN bounds refer to the regions where one of the coupling dominates, either $c_\nu \gg g_{\phi\g}$ or $g_{\phi\g} \gg c_\nu$ (for intermediate region see the main text). When the mass benchmark is in the EFT regime for the corresponding process, we use Eq.~\eqref{eq:matching} for the matching.  
}
\label{table:bounds_summary_PHI:ev,kev}
\end{table}

\begin{table}[h!]
\renewcommand*{\arraystretch}{1.2}
\centering
\begin{tabular}{ c c c c } 
 \hline\hline
 Process &  ~ $m_\phi = $ MeV ~ & ~ $m_\phi = $ GeV ~ & Sec.  \\  
 \hline
 BBN & $c_\nu\lesssim 5\times10^{-9}$ & $-$ & \ref{sec:BBN}  
 \\ 
  $\nu$ self-interaction & $c_\nu<2.8\times10^{-1}$ & $c_\nu<2.8\times10^{2}$ & \ref{sec:planck}  
 \\
 HB star & $-$ & $-$ & \ref{sec:HBstars}
   \\ \hline
   \multirow{2}{4em}{SN1987a} & $g_{\phi\g}\in[10^{-2},5\times10^{-6}] $ & $-$ & \multirow{2}{2em}{\ref{sec:SN} }
 \\ 
   & $c_\nu\in[10^{-9}, 10^{-5}] $ & $-$ & 
 \\ \hline
 Borexino & $c_\nu g_{\phi\g}<2.5\times10^{-6}$ & $c_\nu g_{\phi\g}<1.8$ & \ref{sec:Borexino}
  \\ 
 Xenon-nT & $c_\nu g_{\phi\g}<9.2\times10^{-7}$ & $c_\nu g_{\phi\g}< 0.6$ & \ref{sec:XenonnT} 
 \\  
 MiniBoone & $c_\nu g_{\phi\g}<4\times10^{-6}$ & $c_\nu g_{\phi\g}<1.2\times10^{-5}$ & \ref{sec:MiniBoone} 
 \\ 
  $M/\tau$ rare dec. & $c_\nu <4\times10^{-3}$ & $c_\nu < 0.3$ & \ref{sec:ColliderConstraints} 
  \\
    $0\nu2\beta$ & $c_\nu <2\times10^{-5}$ & $-$ & \ref{sec:ColliderConstraints} 
 \\ 
    Beam dump & $g_{\phi\g} <10^{-5}$ & $-$ & \ref{sec:ColliderConstraints} 
 \\ 
  $e^+e^-\to3\g$ & $-$ & $g_{\phi\g} <10^{-2}$ & \ref{sec:ColliderConstraints} 
 \\ 
 $\pi^0\to\g\g\to\nu\nu$ & $c_\nu g_{\phi\g}< 2\times10^{-2}$ & $c_\nu g_{\phi\g}< 0.9$ & \ref{sec:ColliderConstraints} 
 \\ 
  $B^0\to\g\g\to\nu\nu$ & $c_\nu g_{\phi\g} < 180$ & $c_\nu g_{\phi\g} < 180$ & \ref{sec:ColliderConstraints} 
 \\ 
 BaBar & $g_{\phi\g}<1.5\times10^{-4}$ & $g_{\phi\g}<1.5\times10^{-4}$ &  \ref{sec:ColliderConstraints}
 \\
  \hline 
  $h\to\g\g\to\nu\nu$ & $c_\nu g_{\phi\g} < 2.4$ & $c_\nu g_{\phi\g} < 2.4$ & \ref{sec:ColliderConstraints}
  \\
 \hline\hline
\end{tabular}
\caption{Same as Table~\ref{table:bounds_summary_PHI:ev,kev}, but for mass benchmarks, $m_\phi = 1\,$MeV, $1\,$GeV.
}\label{table:bounds_summary_PHI:mev,gev}
\end{table}

\section{Cosmological constraints}
\label{sec:cosmology}

 It is well known that the precision cosmological data impose some of the strongest constraints on the light mediator models, such as ALPs and majoron models \cite{Cadamuro:2011fd,Millea:2015qra,Depta:2020wmr,Blinov:2019gcj}. 
  These constraints come from a variety of cosmological measurements including those from the measurements of the CMB and the abundances of heavier nuclei. In this section, we apply such  constraints to the case where the light mediator can couple to both photons and neutrinos, Eq. \eqref{eq:lagr:phi}.

 The four most relevant processes that determine the evolution history of $\phi$ are pair annihilation of neutrinos ($\nu\nu \to \phi \phi$), neutrino coalescence ($\nu \nu \to \phi$), Primakoff conversion ($\gamma\to \phi$) and photon coalescence ($\gamma \gamma \to \phi$).
 The pair annihilation processes are dominant at early times when the number densities are large, whereas coalescence processes are dominant at temperature $T\sim m_\phi$.

 \begin{figure}
	\centering
	\includegraphics[width=0.45\linewidth]{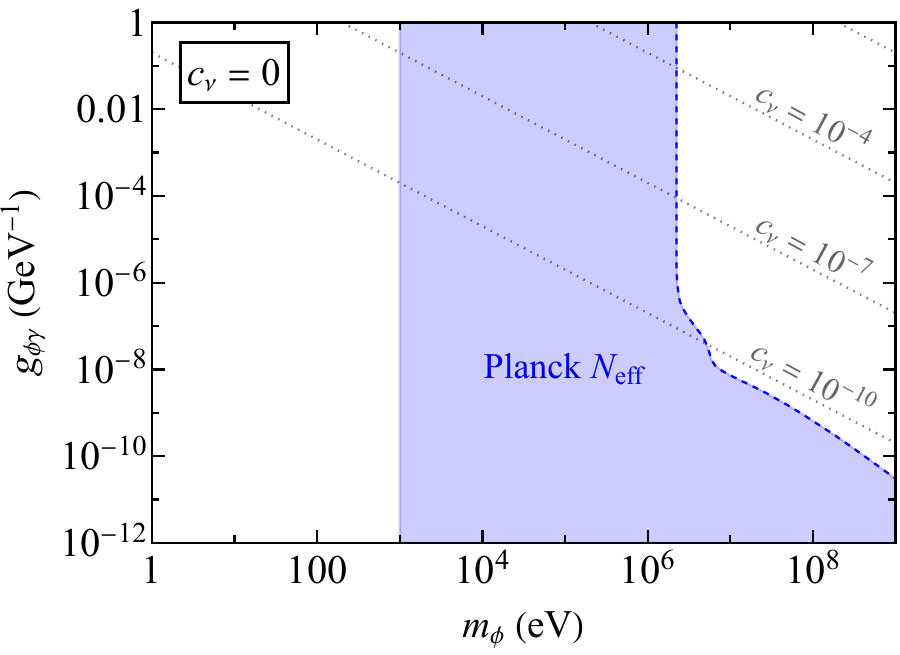}
	~~\quad
	\includegraphics[width=0.45\linewidth]{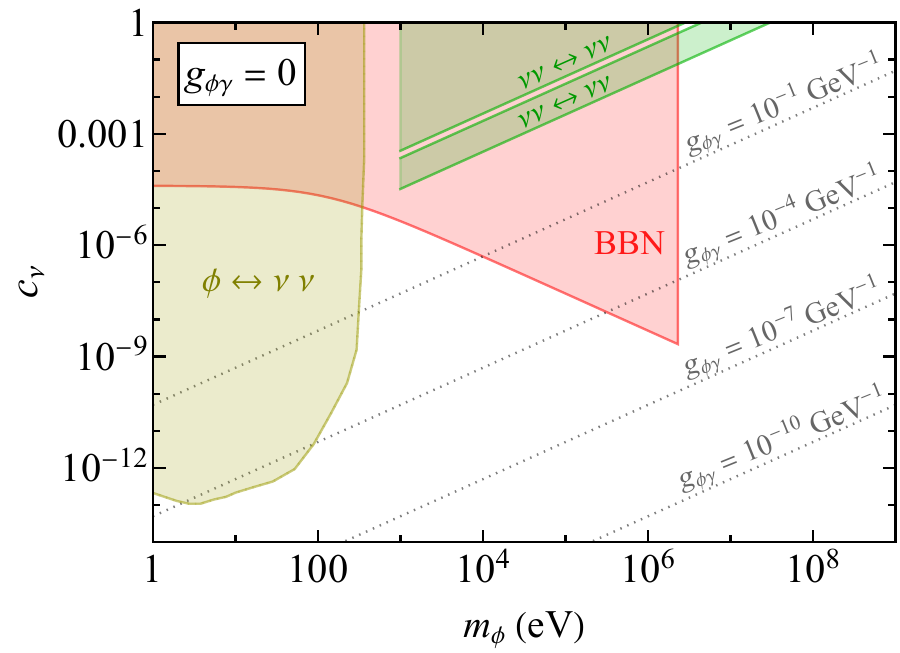}
	\caption{Left: Constraints on $g_{\phi\gamma}$ from Planck bounds on $N_{\rm eff}$ (blue), setting $c_\nu=0$. Right: Constraints on $c_\nu$ from BBN (red), $\nu$SI (green) and $\phi \leftrightarrow \nu \nu$ (brownish green) from Planck data, setting $g_{\phi\gamma}=0$.
	Dotted lines correspond to constant values of $c_\nu$ (left) and $g_{\phi\gamma}$ (right) for which $\Gamma (\phi \to \gamma \gamma) = \Gamma (\phi \to \nu \nu)$.
	}
	\label{fig:only_nu_or_gamma}
\end{figure}

 \begin{figure}
	\centering
	\includegraphics[width=0.48\linewidth]{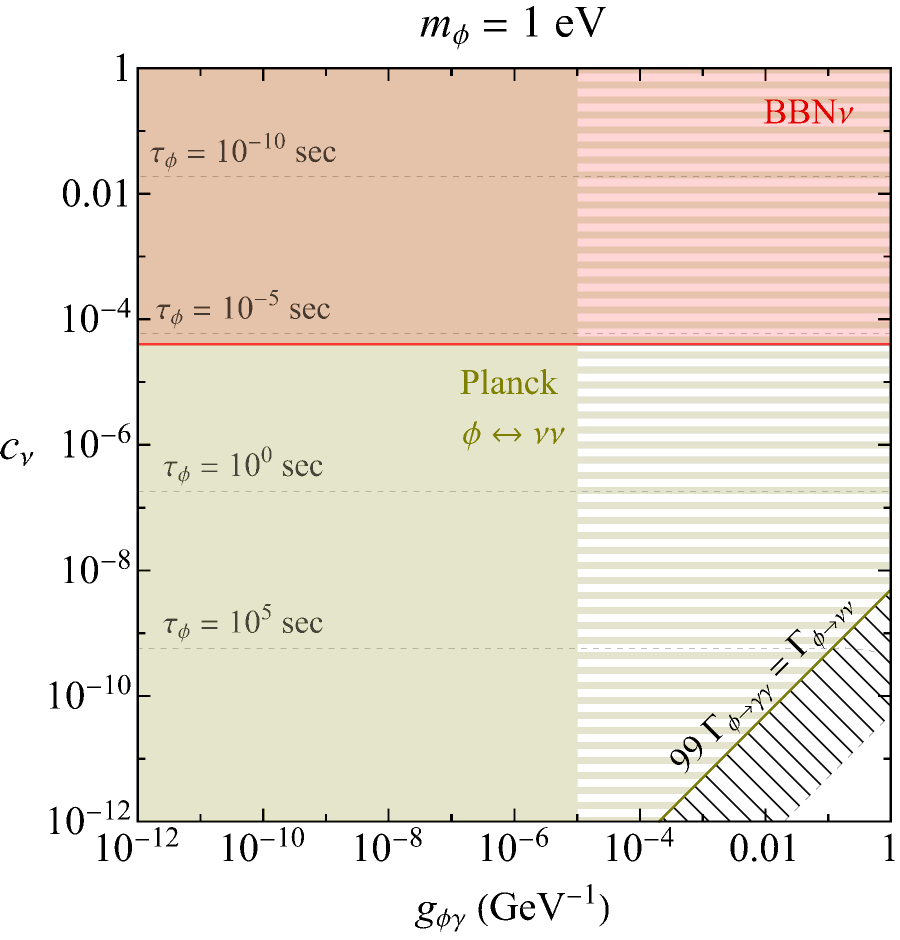}
	\quad
	\includegraphics[width=0.48\linewidth]{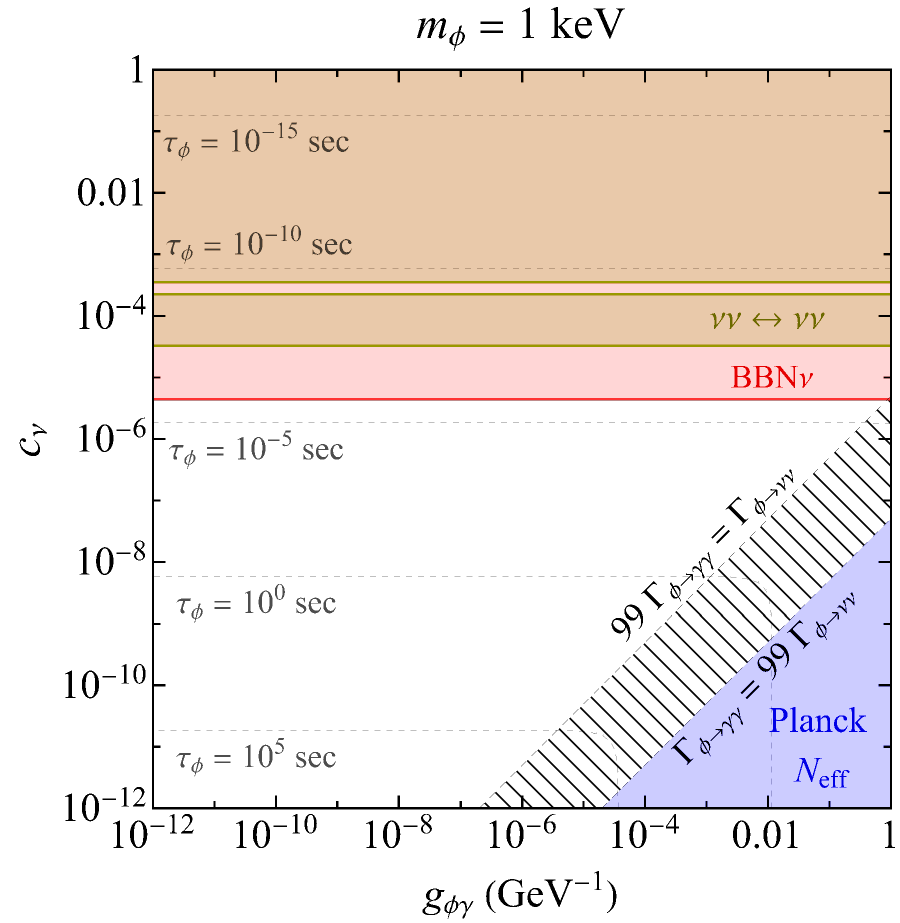}
	\includegraphics[width=0.48\linewidth]{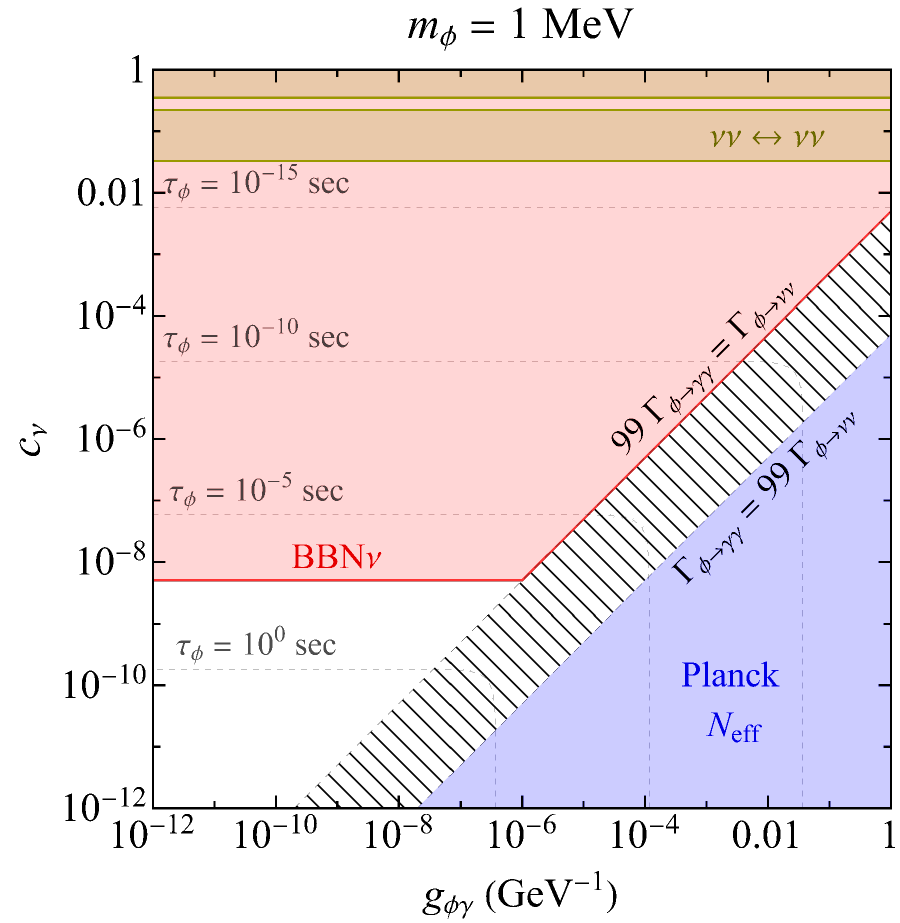}
	\caption{Constraints from cosmology on pseudoscalar mediator couplings to neutrinos, $c_\nu$, and photons, $g_{\phi\gamma}$, for three mass benchmarks, $m_\phi = 1$ eV, keV, MeV, as denoted (for the considered parameter ranges there are no cosmological bounds on $m_\phi = 1$ GeV). The red, blue and green regions are excluded from BBN (from $\nu-\phi$ coupling), from Plank measurement of $N_{\rm eff}$, and by CMB bounds, respectively. The brown bands, labeled $\nu\nu\to\nu\nu$, are excluded by neutrino self-scattering bounds, derived in the EFT regime, $m_\phi\gg100$ eV, see the text for details. The dotted lines show iso-countours of $\phi$ lifetimes, while the hatched transition region is defined as in Fig.~\ref{fig:Summary:AllBounds}.}
	\label{fig:CosmoConstraints}
\end{figure}

\subsection{Constraints from Planck} \label{sec:planck}
 
The decay of $\phi$ to photons and neutrinos can change the number of relativistic degrees of freedom in the Universe, $N_{\rm eff}$. If the decay happens before photon-neutrino decoupling, $ T \sim 1~\MeV$, the $\phi$ energy and entropy are quickly distributed between the different species present in the thermal bath; the standard cosmology scenario is thus preserved. However, if $\phi$ decays after decoupling, there is no energy and entropy exchange between neutrinos and photons. As a consequence, the value of \Neff{} during recombination can be different from the $\Lambda$CDM prediction, $N_{\rm eff}^{\rm SM} = 3.044$~\cite{Gariazzo:2019gyi}. The effect can be two-fold: the $\phi\to\nu\nu$ decays will increase \Neff{}, whereas the $\phi\to\g\g$ decays will decrease it. This happens because, as $\phi$ decays to photons, the photon temperature increases, while the effective neutrino energy density is diluted.

 The Planck measurement of CMB observables limits the allowed range of relativistic degrees of freedom to $N_{\rm eff} = 2.99^{+0.34}_{-0.33}$ at 95\% C.L.~\cite{Planck:2018vyg}. 
 If $\phi$ decays only to neutrinos, one finds $N_{\rm eff} = 3.57$ using conservation of entropy~\cite{Huang:2017egl}, while in the opposite regime, $\phi$ decaying only to photons, one finds $N_{\rm eff} = 2.4$~\cite{Cadamuro:2011fd}.
 However, the latter case presents additional complications. Firstly, the presence of \textit{extra photons} can also shift the time of matter-radiation equality. Secondly, if $\phi$ decays close to or after recombination, $z_{\rm rec} \sim 1089$, the standard assumptions of cosmology no longer hold. Lastly, additional complications may occur when photons keep $\phi$ in equilibrium, but $\phi$ predominantly decays to neutrinos. In these cases, the Planck bound on \Neff{} is not reliable and cannot be used. In this work we do not attempt to compute bounds in parts of the parameter space which can lead to these effects, and only employ Planck's \Neff~bounds when $\phi$ predominantly decays to neutrinos after decoupling from the thermal bath.

 Another effect that can change \Neff~is the re-thermalization of $\phi$. 
 A light $\phi$, $m_\phi \lesssim \rm{MeV}$, can thermalize again with photons (neutrinos) due to the inverse decay process, $\gamma\gamma\to\phi$ ($\nu\nu\to\phi$). 
 If this rethermalization occurs at $T \gtrsim m_\phi$, the abundance of $\phi$ will increase to its thermal equilibrium value. 
 However, as the Universe cools down, $\phi$ becomes non-relativistic at $T\sim m_\phi/3$ and decays out of equilibrium to photons (neutrinos). 
 As shown in \cite{Cadamuro:2011fd}, this effect is subdominant for ALPs that only couple to photons, we expect the same for $\phi$ as well.

In Fig.~\ref{fig:only_nu_or_gamma} (left), we show the constraints on $\phi$ coupling to photons (setting $c_\nu=0$) as a function of $\phi$ mass that were obtained for ALPs using Planck data  in Ref.~\cite{Depta:2020wmr} (see also Ref.~\cite{Millea:2015qra}). Note that there are no constraints for $m_\phi < 1\,\text{keV}$: for these masses $\phi$ decays to photons close to the time of recombination. For masses larger than a few MeV, $\phi$ decays before photon-neutrino decoupling and therefore only very small couplings are constrained. The dotted lines in Fig.~\ref{fig:only_nu_or_gamma} (left) show constant values of $c_\nu$ for which ${\cal B}(\phi \to \gamma\gamma)={\cal B}(\phi \to \nu\nu)$. The couplings to neutrinos need to be below these values for the $N_{\rm eff}$ bounds to apply unchanged.

We discuss next the CMB bounds on $\phi$ couplings to neutrinos. The $\nu - \phi$ coupling suppresses the neutrino anisotropic stress energy tensor, which leads to distortions in the CMB power spectrum. In the $\Lambda$CDM model, neutrinos are free-streaming particles and this description agrees very well with the Planck measurements. The new coupling with $\phi$ induces neutrino self-scattering ($\nu$SS), $\nu\nu\to\nu\nu$. Self-scattering neutrinos behave like a fluid, rather than free-streaming radiation, and thus leave their imprints in the CMB power spectrum. The processes that define the latter spectrum happen at the typical temperature $T\sim100$ eV, which defines two different $m_\phi$ regimes.

 For  $m_\phi \gg100$ eV, the flavor universal $\nu$SS mediated by $\phi$ is effectively described by a dimension 6 scalar operator
 \begin{equation}
     \mathcal{L}_{\nu{\rm SS}} \supset \frac{1}{4} \frac{c_\nu^2}{m_\phi^2} (\bar\nu_i P_L \nu_i)(\bar \nu_j P_L \nu_j)
     \equiv G_{\nu} (\bar \nu \nu) (\bar \nu \nu)\,,
 \end{equation}
where we already assumed that the process is flavor-universal. Fits to cosmological data find that $\log_{10} (G_\nu \times\rm{MeV}^2) < -3.57$ at 95\% C.L. or $\log_{10} (G_\nu\times \rm{MeV}^2) = -1.711 \pm 0.099$ at 68\% C.L.~\cite{Lancaster:2017ksf}. Stronger neutrino self-interactions are ruled out by terrestrial experiments: from meson decays, $\tau$ decays, and double beta decays~\cite{Blinov:2019gcj}. We show these bounds in the right panel of Figure~\ref{fig:only_nu_or_gamma}.

In the opposite regime, $m_\phi\ll100$ eV, $\phi$ is a relativistic degree of freedom at the CMB formation temperature and one needs to include $\nu-\phi$ interactions in the Boltzmann equation. Ref.~\cite{Escudero:2019gvw} computed such constraints on light $\phi$ from the Planck 2018 data, which we show as a green band in Fig.~\ref{fig:CosmoConstraints}. The green horizontally hatched region indicates the parts of the parameter space in which $\phi$ and $\g$ decouple around the time of recombination and the standard assumptions of cosmology may not hold.
 
Finally, we comment on the effect of the $\nu\nu\g\g$ interaction due to Rayleigh operators, i.e., for $m_\phi$ induced interactions, but in the EFT regime, cf. Eq.~\eqref{eq:matching}. The Rayleigh operators can keep neutrinos and photons in thermal equilibrium and modify the CMB power spectrum measured by Planck. As already mentioned, the relevant scale for CMB is $T \sim 100$~eV. Thus we need to estimate if thermal equilibrium can be achieved at $T \lesssim 100$~eV. The scattering rate for the process can be approximated as
 \begin{equation}
    \Gamma_{\nu\nu\gamma\gamma} 
    \sim\left(\frac{\alpha}{2\pi}\frac{\C_2^{(7)}}{\Lambda^3}\right)^2 T^7. 
 \end{equation}
At the decoupling temperature $T_{\rm fo}$, the scattering rate should satisfy the condition 
 \begin{equation}
     \frac{\Gamma_{\nu\nu\gamma\gamma}}{H} \lesssim 1 \quad \implies 
     \left(\frac{\alpha}{2\pi}\frac{\C_2^{(7)}}{\Lambda^3}\right)^2 T_{\rm fo}^5 M_{\rm Pl} \lesssim 1\,,
 \end{equation}
where $H$ is the Hubble rate at decoupling. 
Setting $T_{\rm fo} = 100 \rm{~eV}$ leads to the following bound on the neutrino polarizability operator
\beq
\label{eq:bound:Lambda:CMB}
\Lambda^3/\C_2^{(7)} \gtrsim (0.2\,\text{MeV})^3,
\eeq
or in terms of the $\phi$ mediator model,
\begin{equation}
\label{eq:bound:scalar:EFT:bound}
     c_\nu^2\, (g_{\phi\gamma}\times \text{GeV})^2 \left(  \frac{\rm{keV}}{m_\phi}\right)^4 \lesssim 10^{-7}\,.
\end{equation}
The bound in \eqref{eq:bound:Lambda:CMB} is model independent as long as the mediators generating the Rayleigh operator are heavier than about 100 eV. However, this EFT bound is also relatively weak. For instance, if the neutrino polarizability is induced by the pseudoscalar mediator,  Eq. \eqref{eq:lagr:phi}, the other cosmological bounds for $m_\phi > 100$ eV are more stringent  than Eq.  \eqref{eq:bound:scalar:EFT:bound},  cf.~Figs.~\ref{fig:Summary:AllBounds},~\ref{fig:only_nu_or_gamma}. The neutrino-photon interaction  in this case therefore freezes-out much before recombination and does not lead to any new constraint.
 
\subsection{Constraints from BBN}
\label{sec:BBN}
 
In order to estimate the impact of $\phi$ on the abundances of primordial elements produced during the BBN, we consider three parameter regimes. Below, we compute the BBN bounds for the case when decays to neutrinos dominate, by considering the extreme case of no coupling to photons, $g_{\phi\g}=0$. In the opposite regime, $c_\nu=0$, the bounds from Planck data dominate and the effects on BBN can be neglected~\cite{Depta:2020wmr}. The constraints in the intermediate regime, shown as the diagonally hatched regions in Figs.~\ref{fig:Summary:AllBounds} and \ref{fig:CosmoConstraints}, are more involved to estimate and go beyond the scope of this paper. The upper (lower) boundary of the intermediate regime region are defined by requiring that the neutrino (photon) channel accounts for 99\% of the total width.
 
The BBN bounds on $\phi-\nu$ coupling, in the limit of no couplings to photons, is shown in red color in Fig.~\ref{fig:CosmoConstraints}, indicated as ``BBN$\nu$'', and is obtained as follows. The abundance of $\phi$ during BBN can increase \Neff{}, which in turn modifies the expansion rate of the Universe and thus the abundance of heavy elements. The two processes that can keep $\phi$ in thermal equilibrium are the neutrino pair annihilation ($\nu \nu \to \phi \phi$) and the neutrino coalescence ($\nu\nu\to\phi$). 
 For very light $\phi$, $m_\phi \ll 1~\mev$, the pair annihilation process dominates and keeps $\phi$ in thermal equilibrium during BBN, whereas for $m_\phi \lesssim 1~\mev$, the inverse decay dominates.
The BBN bounds on $\phi-\nu$ coupling can be written as~\cite{Escudero:2019gvw}
\begin{equation}
c_\nu<\frac{1}{\left(5 \times 10^{-9} \frac{\mathrm{MeV}}{m_{\phi}}\right)^{-1}+\left(4 \times 10^{-5}\right)^{-1}}\,,
\end{equation}
where the two parts in the denominator come from the two aforementioned processes. The above bound is shown as excluded red regions in Fig.~\ref{fig:CosmoConstraints}.

Both contributions can be obtained by requiring that $\phi$ is not in thermal equilibrium at the photon-neutrino decoupling temperature, $T_D\sim 1~\MeV$, as this would otherwise result in $\Delta N_{\rm eff} \sim 0.5$ at the time of BBN. We then require
 \begin{eqnarray}
 	\frac{\Gamma_{\nu\nu\to\phi\phi} (T_D)}{H(T_D)} \lesssim 1 \implies c_{\nu}^4 \lesssim \frac{T_D}{M_{pl}} \sim 10^{-5}, \rm{~~~for~m_\phi \ll 1\mev}\,,
 \end{eqnarray}
while for $m_\phi \lesssim 1~\mev$,
\beq
 	\frac{\Gamma_{\nu \nu \to \phi}(T_D)}{H(T_D)} \sim \frac{c_{\nu}^2 m_\phi^2 M_{pl}} { T_{D}^3} \lesssim 1\implies c_\nu m_\phi \lesssim 10^{-10}~\mev.
\eeq

In the opposite limit ($c_\nu = 0$), where $\phi$ behaves like an ALP coupling to photons, Ref.~\cite{Depta:2020wmr} found that the BBN constraints on ALPs are weaker than those from the Planck data. Therefore, in the parameter regime where decays to photons dominate, we indicate in Fig.~\ref{fig:CosmoConstraints} only the Planck constraints.  

\subsection{Neutrino decay}
\label{sec:NeutrinoDecay}
 
The Rayleigh operators can induce the decay of a neutrino into a lighter mass eigenstate, along with two photons, $\nu_i \to \nu_j \gamma \gamma$, where $m_{\nu_i} > m_{\nu_j}$. 
The sum of the neutrinos masses is bounded from CMB Planck data to be $\sum_i{m_{\nu_i}} <  \rm{0.12~eV}$~(95\%~CL, Planck TT,TE,EE+lowE +lensing+BAO~\cite{Planck:2018vyg}).
Therefore, the photon energy spectrum will follow the typical 3-body decay distribution, with a maximum energy $\lesssim 0.1~\rm{eV}$. Depending on the time of their injection, these photons may leave their imprints in the CMB power spectrum measured by Planck~\cite{Planck:2018vyg} or the CMB blackbody spectrum measured by COBE/FIRAS~\cite{Fixsen:2009ug, Fixsen_2002}.

We assume for simplicity that the final state neutrino is massless, $m_{\nu_j} = 0$, and that the mediator is heavy, $m_\phi \gg m_{\nu_i}$, which is true for all the mass benchmarks considered. For our purposes it suffices to estimate the decay width using naive dimensional analysis,
\beq 
 \Gamma_{\nu_i \to \nu_j \gamma \gamma} \sim \frac{1}{8\pi}\lp \frac{g_{\phi\g} c_\nu^{ij}}{4 \pi} \rp^2 \frac{m_\nu^7}{m_\phi^4}\,,
 \eeq
which corresponds to a lifetime
 \beq 
 \tau_{\nu_i} \simeq 8
 \times 10^{17} \lp \frac{1}{c_\nu \lp g_{\phi\g}\cdot{\rm GeV}\rp} \rp^2 \lp \frac{m_\phi}{{\rm keV}} \rp^4 \lp \frac{0.1~{\rm eV}}{m_\nu} \rp^7~{\rm years}\,.
 \eeq
If the neutrino lifetime becomes comparable or smaller than the age of the Universe, $t_0\simeq 1.4\times 10^{10}$ years, the emitted photon would affect the observed CMB spectrum. We then require 
\beq 
c_\nu^{ij} g_{\phi\g}  \leq 8 \times 10^3 \times\lp \frac{m_\phi}{{\rm keV}} \rp^2 \lp \frac{0.1~{\rm eV}}{m_\nu} \rp^{7/2}~{\rm GeV}^{-1}\,.
\eeq
Numerically, taking $m_\nu = 0.1$ eV, and $m_\phi = 1$ eV, this gives $c_\nu^{ij} g_{\phi\g} \lesssim  8 \times 10^{-3}~{\rm GeV}^{-1}$, while for $m_\phi = 1$ GeV, the constraint is $c_\nu^{ij} g_{\phi\g} \lesssim  8 \times 10^{15}~{\rm GeV}^{-1}$.

\section{Stellar cooling constraints} 
\label{sec:StarCooling}

\begin{figure}
	\centering
	\includegraphics[width=0.48\linewidth]{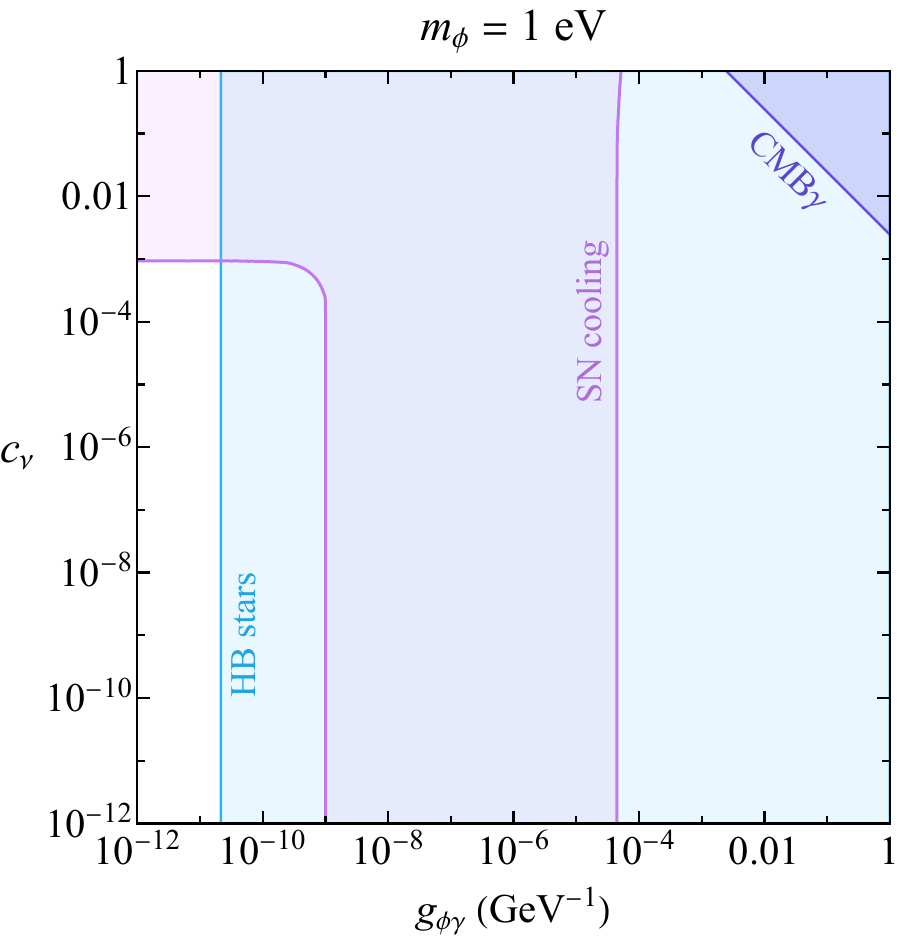}
	\quad
	\includegraphics[width=0.48\linewidth]{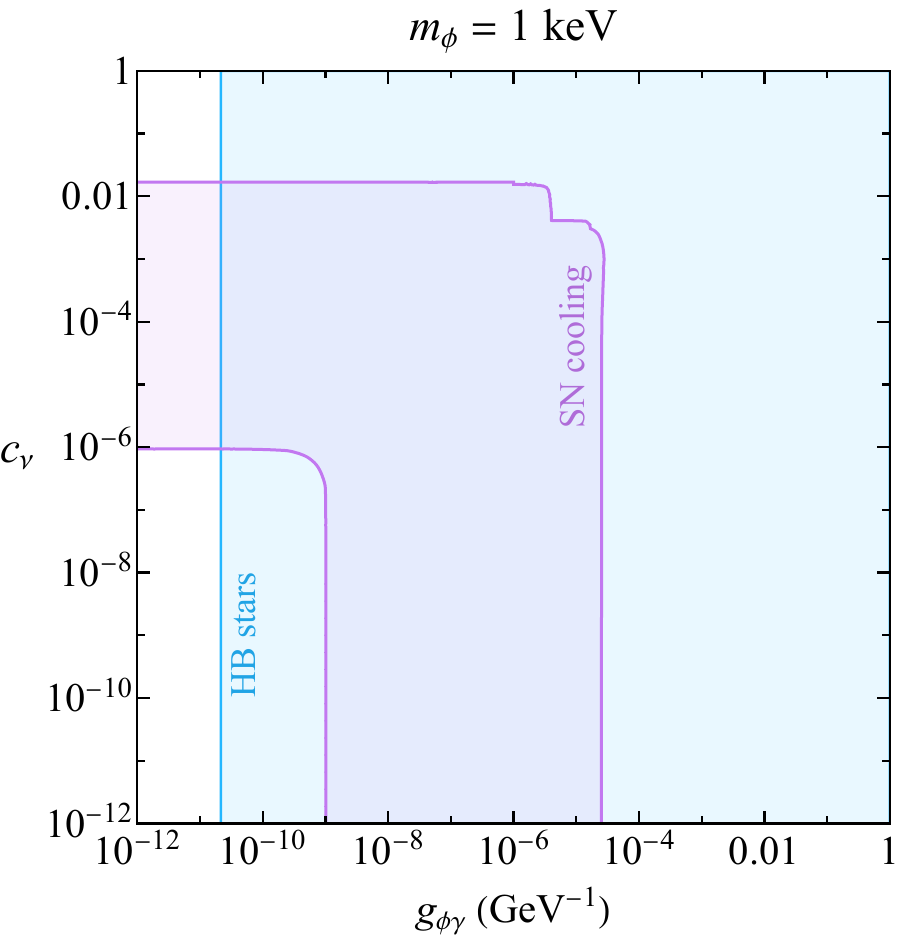}
	\includegraphics[width=0.48\linewidth]{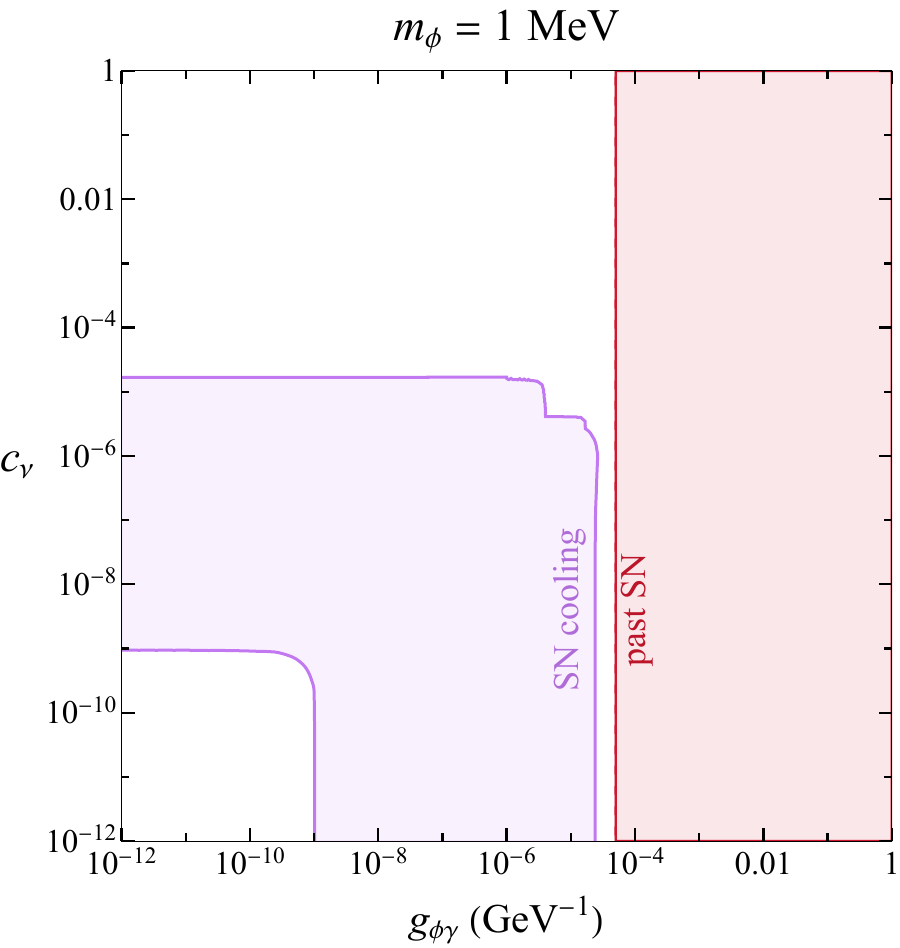}
	\caption{ Stellar cooling bounds on pseudoscalar couplings to neutrinos, $c_\nu$ and photons, $g_{\phi\gamma}$, for three mass benchmarks, $m_\phi = 1$ eV, keV, MeV, while there are no bounds for $m_\phi = 1$ GeV. The cyan and purple regions show the bounds from HB stars and Supernovae, respectively. The dark blue region in the upper left plot indicates possible bounds from scattering of neutrinos emitted from SN1987a on CMB photons. The red region in the bottom plot shows bounds from the diffuse $\g$ ray spectrum of past SN.}
	\label{fig:StarCooling}
\end{figure}

If stellar dynamics is able to produce light new physics states that efficiently escape from its core, it can lead to excessively large stellar cooling rates. Requiring that the additional cooling  does not  exceed the standard model one, typically leads to very stringent bounds on light new physics sectors. In this section we evaluate the stellar cooling bounds for the light scalar $\phi$ that couples to photons and neutrinos, Eq. \eqref{eq:lagr:phi}, with the results summarized in Fig. \ref{fig:StarCooling}. The stellar cooling rates are controlled by the $\phi$ production rates, as well as its decay length and/or mean free path. 

In the analysis we distinguish two cases. In the first category are the cooling rates for Horizontal Branch stars (HB), Red Giants (RG) and White Dwarves (WD), for which the core temperature is low,  $T\sim$ few keV. In the second category are the Supernova (SN) cooling constraints, for which the core of the proto-neutron star is much denser and hotter,  $T_{\rm SN}\sim30$ MeV. 
The $\phi$ production mechanisms are the Primakoff conversion, $\gamma \to \phi$, the photon coalescence, $\gamma\gamma \to \phi$, and for SN also the neutrino coalescence, $\nu\nu\to \phi$. 
The rates for these processes are given in Appendix \ref{sec:app:cooling:rates}.

 The inverse reactions constitute the main decay channels: the decays to two neutrinos, $\phi\to\nu\nu$, Eq. \eqref{eq:phidecayrate:gg}, the decays to two photons, $\phi\to\g\g$,  Eq.~\eqref{eq:phidecayrate:gg}, and the inverse Primakoff process, $\phi \to\g $, Eq.~\eqref{eq:InversePrim}. The $\phi$ decays to photons, as well as the Primakoff process, reduce the stellar cooling rates, since they reduce the number of $\phi$ particles that escape the interior of the star. The same is true for $\phi \to \nu\nu$ decays, whenever these occur inside the SN, since the neutrinos get trapped inside the proto-neutron star. The HB, RG and WD cooling rates, on the other hand, are not affected by  the $\phi\to \nu\nu$ decays, since the neutrinos escape from these types of stars in the same way as the undecayed $\phi$ would have. 

For processes that involve photons we need to take into account finite temperature effects. To first approximation, the effect of a photon propagating in plasma instead of in a free space can be taken into account by replacing the dispersion relation for a massless photon with the one for a collective excitation -- plasmon, i.e., a massive photon with the mass 
equal to the plasma frequency 
\beq
\label{eq:omegaP}
\omega_P^2 = \frac{4\pi\alpha n_e}{E_F}\,,
\eeq
where $n_e$ is the electron number density in the core and $E_F$ the Fermi energy of the electron gas. For rough numerical estimates we can use the non-relativistic Fermi gas expression for the plasma frequency $\omega_P \sim 28.7\,{\rm eV}\big[{\rho}/({{\rm g/cm^3}})\big]^{1/2}$.

The $\phi$ emissivity, i.e., the energy emitted in the production of the final state $\phi$ per unit volume and time, is given by 
\beq\label{eq:emissivity:general}
Q_\phi = g_I \int \frac{{d}^3 p}{(2\pi)^3} \Gamma_{I\to \phi} E_\phi f(E_\phi) = \int_{m_\phi}^{\infty} dE_\phi E_\phi \frac{{d}^2 N_\phi}{{d}E_\phi{d}t}.
\eeq
Here, $p$ is the $\phi$ momentum, $\Gamma_{I\to \phi}$ the $\phi$ production rate from initial state $I$, with $g_I$ number of degrees of freedom (the number of polarization states) of the state $I$, and $f(E_\phi)$ the thermal distribution of particles $\phi$ in the stellar core. 
The number of emitted $\phi$ particles per unit of energy and time, ${{d}^2 N_\phi}/{{d}E_\phi{d}t}$ is a sum of  different $\phi$ production mechanisms: Primakoff conversion, photon coalescence, and neutrino coalescence (for SN), see Appendix \ref{sec:app:cooling:rates}.

\subsection{Horizontal branch stars}
\label{sec:HBstars}
For cold cores, i.e., for WD and HB stars, we can safely approximate the stellar core with a homogeneous sphere of radius $R_c$. The total luminosity is then given by
\beq\label{eq:luminosity:newparticle}
L_\phi = V_c \left[ g_I \int \frac{{d}^3 p}{(2\pi)^3} \Gamma_{I\to \phi} E_\phi f(E_\phi) S(R_c) \right] \,,
\eeq
where $V_c = 4\pi R_c^3/3$ is the volume of the stellar core, and we used Eq.~\eqref{eq:emissivity:general} for the emissivity. The factor $S(R_c)$  takes into account the suppression of the $\phi$ luminosity due to decays of $\phi$ into photons. Namely, if $\phi$ decays via $\phi\to\g\g$ while still inside the stellar core, then its energy is reabsorbed by the plasma. If  $\phi$ decays instead via the $\phi\to\nu\nu$ channel, its energy is still carried away by the two neutrinos, thus contributing to the exotic cooling of the star. The suppression factor $S(R_c)$ is then given by
\beq\label{eq:Sfactor}
S(R_c) = 1-\exp\lp -\Gamma_\nu {R_c}\rp +\exp\lp -\Gamma_{\rm tot} {R_c}\rp\,,
\eeq
where 
\beq\label{eq:decrate:star}
\Gamma_\nu=\frac{1}{\beta_\phi} \frac {\Gamma_{\phi\to \nu\nu}} { \gamma_\phi}, \qquad
\Gamma_{\rm tot}=\frac{1}{\beta_\phi}\Big(\frac { \Gamma_{\phi\to \gamma\gamma}+ \Gamma_{\phi\to \nu\nu}} { \gamma_\phi}+{ \Gamma_{\phi\to \gamma}}\Big)\,,
\eeq
with $\gamma_\phi = E_\phi/m_\phi$ the Lorentz factor, and $\beta_\phi$ the velocity of $\phi$. In writing \eqref{eq:Sfactor} we approximated the suppression factor for $\phi$ originating from any given point inside the star to be the same as when traversing distance $R_c$, i.e., the typical linear dimension. 
Beside total luminosity, $L_\phi$, the quantity often considered in the literature is the luminosity per unit mass,  $\epsilon_\phi = L_\phi/M$, where $M$ is the mass of the stellar core. 

For cold cores the main energy-loss process, within the standard model, is the neutrino emission via plasmon decay into two photons \cite{Haft:1993jt}. 
The measurements of RG and HB cooling rates give a typical value for the  observed luminosity per unit mass of $| \epsilon_\nu | \simeq 100 ~ {\rm erg} ~ {\rm g}^{-1} ~{\rm s}^{-1}$, with a $\sim 10\%$ uncertainty.
Requiring that the extra cooling due to emission of $\phi$ is smaller than the experimental error on the measurement gives~\cite{PhysRevD.37.549}
\beq\label{eq:HBcoolingBound}
| \epsilon_\phi | \leq 10 ~ {\rm erg} ~ {\rm g}^{-1} ~{\rm s}^{-1}\,.
\eeq
The excluded regions in the $c_\nu -g_{\phi\g}$ plane are shown in Fig.~\ref{fig:StarCooling} as blue bands for two mass benchmarks, $m_\phi = 1\,$eV, $1\,$keV, while for  $m_\phi = 1\,$MeV, $1\,$GeV benchmarks $\phi$ is too heavy to be produced in a cold stellar core. We only show the bounds obtained from HB stars, due to their higher density and thus higher luminosity than RG stars. The benchmark values are $T_{\rm HB} \sim 10^8 {\rm K} \sim 8.6$ keV, $\rho_{\rm HB} \sim 10^4 {\rm g/cm^3}$ and $R_{\rm HB} \sim 0.03 R_\odot$ for temperature, density and helium-burning core radius respectively, where $R_\odot = 6.96\times10^5$ km is the radius of the Sun. The associated plasma frequency is $\omega_{\rm HB}\sim3$ keV.

The lower boundary of excluded $g_{\phi\g}$ values in Fig.~\ref{fig:StarCooling}, $g_{\phi\g}\sim10^{-11}~{\rm GeV}^{-1}$, indicates the value at which $\phi$ starts to be produced efficiently in the star, thus exceeding the bound on exotic cooling, Eq.~\eqref{eq:HBcoolingBound}. The upper boundary of $g_{\phi\gamma}$ exclusion, on the other hand, indicates the onset of parameter region for which $\phi$ is trapped: the coupling to photon is strong enough that $\phi$ always decays inside the  stellar core. In the parameter space scanned in this work, the trapping regime is never reached; the decays to neutrinos compensate the exponential suppression from photon decay, Eq.~\eqref{eq:Sfactor}, and thus the production of $\phi$ still contributes to stellar cooling. For very small values of couplings to neutrinos, below the values shown in the figure, there are however regions where neutrino coupling is not able to overcome the trapping.

In the EFT regime, where $\phi$ is too heavy to be produced on-shell, the cooling mechanism is due to a Primakoff conversion $\g^*\g_L\to\nu\nu$, where $\g^*$ is the plasmon and the transition is induced by the longitudinal plasmon of the external electric field of the plasma, $\g_L$. In the limit where the latter is taken as static, the transition can be approximated as the two body decay~\cite{Raffelt:1996wa}, and the  rate scales roughly as 
$\Gamma_{\gamma^* \gamma_L\to \nu\nu}\sim\omega_{HB}^7/\Lambda^6$. Demanding that the cooling is smaller than the error on the measured rates gives the lower bound on the effective suppression scale of the Rayleigh operator 
 \beq
 \Lambda^3/\C_2^{(7)}\gtrsim \big(8\,\text{MeV}\big)^3.
 \eeq

\subsection{Supernova cooling}
\label{sec:SN}

There are several important differences between cooling rates deduced from HB stars and the SN. Firstly, at proto-neutron star densities and temperatures the neutrinos produced inside the core are efficiently trapped, 
leading to a thermal population of neutrinos with a chemical potential $\mu_\nu\sim200\,$MeV. 
The decays of a propagating $\phi$  into neutrinos, therefore no longer lead to enhanced cooling rates.
Secondly, the total luminosity is quite sensitive to the exact radial profile of the SN core after the start of the explosion. The SM temperature and density profiles as a function of the distance from the center, $r$, can be obtained via numerical simulations, and depend both on the initial conditions of the progenitor star and the explosion mechanism. In the numerical analysis  we use the profile from Ref.~\cite{Fischer:2018kdt} at the benchmark time $t=1$ s after the start of the explosion.  

The production of $\phi$ is $r$ dependent, since the production rates depend on $T(r), \omega_P(r),\ldots$ After the production, $\phi$ propagates inside the core and contributes to the SN cooling, if it escapes the neutrino-sphere of radius $R_\nu\sim23\,$km, i.e., the region where neutrino production rate is higher than the absorption. In the opposite case, the energy taken by the $\phi$ is re-deposited into neutrinos. The probability that $\phi$ reaches a distance $R_\nu$ is controlled by the optical depth~\cite{Lucente_2020},
\beq
\tau_\phi\lp r, E_\phi, R_\nu \rp = \int_r^{R_\nu} \Gamma_{\rm tot} (\tilde r) \frac{{\rm d}\tilde r}{d_\phi(\tilde r)}\,,
\eeq
with $\Gamma_{\rm tot}$ given in \eqref{eq:decrate:star} now depends on the radial distance. While the use of optical depth to derive SN cooling bounds can lead to appreciable difference relative to a more systematic treatment, see, e.g., Ref.~\cite{Caputo:2022rca}, the precision suffices for our purposes. 

The total luminosity is then given by
\beq
L_\phi = 4\pi \int_0^{R_\nu} {\rm d}r ~ r^2 \lp g_I \int \frac{{\rm d}^3 p}{(2\pi)^3} \Gamma_{I\to \phi} E_\phi f(E_\phi) ~ \exp\left[-\tau_\phi\lp r, E_\phi, R_\nu \rp\right] \rp \,.
\eeq
The bounds on neutrino and photon couplings can be obtained by imposing that $L_\phi$ does not exceed the measured neutrino luminosity,
\beq
L_\phi \lesssim L_{SN}^\nu\sim3\times10^{52}~{\rm erg}~{\rm s}^{-1}\,,
\eeq
i.e., which is the usual rule of thumb prescription that translates the absence of large cooling effects in the observed neutrino flux from SN1987a to a bound on the production of light particles. 

The resulting excluded region is shown as a purple band in Fig.~\ref{fig:StarCooling}. In contrast to HB stars, the SN core is hot enough to produce new particles with masses up to $m_\phi\sim100$ MeV. In the parameter range we are interested in, the bound is affected both by $\phi$ couplings to photon and neutrinos. The excluded region in Fig.~\ref{fig:StarCooling} is a horizontal band when $c_\nu$ dominates, and is a vertical band when $g_{\phi\g}$ is more important. The change from one regime to the other can be roughly understood through Eq. \eqref{eq:Brfraction}.

For $\phi$ heavier than $m_\phi\gtrsim100$ MeV the $\phi$ particle is not produced on-shell in the proto-neutron star, and thus does not contribute to cooling. The situation is different from HB stars, where the Rayleigh operator created by integrating out a heavy $\phi$ can still enhance the cooling rates via the production of neutrinos through the Primakoff transition, $\g^*\g_L\to\nu\nu$. The neutrinos then escape and lead to enhanced stellar cooling rates. For SN, the neutrinos are instead trapped inside the dense SN core. Increased coupling between photons and neutrinos, due to a new off-shell degree of freedom, therefore has no visible effect. 

So far we focused on constraints that arise from SN cooling (using optical depth approximation, for a more detailed treatment see \cite{Fiorillo:2022cdq}). For the $m_\phi = 1$ MeV benchmark, the strongest SN constraint, however, is due to the absence of observed $\g$ rays during the SN1987a explosion~\cite{Caputo:2021rux}, resulting in $g_{\phi\g} \lesssim10^{-11}~{\rm GeV}^{-1}$ for $\phi$ coupling just to photons and in a free-streaming regime. 
For our case this bound needs to be rescaled by the neutrino branching ratio, Eq.~\eqref{eq:rescaling}, to account for the additional $\phi\to\nu\nu$ decay channel. We find that for all benchmarks this bound is then weaker or comparable to HB and SN cooling bounds for lighter benchmarks, and thus we do not show it. Note that the SN $\g$ ray bound extends up to $\phi$ masses of $\sim100-200$ MeV. In the trapping regime, there is an additional constraints from $\phi$ decaying to photons inside the proto-neutron star and contributing to the diffuse $\g$ ray background from the past SN. This gives an upper bound $g_{\phi\g} \lesssim 5\times10^{-5}~{\rm GeV}^{-1}$ for $\phi$ masses in the $\sim1 - 100$ MeV range, shown as the red line in the bottom panel in Fig.~\ref{fig:StarCooling}.

Finally, we comment on the possibility that the neutrinos produced in the SN1987a core would interact with CMB photons and modify the observed CMB spectrum. The scattering length is given by   $\lambda_{\nu\g}\sim (n_\g \sigma_{\nu\g})^{-1}$, where $\sigma_{\nu\g}$ is the  cross section for $\nu\g\to\nu\g$ scattering, mediated by the $s-$ and $u-$channel tree level $\phi$ exchange.  Taking the neutrinos to have fixed energy $E_\nu\sim30\,$MeV, and the CMB photon the typical energy $E_\g\sim T_{\rm CMB}\sim2\times10^{-13}\,$GeV, with the number density $n_\g\sim2.2\times10^{8}\,{\rm m}^{-3}$, we find 
\beq 
\lambda_{\nu\g}\sim1.2\times10^{53} \lp \frac{m_\phi}{{\rm GeV}} \rp^4 \Big( \frac{1}{c_\nu \lp g_{\phi\g}\times{\rm GeV} \rp} \Big)^{2}~{\rm m}\,.
\eeq
The condition that the scattering length $\lambda_{\nu\g}$ is less than the distance of SN1987a from Earth, $d_{\rm SN}\sim2\times10^{21}\,$m, is achieved for 
\beq 
c_\nu \lp g_{\phi\g}\times{\rm GeV} \rp \gtrsim 2.5\times10^{-3} \lp \frac{m_\phi}{{\rm eV}} \rp^2\,.
\eeq
This can exclude part of the parameter space we are intersted in for the mass benchmark $m_\phi = 1$ eV. The bound is shown as a red region in Fig.~\ref{fig:StarCooling}. For $m_\phi>{\mathcal O}(\text{eV})$ only relatively large values of $c_\nu, g_{\phi\g}$ are covered, which are already well excluded by cosmology and laboratory searches, and thus do not appear in the plots.

\section{Bounds from terrestrial experiments }
\label{sec:NeutrinoScatteringExperiments}

\begin{figure}
	\centering
	\includegraphics[width=0.48\linewidth]{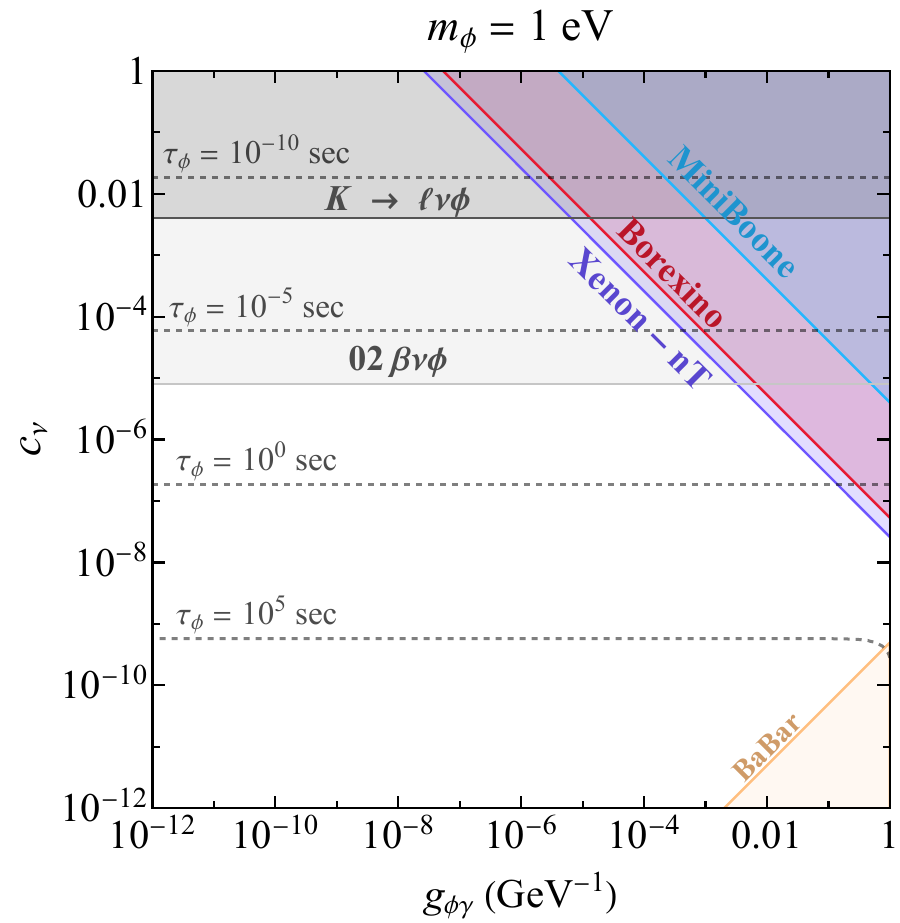}
	\quad
	\includegraphics[width=0.48\linewidth]{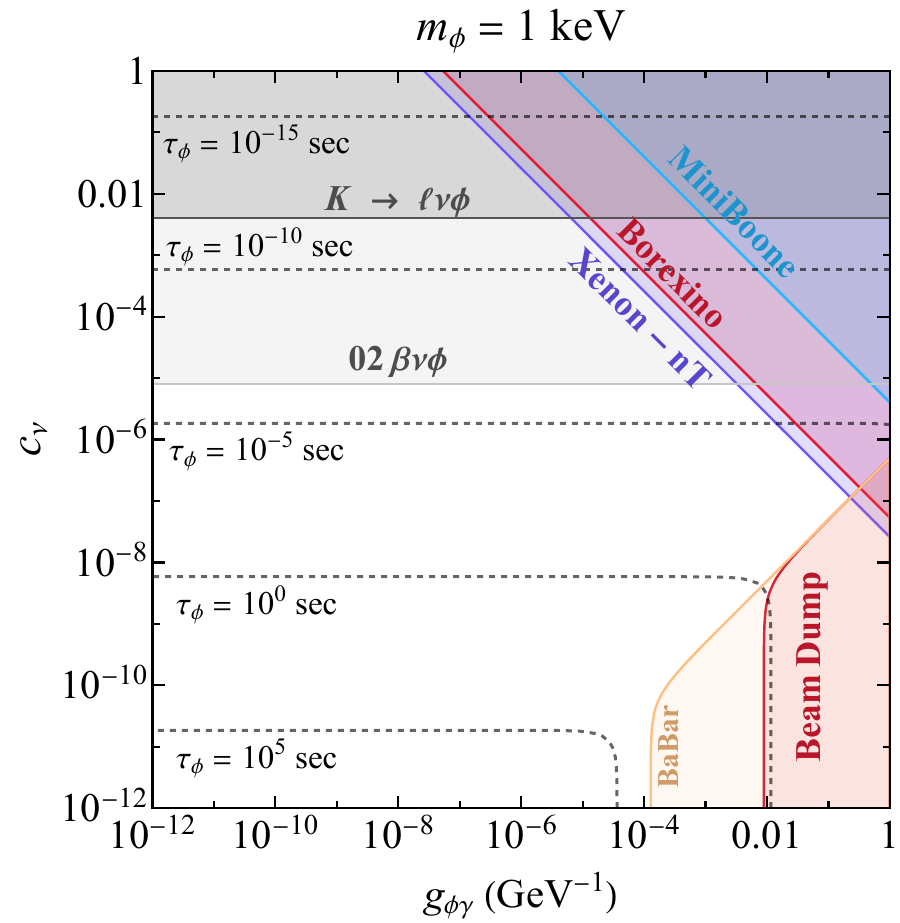}
	
	~

	\includegraphics[width=0.48\linewidth]{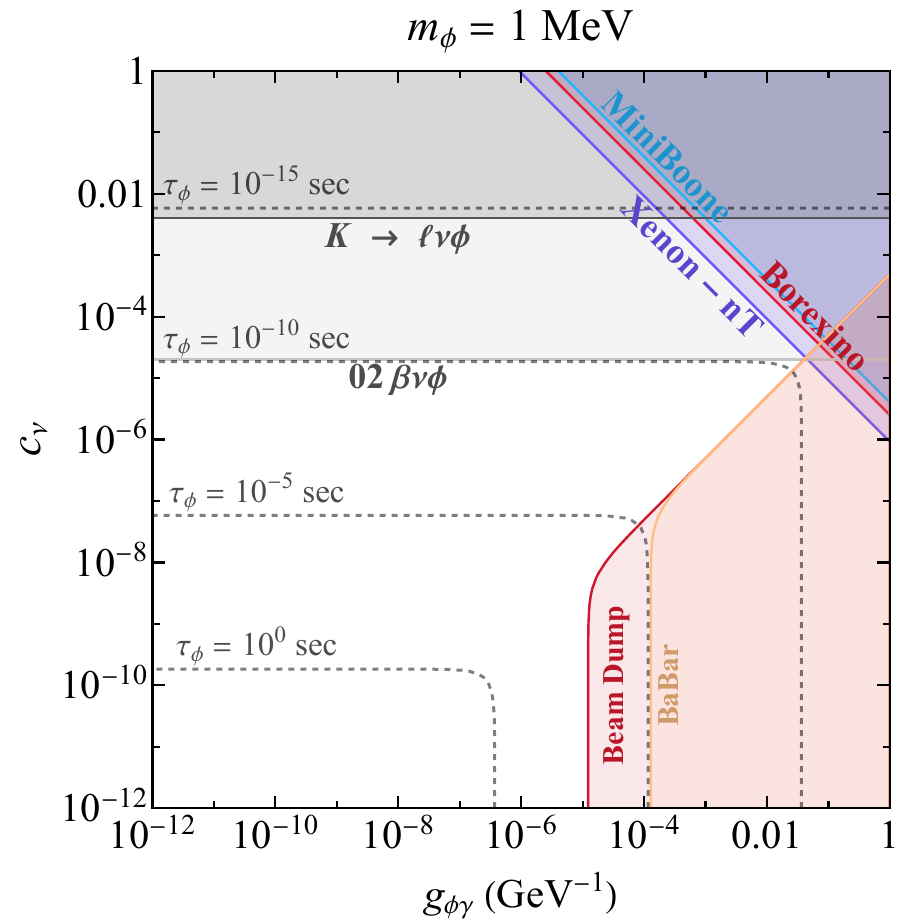}
	\quad
	\includegraphics[width=0.48\linewidth]{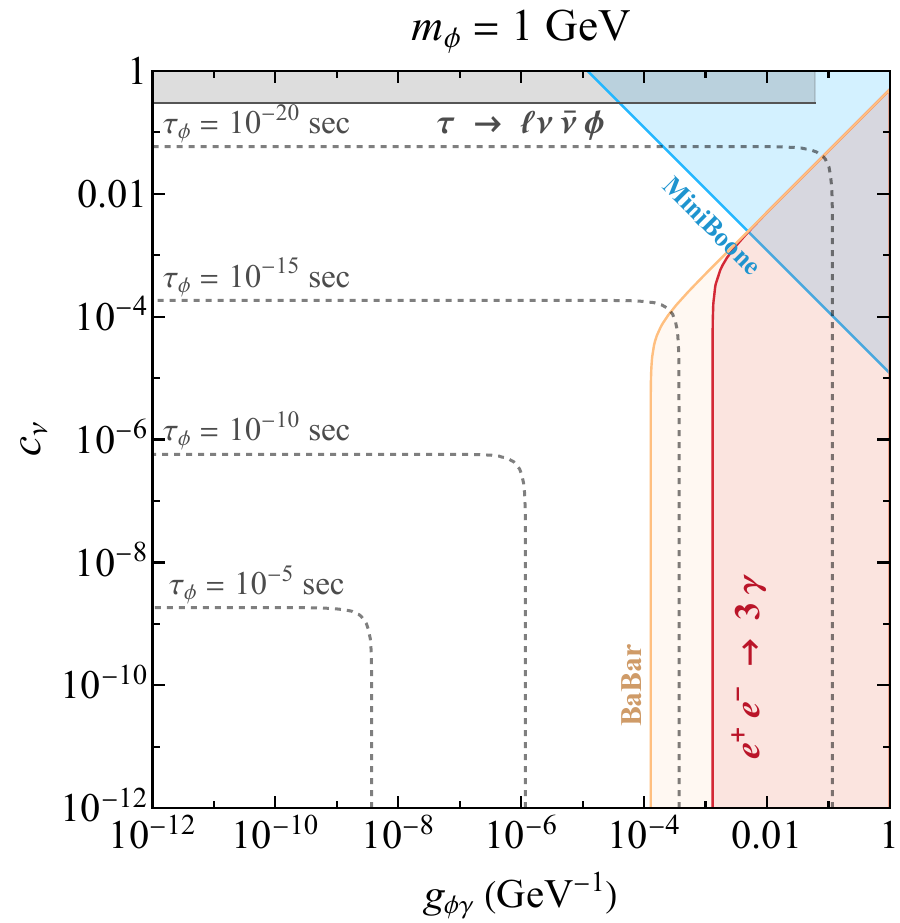}
	\caption{ Constraints on neutrino and photon couplings of pseudoscalar mediator from terrestrial experiments: from Xenon-nT (purple), Borexino (red), MiniBoone (blue), BaBar monophoton (orange), Belle II $e^+e^-\to 3\gamma$ (light red) and beam dump (light red) searches, for the four mass benchmarks $m_\phi = 1\,$eV, $1\,$keV, $1\,$MeV, $1\,$GeV.  The dotted lines are iso-contours of $\phi$ lifetimes.}
	\label{fig:TerrestrialBounds}
\end{figure}

Next we discuss the bounds on neutrino polarizability from terrestrial detectors. 
In neutrino and dark matter experiments, the incoming neutrinos can scatter on electrons or nuclei in the detector. The Rayleigh operators induce at 1-loop the $\nu X\to \nu X$ scattering, where $X=e,N$ is either an electron or a nucleon, and at tree level $\nu X\to \nu X \gamma$, i.e., neutrino scattering with an emission of an extra photon. The scatterings on nucleons, such as the coherent neutrino nucleus scattering \cite{COHERENT:2017ipa,COHERENT:2020iec,CONUS:2020skt,CONUS:2021dwh,nuGeN:2022bmg}, leads to less stringent bounds than scattering on electrons \cite{Altmannshofer:2018xyo}. The resulting bounds from Borexino, Xenon-nT and MiniBoone are given in Sections~\ref{sec:Borexino}, \ref{sec:XenonnT}, and \ref{sec:MiniBoone}, respectively. In Section  \ref{sec:ColliderConstraints} we discuss collider constraints. Summary of the terrestrial constraints on the pseudoscalar coupling to neutrinos and photons is given in Fig.~\ref{fig:TerrestrialBounds}, for the four mass benchmarks, $m_\phi = 1$\,eV, 1\,keV, 1\,MeV, 1\,GeV.

\subsection{Bounds from Borexino}
\label{sec:Borexino}
Borexino measured the scattering of solar neutrinos on electrons \cite{Borexino:2017fbd}, where the detector response does not distinguish between $\nu e\to \nu e$ and $\nu e\to \nu e \gamma$ scattering events, and thus the two need to be added  when comparing with the measured rates. 
 We calculate the tree-level  $\nu e\to \nu e \gamma$ scattering numerically by implementing the Rayleigh operator in {\tt MadGraph} \cite{Alwall:2011uj}. For the much smaller 1-loop induced $\nu e\to \nu e$ scattering we use the NDA estimate,  $\sigma_{i} \simeq \big( \big|\hat \C_{1,i}^{(7)} \big|^2 +3\big|\hat \C_{2, i}^{(7)}\big|^2/2\big) \alpha^4 s^2/(48 \pi^4)$, where $s = m_e^2 + 2 m_e E_\nu$ is the center of mass energy of the scattering process, and assume flavor diagonal couplings, cf. Eq. \eqref{eq:C12:flav:diag}. 

The event rate per day per 100 tons of detector is given by 
\beq
 \begin{split}
 	\label{eq:Rirate:C17}
 	R_i(\hat \C_1^{(7)},\hat \C_2^{(7)}) 
 	= T N_e \int_{E_{\nu, {\rm min}}}^{E_{\nu,{\rm max}}} dE_\nu\, \phi_i(E_\nu) &\Big[ P_i^e \sigma_{\nu_e e}(E_\nu,\hat \C_{1,e}^{(7)},\hat \C_{2,e}^{(7)}) 
 	\\
 	&+ P_i^\mu \sigma_{\nu_{\mu} e}(E_\nu,\hat \C_{1,\mu}^{(7)},\hat \C_{2,\mu}^{(7)})\Big]\,,
 \end{split}
 \eeq
 with $T = 1\,{\rm day} = 8.64\cdot10^4 \,{\rm s}$ the exposure time and $N_e=3.307\cdot10^{31}$ the number of target electrons in 100 tons of detector mass, while $\ell=e, \mu$ are the incoming neutrino flavors. 
 The label $i$ in \eqref{eq:Rirate:C17} denotes the main components of the solar neutrino flux  on Earth, $\phi_i(E_\nu)$, Ref.~\cite{Vitagliano_2020}; due to proton-proton fusion ($i=pp$), Berillium 7 electron capture ($i={}^7{\rm Be}$), and proton electron capture ($i=pep$). The $\nu_e$ from $pp$ have a continuous energy spectrum with the maximal energy $E_{\nu, {\rm max}}= 0.423$ MeV, while $^7$Be and $pep$ neutrinos are monochromatic, with energies $E_{^7{\rm Be}} = 0.863$ MeV and $E_{pep} = 1.445$ MeV, respectively.
 The minimal incoming neutrino energy that can still produce the threshold $\sim50$ keV  recoil in Borexino is $E_{\nu,{\rm min}} = 0.139$ MeV.
 The $\nu_e$ neutrinos produced in the Sun undergo flavor oscillations while propagating to Earth.  The $\nu_e$ survival probabilities are $P_{i}^{e}=\{0.554\,,0.536\,,0.529\,\}$ for $i=\{pp,^7{\rm Be},pep\}$,
  once matter effects are taken into account \cite{Khan_2020}, while $P_i^\mu = (1 - P_i^e)/2$,  assuming maximal $\theta_{23}$ for simplicity.
 
The bounds on $\hat \C_{j,\ell}^{(7)}$ are obtained using the following chi-squared function
 \beq
 \chi^2(\vec\alpha,\hat \C_{j,\ell}) = \sum_i \frac{\Big[ R_{{\rm meas},i} - R_i(\hat \C_{1,\ell}^{(7)},\hat \C_{2,\ell}^{(7)}) (1+\alpha_i) \Big]^2}{\sigma_i^2} + \lp \frac{\alpha_i}{\sigma_{\alpha_i}} \rp^2,
 \eeq
 where the sum is over the three types of solar neutrino fluxes. 
The measured event rates in Borexino phase-I and their statistical uncertainties are~\cite{Khan_2020} $R_{{\rm meas},i}\pm \sigma_i=\{134\pm10, 48.3\pm1.1, 2.43\pm0.36\}$, $i=\{pp,^7{\rm Be},pep\}$, to be compared with the SM rates $R_{i}(0) = \{131.4, 48.1, 2.8\}$, where the theoretical errors on the predictions are accounted for by marginalizing over the parameters $\alpha_i$, with  $\sigma_{\alpha_{i}}=\{1.1\%, 5.8\%, 1.5\%\}$ \cite{Khan_2020}.
The resulting $1\sigma$ allowed ranges on Rayleigh operators are
\beq
	\left| \frac23 \hat \C_{1,\ell}^{(7)} + \hat \C_{2,\ell}^{(7)} \right| \leq \left\{1.5,~ 5.7,~1.5 \right\}\times10^{3}~ \text{GeV}^{-3}\,, \quad \text{for}\quad \ell=\{e, \mu, \text{univ.}\},
\eeq
assuming photon couplings to either only $\nu_e$ or $\nu_\mu$ or both (with universal couplings).
 In Table~\ref{table:bounds_summary_EFT} we list the result for the universal couplings, assuming only the CP-odd Rayleigh operator is nonzero.

For light mediators, with mass much lower than a typical momentum exchange in Borexino, $m_\phi^2 \ll |q^2|$, where $|q|\sim100$ keV, the EFT framework no longer applies, and we include the full $\phi$ propagator in the {\tt MadGraph} calculation of the cross sections. Comparison with the measurements then gives for light $\phi$
\beq
c_\nu g_{\phi\g} \leq \left\{0.55,~ 1.91,~0.53 \right\}\times10^{-7}~ \text{GeV}^{-1}\,, \quad \text{for}\quad \ell=\{e, \mu, \text{univ.}\}.
\eeq
In Table~\ref{table:bounds_summary_PHI:ev,kev} and \ref{table:bounds_summary_PHI:mev,gev} we quote only the bound for the flavor universal case. The corresponding bounds for the four benchmark masses are shown as excluded red regions in Fig. \ref{fig:TerrestrialBounds}.

\begin{figure}
	\centering
	\includegraphics[width=0.6\linewidth]{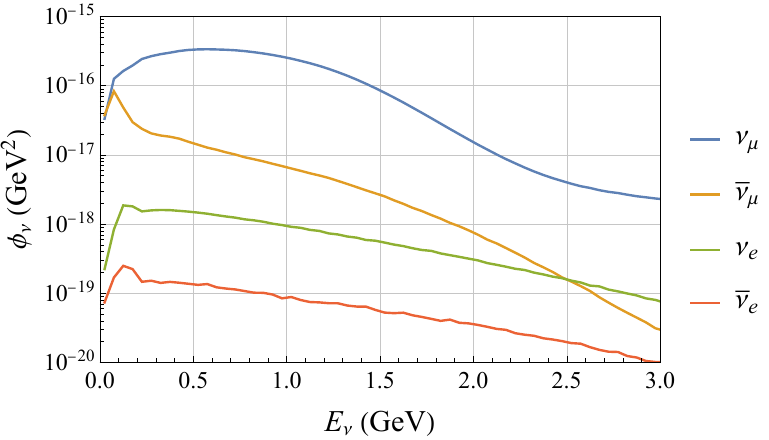}
	\caption{Neutrino flux at the MiniBoone detector for each of the neutrino flavor components, as denoted in the legend.}
	\label{fig:MiniB:flux:numode}
\end{figure}

\subsection{Bounds from dark matter detectors}
\label{sec:XenonnT}
Dark matter direct detection experiments are sensitive to enhanced $\nu e\to \nu e$ scattering rates from solar neutrinos, similar to the Borexino bounds discussed in the previous section, but with a lower recoil energy threshold of $\sim 2\,$keV, which translates to a lower required minimal energy of the incoming neutrinos, $E_{\nu, {\rm min}} \simeq 16$ keV, in Eq. \eqref{eq:Rirate:C17}. The strongest constraints come from the recent measurement of electron recoil events by XENONnT \cite{XENON:2022mpc} with exposure of 1.16\,tonne-years, with no excess observed over the background rate $R_{\rm bg}=(16.1 \pm 1.3,\text{stat})$ events/(t$\cdot$y$\cdot$keV) in the $(1, 30)\,$keV recoil energy search window (see also \cite{CDEX:2019exx,PandaX-II:2020udv,EDELWEISS:2018tde,LUX:2017glr}). Saturating the allowed nonstandard background rate with the neutrino polarizability induced scattering on free electrons translates to the following constraints on Rayleigh operators
\beq
	\left| \frac23 \hat \C_{1,\ell}^{(7)} + \hat \C_{2,\ell}^{(7)} \right| \leq \left\{0.6,~ 2.0,~0.5 \right\}\times10^{3}~ \text{GeV}^{-3}\,,
\eeq
for $m_\phi\gg q^2$, and
\beq
c_\nu g_{\phi\g} \leq \left\{2.6,~ 8.9,~2.5 \right\}\times10^{-8}~ \text{GeV}^{-1}\,,
\eeq
for $m_\phi\ll q^2$. In Tables~\ref{table:bounds_summary_EFT}, ~\ref{table:bounds_summary_PHI:ev,kev}, and \ref{table:bounds_summary_PHI:mev,gev} we list only the constraint for the pseudoscalar case with flavor universal couplings, with the corresponding excluded parameter regions for the four mass benchmarks shown as blue regions in Fig. \ref{fig:TerrestrialBounds}. Note that the use of free electron approximation may be suspect for inner shell electrons, however, we expect the corrections to be subleading due to the steeply rising spectrum, dominated by the largest values or recoil energies. 

In deriving the above bounds we included only $\nu e\to \nu e$ scattering as the signal. Dark matter detectors have in principle the possibility to probe also the subleading $\nu e\to \nu e\gamma$  process, by searching for an extra photon. It would be interesting to explore if this signature can give enhanced sensitivity to neutrino polarizability.  

\subsection{Bounds from MiniBoone}
\label{sec:MiniBoone}
MiniBoone is an electron neutrino appearance experiment in which $\nu_e$ are detected through quasi-elastic charged current interaction, with a typical momentum exchange $q^2\sim-2~{\rm GeV}^2$. $99.4\%$ of the initial neutrino flux is made of $\nu_\mu+\bar \nu_\mu$ and peaks at ${\mathcal O}(500~{\rm MeV})$, with two modes of operation: neutrino and antineutrino modes. For details on neutrino fluxes we use Ref. \cite{MiniBooNE:2008hfu, MiniBooNE:2021bgc}.  The MiniBoone detector is filled with pure mineral oil, CH${}_2$, which acts as both a target and a scintillator.  

The signal of neutrino polarizability interactions is the Rayleigh operator induced $\nu A\to \nu X_A+\gamma$ scattering, where $A$ is the initial nucleus and $X_A$ denotes the final states from either elastic or inelastic scattering. Experimentally, the signature is similar to the radiative up-scattering \cite{Bolton:2021pey,Schwetz:2020xra,Coloma:2017ppo,Atkinson:2021rnp,Acero:2022wqg} and thus the same type of analyses would also be sensitive to neutrino scattering through polarizability operators. In the MiniBoone detector, however, this signature is indistinguishable from the  SM  quasi-elastic charged current scattering of electron neutrino, $\nu_e A\to e X_A$. The only difference is that the Rayleigh induced process leads to a softer deposited energy spectrum due to the final state neutrino that escapes the detector. For scattering on carbon we assume that it is dominated by quasi-elastic scattering, i.e., by neutrino scattering on a single nucleon bound inside the carbon nucleus, which then gets kicked out of the nucleus.  In the calculation of the total scattering rates we also include the scattering on hydrogen, $\nu p \to \nu p +\gamma$, which constitutes a subdominant component of the signal. 

\begin{figure}[t]
	\centering
	\includegraphics[width=0.75\linewidth]{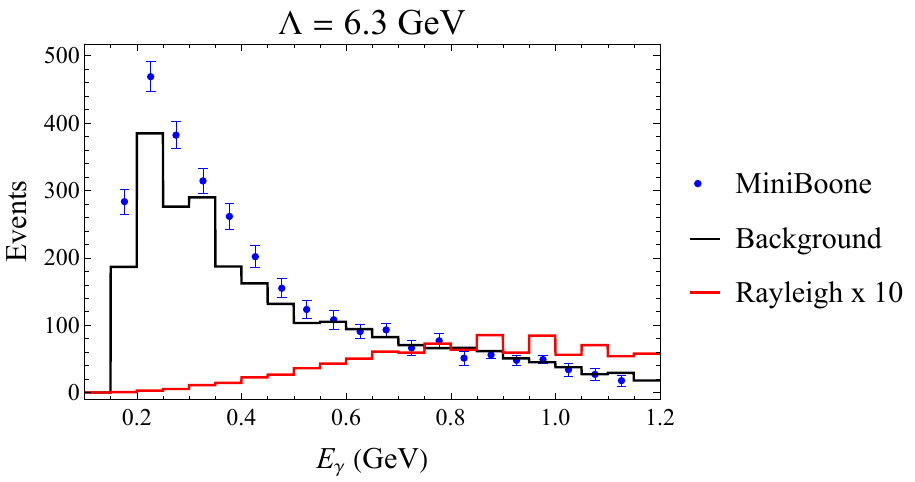}
	\caption{The number of measured MiniBoone events (blue dots and error bars) in each photon$/$ electron energy bin \cite{Aguilar-Arevalo:2020nvw}, compared to the expected SM signal+background (black line). Red line denotes the predicted signal from a Rayleigh operator setting the NP scale to $\Lambda = 6.3$ GeV (and dimensionless coupling to unity), multiplying it by a factor 10 for clarity.
 }\label{fig:MB_bounds}
\end{figure} 

The photon spectrum is given by
\beq
\frac{{d}N}{{d}E_\gamma} = N_p\int~{d}E_\nu \epsilon(E_\gamma) \phi_{\rm tot}(E_\nu) \frac{{d}^2\sigma}{{d}E_\gamma{d}E_\nu}\,,
\eeq
where $N_p = 2.8\times10^{32}$ is the total number of target protons in the detector, $\epsilon(E_\gamma)$ is the acceptance as function of the photon energy and $\phi_{\rm tot}(E_\nu)$ is the total neutrino flux at the detector. The integration is performed over the initial neutrino energies interval, $E_\nu\in (0.05,7)$ GeV, with $\sim98\%$ of the flux below $E_\nu\leq2$ GeV. The differential cross section ${d}^2\sigma/{d}E_\gamma{d}E_\nu$ for a neutrino scattering on a single proton in the nucleus is calculated with {\tt MadGraph}. We use this simple approximation of scattering on single protons to evaluate our bounds; a more refined calculation would involve modelling the nuclear responses of $C$ and $H$ in the detector.

The bounds on neutrino polarizability is obtained using the chi-squared function 
\beq
\chi^2(X) = \sum_i\frac{\lp N_i^{\rm MB} - (N_{i,bg}^{\rm MB} + N_i^{\rm NP})\rp^2}{\sigma_i^2}\,,
\eeq
where the sum is over photon energy bins, with $N_i^{\rm MB}$ the number of measured events in bin $i$, $N_{i,bg}^{\rm MB}$ the expected number of SM background events, and $\sigma_i$ the experimental uncertainty \cite{Aguilar-Arevalo:2018gpe,Aguilar-Arevalo:2020nvw}. The maximal allowed number of NP events, $N_i^{\rm NP}$, 
places the bounds on the  Rayleigh operator Wilson coefficients. 
\beq
\label{eq:bound:MB}
\left| \tfrac23 \hat \C_1^{(7)} + \hat \C_2^{(7)} \right| \lesssim 4\times10^{-3}~{\rm GeV}^{-3}.
\eeq
In Fig.~\ref{fig:MB_bounds} we show the photon energy spectrum induced by the Rayleigh interaction with the saturated bound and enlarged by a factor of 10 (red), compared to the expected SM background (blue). The bound in \eqref{eq:bound:MB} is dominated by the large photon energy "tail", since the photon spectrum from the Rayleigh interaction is broad and peaks at $\sim500$ MeV.

For light mediator we keep the full $\phi$ propagator in the calculation of the scattering cross sections, which then leads to the bounds
\beq
c_\nu g_{\phi\g} \lesssim1.2\times10^{-5}\,,~~m_\phi = 1~{\rm GeV}\,;\qquad c_\nu g_{\phi\g} \lesssim4\times10^{-6}\,,~~m_\phi = 1~{\rm MeV, keV, eV}\,.
\eeq
For the 1 GeV benchmark the $\phi$ mass is comparable to the typical momentum exchange, while to the other three benchmarks the $\phi$ mass can be neglected in the propagator.

While an enhanced $\nu A\to \nu X_A+\gamma$ scattering rate is an intriguing possibility in view of the longstanding MiniBoone anomaly \cite{Aguilar-Arevalo:2018gpe,Aguilar-Arevalo:2020nvw}, we note in passing that for a massless neutrino in the final state the photon spectrum does not match the observed low energy anomaly, see Fig.~\ref{fig:MB_bounds}. For $\nu A\to N X_A+\gamma$ scattering, on the other hand, where $N$ is a heavier sterile neutrino, the final state photon would be softer and could potentially match the MiniBoone measurements (for sample of other new physics explanations of the MiniBoone anomaly, see, e.g., Refs.~\cite{Batell:2022xau,Foppiani:2022qsi,Arguelles:2018mtc,Kamp:2022bpt,Vergani:2021tgc,Dentler:2019dhz,Fischer:2019fbw}). We leave the full investigation of such a possibility for future work.

\subsection{Collider constraints }
\label{sec:ColliderConstraints}
There are a number of constraints on neutrino polarizability from measurements of higher energy processes, mostly from producing the $\phi$ mediator on-shell.

\paragraph{Rare meson and tau decays.}
Couplings of $\phi$ to neutrinos generate the three body $M\to\ell\nu\phi$ decays of mesons $M=K, D, B,\ldots,$ via the emission of $\phi$ from the neutrino leg. The decay is  kinematically allowed for $m_\phi < m_M - m_\ell$.
 For $m_\phi=1\,\text{eV}, 1\,\text{keV}, 1\,\text{MeV}$ benchmarks the most stringent constraint of this type comes from  bounds on $K^+\to e^+\nu_e\phi$ decays, giving $c_\nu \lesssim4\times10^{-3}$, while for $m_\phi=1\,\text{GeV}$ the most stringent bound is from $\tau \to \ell \nu \bar \nu \phi$ decays,  leading to $c_\nu\lesssim0.3$, see Ref.~\cite{Blinov:2019gcj} and references within. 
These constraints are shown as dark gray shaded excluded regions in Fig.~\ref{fig:TerrestrialBounds}. 

\paragraph{Neutrinoless double $\beta$ decay.}
The neutrinoless double $\beta$ decay ($0\nu2\beta$) experiments can be used to also search for $0\nu2\beta\phi$ transitions, where $\phi$ is emitted from one of the neutrino lines. Present experimental bounds translate to a constraint $c_\nu \lesssim10^{-5}$~\cite{Blum:2018ljv} for $m_\phi\lesssim2$ MeV, which we show as light gray shaded excluded region in Fig.~\ref{fig:TerrestrialBounds}.

\paragraph{Beam dump experiments.}
Light pseudoscalars coupling to photons can be produced in electron and proton beam dump experiments via Primakoff process, and are then searched for via their decays to two photons. 
In our case $\phi$ has an additional invisible decay channel $\phi \to \nu\nu$, which dilutes the $\gamma\gamma$ signal, if the decays to neutrinos dominate. 
Rescaling the bounds on ALP couplings to photons, $g_{\phi\g}(m_\phi)$, from Ref.~\cite{Dolan:2017osp} with the diphotonic branching ratio ${\Br}(\phi\to\g\g)$,  Eq.~\eqref{eq:Brfraction}, gives the bound
\beq\label{eq:rescaling}
g_{\phi\g}(m_\phi) < {g_{\phi\g}^{\rm b.d.}(m_\phi)}/{ \sqrt{{\Br}(\phi\to\g\g)} }\,.
\eeq
where $g_{\phi\g}^{\rm b.d.}(m_\phi)$ is the bound quoted in~\cite{Dolan:2017osp}. The constraint in \eqref{eq:rescaling} is shown in Fig.~\ref{fig:TerrestrialBounds} as  the dark red excluded regions. The exception to rescaling rule in \eqref{eq:rescaling} is the newer NA64 analysis \cite{NA64:2020qwq} that included both recoils due to invisible $\phi$ escaping the detector as well as the $\phi\to \gamma\gamma$ events. Assuming $\phi$ only has photon couplings the NA64 analysis is less sensitive than the other beam dump experiments for the four $\phi$ mass benchmarks we consider. For this reason and because the NA64 result is difficult to recast for the more general case of an arbitrary invisible branching ratio, we do not include it in Fig.~\ref{fig:TerrestrialBounds}. 

\paragraph{Peripheral heavy ion collisions.}
The production of  ALP with couplings to photons is $Z^4$ coherently enhanced  in ultra-peripheral ion collisions~\cite{Knapen:2016moh}. The ALP was then searched for in the two photon decay channel \cite{ATLAS:2019azn}.
This leads to a significant bound only for heavy ALPs, with mass $m_\phi \gtrsim 10$ GeV, thus we do not show it in~Fig.~\ref{fig:TerrestrialBounds}.

\paragraph{Search for $e^+e^-\to 3\gamma$.} Belle II collaboration performed a search for ALPs decaying to two photons, $e^+e^-\to\g (\phi\to\g\g)$~\cite{PhysRevLett.125.161806}, and set stringent bounds  on $g_{\phi\g}$ in the mass range $0.2\lesssim m_\phi \lesssim9.5$ GeV, assuming $Br(\phi\to \gamma\gamma)=100\%$. At the mass benchmark $m_\phi = 1$ GeV, the coupling to photons is constrained to be $g_{\phi\g}\lesssim10^{-3}~{\rm GeV}^{-1}$ for values $c_\nu$ small enough that $Br(\phi\to \nu\nu)\ll Br(\phi\to \gamma\gamma)$, while for larger $c_\nu$ we rescale the Belle II bound as in \eqref{eq:rescaling}.
The excluded region is shown with red in the bottom right plot of Fig.~\ref{fig:TerrestrialBounds}. Note that a similar bound on the $m_\phi = 1$ GeV mass benchmark follows from searches for anomalous $2\g$ and $3\g$ signal at LEP~\cite{Jaeckel:2015jla}.

\paragraph{Invisible decays of spin-0 particles.}
The $\nu\nu \g\g$ interaction would induce $S \to \gamma\gamma \to \nu\nu$ decays, i.e., the $S \to \gamma\gamma$ transition leads at one loop to $S \to \nu\nu$ decays, where for the initial spin-0 particle we consider $S=\pi^0, B^0$ and the Higgs boson, $h$, and assume that the EFT limit for $\nu\nu\gamma\gamma$ interaction applies. The SM rates to $S\to \nu \nu$ are negligible \cite{Bhattacharya:2018msv}.
If just one combination of Rayleigh Wilson coefficients, $\hat \C_{1}^{\rm Re}$ or $\hat \C_{2}^{\rm Re}$ in Eq. \eqref{eq:C12:Re}, contributes, then (cf. Appendix \ref{sec:RareDecays}) 
\beq
\label{eq:CRe:bound}
\big|\hat \C_{a}^{\rm Re} \big|\leq k_a \,\frac{128\pi^2}{\alpha}\frac1{m_S^3}\sqrt{\frac{\mbox{Br($S\to$ inv.)}}{\mbox{Br}(S\to\gamma\gamma)}}\,,
\eeq
where $k_1 = 3/2$ and $k_2 = 1$. This is the case for $\pi^0$ and $h$, leading to 
\beq
\pi^0:\quad \big|\hat \C_{2}^{\rm Re} \big|\leq 4.7\cdot 10^3 \mbox{ GeV}^{\,-3},\qquad\qquad
h :\quad\big| \hat \C_{1}^{\rm Re}\big|\leq 1.2 \mbox{ GeV}^{\,-3},
\eeq
from experimental bounds $\text{Br}(\pi^0\to\text{inv})\leq4.4\cdot 10^{-9}$, $\text{Br}(h\to \text{inv})\leq0.19$~\cite{Workman:2022ynf}, along with $\text{Br}(\pi^0\to\gamma\gamma)\approx0.99$ and the SM prediction ${\rm Br}_{\rm SM}( h\to\g\g ) = 2.3\cdot10^{-3}$~\cite{LHCHiggsCrossSectionWorkingGroup:2013rie}, using the fact that the Higgs properties are consistent with the SM. From $\text{Br}(B^0\to\text{inv})\leq2.4\cdot10^{-5}$\cite{Workman:2022ynf} we obtain, on the other hand,
\beq
\label{eq:B0:constr}
B^0:\qquad \big|\tfrac{2}{3}\hat \C_{1}^{\rm Re} + \hat \C_{2}^{\rm Re}\big|\leq 3.7 \cdot 10^4 \mbox{ GeV}^{\,-3},
\eeq
assuming $B^0\to \gamma\gamma$ is as predicted in the SM, see Appendix \ref{sec:RareDecays} for details. The constraint in \eqref{eq:B0:constr} may therefore change, if new physics affects $B^0\to \gamma\gamma$ decays. Note that at present the above bounds are quite weak, and the use of EFT may be questioned.  For light mediator $\phi$, with $m_S\neq m_\phi$, the above results still apply, but with replacement $\hat \C_{1(2)}^{\rm Re}\to 2 \sum_{ij} \Re c_\nu^{ij} c_\gamma^{(')} [f_\phi(m_S^2-m_\phi^2)]^{-1}$ (the case $m_S\simeq m_\phi$ is more involved, and we do not attempt it here).

\paragraph{Monophoton searches.}
\label{sec:BaBar:bounds}
Neutrino polarizability leads to a monophoton signature in $e^+e^-$ collisions. This is either due to $e^+e^-\to\g^*\to\g\,\nu\nu$ scattering, generated by Rayleigh operators in the EFT limit,  or by an on-shell production of the light $\phi$ mediator, $e^+e^-\to\g^*\to\g\,\phi$, where $\phi$ then decays to two neutrinos or escapes the detector. 

We recast the BaBar monophoton search~\cite{BaBar:2017tiz} for the case of light mediator $\phi$. The results in~\cite{BaBar:2017tiz} were interpreted in terms of the bounds on dark photon $A'$  mixing parameter, $\varepsilon$. In Appendix \ref{sec:BaBar:notesMichele} we give the $e^+e^-\to A'\gamma$ differential cross section, $d\sigma_{A'\gamma}/d\cos\theta$, where $\theta$ is the emerging angle of the photon. The BaBar analysis restricted it to $|\cos\theta|<0.6$. For this range we can take the limit $m_e/\sqrt{s}\to 0$ without encountering a singularity at $\sin\theta=0$. We find 
\beq\label{eq:BaBar:diffxsec}
\frac{d\sigma_{A'\gamma}}{d\cos\theta}=\frac{4\pi \alpha^2\varepsilon^2}{s^2(s-m_{A'}^2)}\Big[\frac{s^2+m_{A'}^4}{\sin^2\theta}-\frac{(s-m_{A'}^2)^2}{2}\Big],
\eeq
This result agrees with \cite{Boehm:2003hm,Borodatchenkova:2005ct} and the in the limit $m_{A'}\to 0$, with the known $e^+e^-\to \gamma\gamma$ expression. 

For each benchmark value of $m_\phi$ we compare the above cross section, integrated over $\cos\theta\in[-0.6,0.6]$ and take $m_{A'}=m_\phi$, with the $e^+e^-\to \gamma \phi$ cross section,
\beq
\label{eq:dsigmaphigamma}
\frac{d\sigma_{\phi\gamma}}{d\cos\theta}=\frac{\alpha^3}{256\pi^2}\left(\left|\frac23\frac{c_\gamma} {f_\phi}\right|^2+\left|\frac{c^\prime_\gamma} {f_\phi}\right|^2\right)\left(1-\frac{m_\phi^2}{s}\right)^3 \lp 3 + \cos2\theta\rp\,,
\eeq
integrated over the same range $\cos\theta\in[-0.6,0.6]$. Our expression is twice as large compared to expressions in the literature \cite{Marciano:2016yhf,Dolan:2017osp}. We give the details of the calculation Appendix \ref{sec:BaBar:notesMichele} and encourage the community to reconsider constrains that rely on it.

Using the bound $\varepsilon<9.5 \times 10^{-4}$ from~\cite{BaBar:2017tiz} valid for all four benchmark masses gives
\begin{equation}
\left(\left|\frac23\frac{c_\gamma} {f_\phi}\right|^2+\left|\frac{c^\prime_\gamma} {f_\phi}\right|^2\right){\cal B}(\phi \to \nu\nu)\leq 0.012 \mbox{ GeV}^{\,-2}\,,
\end{equation}
with branching ratio to neutrinos given by Eqs.~\eqref{eq:phidecayrate:gg}, \eqref{eq:phidecayrate:nunu}. The corresponding excluded region is denoted with orange in Fig. \ref{fig:TerrestrialBounds}.

In the EFT limit, neutrino polarizability induces the $2\to3$ scattering, $e^+e^-\to\g\nu\nu$, i.e., in a continuous photon spectrum. Unfortunately BaBar did not provide publicly available measured monophoton rates as a function of the invisible mass. Instead, we use Fig. 1 in  Ref.~\cite{BaBar:2017tiz}, which reports the best fit value of $\epsilon^2$ as a function of $m_{A'}$.  We convert  the best fit $\epsilon^2$ values using Eq.~\eqref{eq:BaBar:diffxsec} (integrated over the angular acceptance)  to the best fit values of the allowed $e^+e^-\to\g A'$ cross section, $\sigma_i$, where $i$ runs over all the $m_{A'}$ bins (and the same for the $1\sigma$ errors on $\epsilon^2$ that get translated to $1\sigma$ errors on the cross sections, $\delta \sigma_i$).   From this we can construct a $\chi^2$ function
\beq 
\chi^2(\hat\C_2^{(7)}) = \sum_i \lp \frac{\sigma_i - \int_i dm_{\nu\nu}\, d\sigma(\gamma\nu\nu)/dm_{\nu\nu}}{\delta \sigma_i} \rp^2\,,
\eeq
where $\int_i dm_{\nu\nu}\, d\sigma(\gamma\nu\nu)/dm_{\nu\nu}$ gives the rate in $i$-th $m_{A'}=m_{\nu\nu}$ bin from neutrino polarizability induced $e^+e^-\to\g\nu\nu$ scattering, and depends on $\hat\C_2^{(7)}$, see details in Appendix \ref{sec:BaBar:notesMichele}.
Requiring  $\chi^2(\hat\C_2^{(7)}) \leq 2.71$, gives the 90$\%$ CL bound
\beq 
\left|\frac23\hat\C_1^{(7)} + \hat\C_2^{(7)}\right| \leq 0.2~{\rm GeV}^{-3}\,.
\eeq

\section{UV models of enhanced neutrino polarizability}
\label{sec:model}

Next, we discuss several UV models that lead to enhanced contributions to the neutrino Rayleigh operators. In Section \ref{sec:minimal:Majoron} we first review the minimal singlet majoron model. This does not predict large neutrino polarizability, but can be used as a useful benchmark. The other models we consider below, the majoron as a QCD axion and the majoron from non-minimal inverse see-saw models, discussed in Sections \ref{sec:Majoron:QCD:axion}-\ref{sec:enhanced:polariz:model},  have enhanced neutrino polarizabitily relative to the minimal majoron model. All of the models rely on spontaneously broken global lepton number, $U(1)_L$, and the associated Goldstone boson, the majoron \cite{Chikashige:1980ui,Gelmini:1980re}, whose tree level exchanges lead to enhanced neutrino polarizability, in the same way as for the simplified model in Section \ref{subsec:toymodel}.

\subsection{Minimal singlet Majoron}
\label{sec:minimal:Majoron}
The minimal singlet majoron model assumes that the SM neutrinos, $\nu_i$, are Majorana fermions, and that their masses are suppressed as the result of the type I see-saw, with spontaneously broken lepton number \cite{Chikashige:1980ui}. The SM is supplemented by three right-handed neutrinos $N_{R,i}$ and a singlet scalar, $S$, that carries lepton number $L=-2$.\footnote{We use a short-handed notation $N_{R,i}=P_R N_i$, where $N_i$ is the four component Majorana fermion field, where as in the rest of the paper we use the notational conventions from Ref. \cite{Dreiner:2008tw}.} The terms in the Lagrangian relevant for the neutrino masses are thus given by
\beq
\label{eq:see-saw}
{\cal L}=-\bar L y N_R H-\frac{1}{2}\bar N_R^c\lambda N_RS+{\rm h.c.},
\eeq
where $y,\lambda$ are $3\times 3$ complex matrices. 
The lepton number is spontaneusly broken once $S$ obtains a vev, $S=(f_\phi+\sigma+i\phi)/\sqrt2$. The radial mode $\sigma$ is assumed to be heavy with mass $m_\sigma\sim{\mathcal O}(f_\phi)$ and not relevant for our discussion. The majoron, $\phi$, is the pNGB of the spontaneously broken lepton number. Its mass term, ${\cal L}\supset -m_\phi^2 \phi^2/2$, represents a (small) explicit breaking of the shift symmetry, where $m_\phi$ is taken as a free parameter \cite{Gu:2010ys,Frigerio:2011in} (it could arise from Planck scale physics since gravity is expected to break global symmetries \cite{Akhmedov:1992hi,Rothstein:1992rh,Alonso:2017avz}).

The SM neutrino masses induced by the interactions \eqref{eq:see-saw} are parametrically  given by $m_{\nu_i}\sim y^2 v^2/\lambda f_\phi$ so that for $\lambda f_\phi\gg v$ (the see-saw limit) the couplings $y$ can be large, which is one of the main motivations for contemplating the see-saw models. The couplings of majoron to the SM fermions are given by \cite{Heeck:2019guh}
\beq
\label{eq:Majoron:SMfermions}
{\cal L}_\phi \supset\frac{i \phi}{2f_\phi} m_{f_i} c_{f_i} \bar f_i \gamma_5 f_i+\cdots. 
\eeq
The majoron--SM-fermion couplings are thus suppressed by the SM fermion masses, $m_{f_i}$, while the dimensionless coefficients are, in the see-saw limit,
\beq
c_{\nu_i}=1, \qquad c_{d_i}=-c_{u_i}=\frac{1}{16\pi^2}\Tr(yy^\dagger),\qquad c_{\ell_i}=\frac{1}{16\pi^2}\big[\Tr(yy^\dagger)-(yy^\dagger)_{ii}\big],
\eeq
for the couplings to the SM neutrinos, down and up quarks, and the charged leptons, respectively.  
The  ellipses in \eqref{eq:Majoron:SMfermions} denote the flavor off-diagonal terms for charged leptons that we do not display, but can be found in \cite{Heeck:2019guh}. The majoron couplings to the SM neutrinos are generated at tree level, while the couplings to quarks and charged leptons are generated at one loop, cf. Eq. \eqref{eq:Majoron:SMfermions}. Still, the couplings to quarks and charged leptons can still be larger if the Yukawa couplings $y$ are sizable. That is, the couplings to quarks and charged leptons are parametrically enhanced for $f_\phi\gg v$ by $\sim \lambda f_\phi m_{f_i}/(4\pi v)^2$ compared to the $\phi$ couplings to neutrinos. 

Couplings of the majoron to photons and gluons are generated at two loops. In the limit of a light majoron, $m_\phi \ll m_{\ell_i, u_i, d_i}$, they match onto the dimension 7 operators $\big(\partial^2 \phi \big)F_{\mu\nu}\tilde F^{\mu\nu}$ and $\big(\partial^2 \phi \big) G^a_{\mu\nu}\tilde G^{a\mu\nu}$ and are thus suppressed both by the two loop factors $\alpha_{{\rm em}}/(16\pi^2)^2$ and $\alpha_{s}/(16\pi^2)^2$, respectively, as well as by the majoron mass, $m_\phi^2/m_{f_i}^2$, for a contribution from a SM charged fermion $f_i$ running in the loop. For a heavier majoron the latter suppression is lifted and in the corresponding transition amplitude $m_\phi^2/m_{f_i}^2$ is replaced by an ${\mathcal O}(1)$ factor. 

\subsection{Majoron as a QCD axion}
\label{sec:Majoron:QCD:axion}
 If the spontaneously broken $U(1)_L$ is  anomalous under QCD, the majoron will act as the QCD axion and solve the strong CP problem, with $U(1)_L$ identified as the Peccei-Quinn symmetry \cite{Mohapatra:1982tc,Langacker:1986rj,Shin:1987xc,Ma:2001ac,Ma:2017vdv}. Recent concrete realization of this idea can be found in \cite{Ma:2017vdv}, where the SM was supplemented by a set of color octet fermions $\Psi_R^A$ (with lepton number charge $L=1$, electroweak singlets), color octet scalars, $\Phi^A$ ($L=0$, doublets of $SU(2)_L$), and an electroweak and color singlet scalar $S$ ($L=2)$. The lepton number is spontaneously broken once $S$ obtains a vev, $\langle S\rangle =f_\phi/\sqrt2$, giving Majorana mass contribution, $M_{\Psi_k}=y_\Psi^k \langle S\rangle$, to the color octet fermion mass matrix as a result of the term  ${\cal L}\supset -\frac{1}{2} y_\Psi^i S \bar \Psi_{iR}^c \Psi_{iR}+{\rm h.c.}$, in the Lagrangian. 

At one loop the breaking of $U(1)_L$ generates the Majorana mass matrix for the SM neutrinos. In the limit of heavy color octet scalars, with almost degenerate mass $M_0\gg M_{\Psi_k}\gg v$, the radiatively generated neutrino mass matrix is given by \cite{Ma:2017vdv},
\beq
\big(m_\nu\big)_{ij}=\frac{1}{4\pi^2} \sum_k h_\Psi^{ik}h_\Psi^{jk} y_\Psi^k \langle S\rangle \frac{\Delta M_0^2}{M_0^2},
\eeq
where the summation is over color octet fermion, and $\Delta M_0^2=2\lambda_5 v^2$ is the mass splitting between the CP-even and CP-odd components of the color octet scalar
 due to the term ${\cal L}\supset - \frac{1}{2}\lambda_5 (H^\dagger \Phi^A)^2+{\rm h.c.}$ in the scalar potential. Note that the neutrino masses are proportional to the $\Delta L=2$ breaking vev, $m_\nu \propto \langle S\rangle$, to two insertions of electroweak vev, $m_\nu \propto v^2$, as well as to the Yukawa interactions between the SM lepton doublets and the new color octet fields, ${\cal L}\supset h_\Psi^{ij}{}^* \bar L_i \Psi_{jR}^A  \Phi^{A\dagger}+{\rm h.c.}$. Taking $f_\phi\sim {\mathcal O}({\rm TeV})$ and all couplings to be ${\mathcal O}(10^{-2})$ the observed neutrino masses are obtained for $M_0\sim {\mathcal O}(10\,{\rm TeV})$.
 
 The majoron $\phi$ is part of the singlet, $S(x)=\frac{1}{\sqrt 2}\big(f_\phi+\sigma(x)\big)\exp\big(i \phi(x)/f_\phi\big)$, and has interactions to neutrinos suppressed by the neutrino masses, 
 \beq
 \label{eq:min:Majoron:nus}
 {\cal L}_\phi\supset - \frac{i}{2}(m_\nu)_{ij} \big(\bar \nu_i P_L \nu_j\big) \frac{\phi}{f_\phi}+{\rm h.c.}.
 \eeq
In the notation of  Eq. \eqref{eq:lagr:phi} the coupling of $\phi$ to neutrinos is  thus given by 
\beq
\label{eq:min:Majoron:gphi}
 c_\nu^{ij}= - i \frac{(m_\nu)_{ij}}{f_\phi}.
 \eeq

 The triangle anomaly induces couplings of the majoron to gluons, 
 \beq
 \label{eq:Majoron:axion:GG}
 {\cal L}_\phi \supset -\frac{3 n_\Psi \alpha_s}{8\pi}  \frac{\phi}{f_\phi}\tilde G_{\mu\nu}^A G^{A\mu\nu},
 \eeq
 where $n_\Psi$ is the number of color octet fermions. 
 The couplings to gluons are not suppressed by the majoron mass, since they are generated from the QCD anomaly, unlike the minimal singlet majoron case, Section \ref{sec:minimal:Majoron}.  At low energies, below QCD confinement, the above interaction induces interactions of $\phi$ with nucleons and pions, and other hadronic states. It also leads to the majoron mass, in the same way as for the standard QCD axion, 
 \beq
 m_\phi\simeq 6 \,{\rm keV}\times \biggr(\frac{1\,{\rm TeV}}{f_\phi/(3n_\Psi)}\biggr).
 \eeq
 as well as to the couplings to photons, 
 Eq. \eqref{eq:lagr:phi}, 
with 
\beq
c_{\gamma}'\simeq 2.0 \times 3 n_\Psi.
\eeq

At energies below $m_\phi$ the majoron can be integrated out, giving rise to the neutrino polarizability with 
\beq
\frac{\C_{2,ij}^{(7)}}{\Lambda^3}=i  \frac{(m_\nu)_{ij}c_{\gamma}' }{f_\phi^2 m_\phi^2}\simeq i \biggr(\frac{1}{81\,{\rm GeV}}\biggr)^3\times \frac{1}{n_\Psi} \times \frac{(m_\nu)_{ij}}{0.1\,{\rm eV}}.
\eeq
Note that the dependence on $f_\phi$  drops out, due to the relation between $m_\phi$ and $f_\phi$, with $m_\phi f_\phi\sim m_\pi f_\pi $ a constant that is fixed entirely by the QCD dynamics apart from  the $n_\Psi\sim {\mathcal O}(1)$ factors that depends on the UV physics. The PQ breaking scale $f_\phi$ thus does not determine directly the effective suppression scale of the Rayleigh operator, $\Lambda$, but rather just determines the range of the validity of the EFT, via the requirement $E,q\gg m_\phi\propto 1/f_\phi$. The effective scale $\Lambda$ is given by a combination of QCD and neutrino mass scales, $\Lambda\sim \big(m_\pi^2 f_\pi^2/m_\nu\big)^{1/3}$, which accidentally turns out to be close to the weak scale. 
 
While the majoron that is the QCD axion is an example of the model that leads to enhanced neutrino polarizability, with the effective scale $\Lambda$ much smaller than the UV scales $f_\phi$ and $M_0$, it does not map straightforwardly onto the phenomenological analyses forming the bulk of the  present paper since  QCD axion couples to gluons, Eq. \eqref{eq:Majoron:axion:GG}, which was not taken into account in our analysis.

In the final two examples: the non-minimal majoron models in Sections \ref{sec:Majoron:inverse:see-saw} and \ref{sec:enhanced:polariz:model}, much lower effective scales $\Lambda$ can be achieved compared to the ones encountered in the minimal see-saw majoron and the majoron as the QCD axion model.

\subsection{Majoron from inverse see-saw with extra triplet fermions}
\label{sec:Majoron:inverse:see-saw}
In this model both the neutrino mass generation sector is enlarged as is the sector that leads to couplings of majoron to the photons. 

\subsubsection{The inverse see-saw sector}
The sector relevant for the generation of the SM neutrino masses contains three generations of left-handed and right-handed Weyl fermions, $N_{R,i}$ and $N_{L,i}$, $i=1,2,3$, singlets under the SM gauge group, and carry a global lepton number $L=+1$. The interaction Lagrangian is given by (suppressing generation indices)
 \beq
 \label{eq:calLy'}
 -{\cal L}_Y=y_\nu \bar L \tilde H^\dagger N_R +\bar N_L M_N N_R +\frac{\lambda_R}{2}\bar N_R^c S N_R+\frac{\lambda_L}{2} \bar N_L S N_L^c+{\rm h.c.},
 \eeq
 where $L_i$ are the SM lepton doublets, and $H$ is the SM Higgs doublet.
The Yukawa coupling $y_\nu$ and neutrino Dirac mass matrix $M_N$ are general $3\times 3$ complex matrices, while $\lambda_{L,R}$ are symmetric $3\times 3$ complex matrices.  

The lepton number is spontaneusly broken once $S$ obtains a vev, 
\beq
\label{eq:S:vev}
S=(f_\phi+\sigma+i\phi)/\sqrt2.
\eeq
This then gives the following neutrino mass matrix in the basis $\{\nu_L^c, N_R, N_L^c\}$, see, e.g. \cite{Dev:2012sg}, 
\beq
{\cal M}=
\begin{pmatrix}
0 &M_D & 0 \\
M_D^\top & \mu_R & M_N^\top \\
0 & M_N & \mu_L
\end{pmatrix},
\eeq
where we shortened the notation to $M_D=y_\nu v/{\sqrt 2}$, $\mu_{L,R}=\lambda_{L,R} f_\phi/\sqrt{2}$. Without loss of generality we can work in the basis, where $M_N$ is diagonal. We will assume the hierarchy $\mu_{R,L}\ll M_D\ll M_N$, to be understood as the hierarchy among all the eigenvalues of the corresponding $3\times 3$ complex matrices. 

We start the analysis with the one-generation case where the three parameters $\mu_{L}, M_D$, and $M_N$ can be made real via phase redefinitions of $\nu_L^c, N_R, N_L^c$, while $\mu_R$ is a complex parameter. Up to corrections of higher order in $\mu_{L,R}/M_{N,D}$ the lightest neutrino mass eigenstate is given by
\beq
\nu=c_\alpha \nu_L -s_\alpha N_L, \qquad t_\alpha\simeq \frac{M_D}{M_N},
\eeq
where we abbreviated $c_\alpha\equiv \cos\alpha$, $s_\alpha \equiv \sin \alpha$, $t_\alpha\equiv \tan\alpha$. In order for $\nu$ to be predominantly composed of  the neutral component of the electroweak doublet with only a small admixture of the sterile neutrino, we require $M_D\ll M_N$. The corresponding mass is
\beq
m_\nu\simeq \mu_L s_\alpha^2.
\eeq
The two heavy neutrinos are mass degenerate, with masses equal to $\sqrt{M_N^2+M_D^2}$, up to $\mu_{L,R}$ suppressed corrections. Note that the SM neutrino masses are proportional to the lepton number breaking parameter $\mu_L=\lambda_L f_\phi/\sqrt2$, and vanish in the limit $f_\phi\to 0$, as expected. The neutrino masses can now be small either due to a small value of $f_\phi$, the smallness of the mixing angle $s_\alpha$ (i.e., $M_D /M_N\ll1$), or a combination of the two. 

These results extend trivially to the case of three generations. To linear order in $M_D/M_N$ the mass eigenstates, the left-handed fields $\nu, N_1, N_2$, are expressed in terms of the initial states as
\begin{align}
\nu_L^c&=\nu^c+ \big(M_N^{-1}\big)^T M_D \frac{1}{\sqrt{2}}\big(N_1^c+N_2^c\big),
\\
N_L^c&=-M_N^{-1} M_D^T\nu^c+ \frac{1}{\sqrt{2}}\big(N_1^c+N_2^c\big),
\\
N_R&= \frac{1}{\sqrt{2}}\big(N_1^c-N_2^c\big).
\end{align}
The light neutrino mass terms are then given by
\beq
{\cal L}\supset \frac{1}{2} \bar \nu M_D \big(M_N^{-1}\big)^T \mu_L M_N^{-1} M_D^T P_R \nu^c +{\rm h.c.}
\eeq
and are proportional to $\mu_L$ lepton number violating parameter, while the dependence on $\mu_R$ only enters at higher orders. 

The interactions with the majoron can be obtained by replacing $\mu_L\to i \mu_L \phi/f_\phi$ in the mass terms, which then gives for the majoron couplings to neutrinos, 
 \beq
 \label{eq:inv:Majoron:nus}
 {\cal L}_\phi\supset - \frac{i}{2}(m_\nu)_{ij} \big(\bar \nu_i P_L \nu_j\big) \frac{\phi}{f_\phi}+{\rm h.c.}\,.
 \eeq
In the notation of  Eq. \eqref{eq:lagr:phi} the coupling of $\phi$ to neutrinos is  thus given by 
\beq
 c_\nu^{ij}= - i \frac{(m_\nu)_{ij}}{f_\phi}.
 \eeq
On the face of it, this is the same result as for the minimal majoron, cf. Eqs. \eqref{eq:min:Majoron:nus}, \eqref{eq:min:Majoron:gphi}. However, there is a major difference between the two, namely that in the inverse see-saw model the smallness of neutrino masses, $m_\nu$, can be due to the smallness of $f_\phi$. In principle, $f_\phi$ can be as small as $m_\nu$ and thus $c_\nu$ as large as $c_\nu \sim {\mathcal O}(1)$. Numerically, the bounds on self-interactions of neutrinos limit the value of $c_\nu$  to be well below 1, cf. Section \ref{sec:cosmology}.

\subsubsection{Couplings to photons via heavy electroweak triplets}
\label{sec:ew:triplets}
We assume that the field content of the theory contains a set of heavy $SU(2)_L$ triplet fermions $\Psi_R^a$, charged under lepton number, $L=-1$. They obtain their masses through interactions with the scalar $S$,
\beq
{\cal L}_{\Psi_R}=-\frac{1}{2} y_\Psi S \overline{\big(\Psi_R^a\big)^c} \Psi_R^a+{\rm h.c.},
\eeq
after $S$ obtains a vev, Eq.~\eqref{eq:S:vev}, giving $M_{\Psi}=y_\Psi \langle S\rangle=y_\Psi f_\phi/\sqrt 2$. This generates a coupling of PNGB $\phi$ with the $SU(2)_L$ gauge fields through anomaly, 
\beq
 \label{eq:see:saw:WW}
 {\cal L}_\phi \supset -\frac{9 \alpha}{64\pi}   n_\Psi \frac{\phi}{f_\phi}\tilde W_{\mu\nu}^a W^{a\mu\nu},
 \eeq
 where $n_\Psi$ is the number of $\Psi_R^a$ generations, and $W^a_{\mu\nu}$ is the $SU(2)_L$ field strength. This gives couplings of $\phi$ to $W^\pm$, $Z^0$ and photons, where for the latter
 \beq
 \label{eq:see:saw:FF}
 {\cal L}_\phi \supset -\frac{9 \alpha}{64\pi} n_\Psi  s_w^2 \frac{\phi}{f_\phi}\tilde F_{\mu\nu} F^{\mu\nu},
 \eeq
 where $s_w=\sin \theta_w$, with $\theta_w$ the weak mixing angle. 
 
 Since $\Psi_R^a$ carry electroweak charges they could be produced in $e^+e^-$ collisions at LEP or in $pp$ collisions at the LHC. The bounds on their mass  depends on the $\Psi_R^a$ decay channels, and is thus model dependent. In general, we expect the bound to be in the range of a few 100 GeV. Taking this as the typical lower bound on $f_\phi$, this would then translate to the following typical size of the Rayleigh operator, 
 \beq
 \begin{split}
\frac{\C_{2,ij}^{(7)}}{\Lambda^3}&=i  \frac{(m_\nu)_{ij}c_{\gamma}' }{f_\phi^2 m_\phi^2}=i  \frac{9}{8}\frac{(m_\nu)_{ij}n_\Psi s_w^2 }{f_\phi^2 m_\phi^2}
\\
&\simeq i \biggr(\frac{1}{7.3\,\text{GeV}}\biggr)^3 \biggr(\frac{100\,\text{GeV}}{f_\phi}\biggr)^2\times \biggr(\frac{1\,\text{keV}}{m_\phi}\biggr)^2 \times n_\Psi\times \frac{(m_\nu)_{ij}}{0.1\,{\rm eV}}.
\end{split}
\eeq
That is, for majoron mass $m_\phi\sim {\mathcal O}(\text{keV})$ the effective suppression scale of the Rayleigh operator is only $\Lambda\sim \mathcal{O}(10\,\text{GeV})$, and is parametrically smaller than the $U(1)_L$ breaking scale, $f_\phi$. For $m_\phi\sim \mathcal{O}(1\,\text{eV})$ the effective suppression scale would be $\Lambda\sim \mathcal{O}(0.1\,\text{GeV})$. The majoron mass is an explicit $U(1)_L$ breaking term and is treated as a free parameter. On general grounds one expects $m_\phi\ll f_\phi$ so that the explicit breaking is smaller than the spontaneous symmetry breaking, and thus $U(1)_L$ is a good approximate symmetry.

\subsection{Enhanced neutrino polarizability from $U(1)_L\times U(1)'$}
\label{sec:enhanced:polariz:model}
In non-minimal versions of the above model one can obtain even larger parametric enhancements of the neutrino polarizability. Let us consider an example of a model with two global $U(1)$ factors, $U(1)_L\times U(1)'$, a simple modification of the inverse see-saw model in Section \ref{sec:Majoron:inverse:see-saw}, but with two different scalars in the inverse see-saw sector and the electroweak triplet sector. That is, the model contains  two scalar SM gauge singlets, $S$ and $S'$, where the scalar $S$ carries a global charge $L=2$ under $U(1)_L$, while $S'$ carries a charge of $+2$ under $U(1)'$. Otherwise the field content is the same as in Section \ref{sec:Majoron:inverse:see-saw}. The SM is extended by 
three generations of left-handed and right-handed sterile neutrinos, $N_{R,i}$ and $N_{L,i}$, $i=1,2,3$, singlets under the SM gauge group, with global lepton number $L=+1$, and a set of $n_\Psi$ weak triplets that carry a charge $-1$ under $U(1)'$. 

 The $U(1)_L$ and $U(1)'$ are broken by $S$ and $S'$ once these obtain vevs, $S=(f_\phi+\sigma+i\phi)/\sqrt2$ and $S'=(f_\phi'+\sigma'+i\phi')/\sqrt2$. In general, the two vevs can be very different, which can be phenomenologically beneficial.  In particular, it is possible to have $f_\phi\ll f_\phi'$, which would explain the smallness of neutrino masses.\footnote{We set aside the question of a hierarchy problem in the scalar potential. In general one would need to assume that some terms, such as $S^\dagger S S'^\dagger S'$ or $S^\dagger S H^\dagger H$, are suppressed by small couplings.}
 Such a hierarchy of vevs would then also lead to an enhancement of Rayleigh operators.
  
More explicitly, the Lagrangian of the model is the same as in Section \ref{sec:Majoron:inverse:see-saw}, except that in subsection \ref{sec:ew:triplets} we should replace $S\to S'$, $\phi\to \phi'$, $f_\phi\to f_\phi'$. The interactions of $\phi$ and $\phi'$ with the SM particles are thus given by
 \beq
 \label{eq:inv:Majoron:nus:prime}
 {\cal L}_{\rm int}\supset - \frac{i}{2}(m_\nu)_{ij} \big(\bar \nu_i P_L \nu_j\big) \frac{\phi}{f_\phi}+{\rm h.c.},
 \eeq
and
\beq
 \label{eq:see:saw:WW:prime}
 {\cal L}_{\rm int} \supset -\frac{9 \alpha}{64\pi}   n_\Psi \frac{\phi'}{f_\phi'}\tilde W_{\mu\nu}^a W^{a\mu\nu}\to -\frac{9 \alpha}{64\pi} n_\Psi  s_w^2 \frac{\phi'}{f_\phi'}\tilde F_{\mu\nu} F^{\mu\nu}.
 \eeq
 
 As in Section \ref{sec:Majoron:inverse:see-saw},  we assume that the dominant explicit breaking of the global symmetry  $U(1)_L\times U(1)'$ is given by the masses of the two PNGBs. Completely generally, the mass term is given
 \beq
 \label{eq:Lm:break}
 {\cal L}_m=-\frac{1}{2}m_\phi^2 \phi^2 -\frac{1}{2}m_\phi'^2 \phi'^2-m_{\phi\phi'}^2 \phi \phi',
 \eeq 
 leading to two mass eigenstates, $m_{\phi_1,\phi_2}$,  with $\phi_1=c_\theta \phi +s_\theta \phi'$, $\phi_2=-s_\theta \phi +c_\theta \phi'$, where $c_\theta=\cos\theta$, $s_\theta=\sin_\theta$, and $\theta$ the mixing angle. In principle, all the explicit symmetry breaking terms in \eqref{eq:Lm:break} can be of comparable size, and thus the mixing angle large. 
 
 It is instructive to calculate the $\nu\nu\to \gamma\gamma$ amplitude due to tree level exchanges of  $\phi_{1,2}$,
 \beq
 {\cal M}_{\nu\nu\to \gamma\gamma}=i  \frac{9}{8}\frac{(m_\nu)_{ij} }{f_\phi  f_\phi'} n_\Psi s_w^2 c_\theta s_\theta
 \biggr(\frac{1}{q^2-m_{\phi_1}^2}-\frac{1}{q^2-m_{\phi_2}^2}\biggr),
 \eeq
 where $q^\mu$ is the sum of the initial neutrino momenta. In the center of mass of the  $\nu\nu$ collision it is given by $q^\mu=(2E_\nu,\vec 0)$, where $E_\nu$ is the neutrino  energy.  For high energy collisions, $q^2\gg m_{\phi_{1,2}}^2$, the scattering amplitude scales as ${\cal M}\propto m_{\phi_{1,2}}^2/q^4$, while for low energy processes, $q^2\ll m_{\phi_{1,2}}^2$, it matches onto the neutrino polarizability operator, with
\beq
\frac{\C_{2,ij}^{(7)}}{\Lambda^3}=i  \frac{9}{8}\frac{(m_\nu)_{ij} }{f_\phi  f_\phi'} n_\Psi s_w^2 c_\theta s_\theta
 \biggr(\frac{1}{m_{\phi_2}^2}-\frac{1}{m_{\phi_1}^2}\biggr).
\eeq
Numerically, since $f_\phi$ and $f_\phi'$ can take very different values, the Rayleigh operator can be suppressed by a light effective scale. For instance, taking $m_{\phi_1}\ll m_{\phi_2}$ for simplicity, 
\beq
\frac{\big|\C_{2,ij}^{(7)}\big|}{\Lambda^3}\simeq \biggr(\frac{1}{16\,\text{MeV}}\biggr)^3 \biggr(\frac{100\,\text{GeV}}{f_\phi'}\biggr)  \biggr(\frac{1\,\text{keV}}{f_\phi}\biggr)\biggr(\frac{1\,\text{keV}}{m_{\phi_1}}\biggr)^2 \times n_\Psi c_\theta s_\theta \times \biggr(\frac{(m_\nu)_{ij}}{0.1\,{\rm eV}}\biggr).
\eeq
Even lower effective scales than shown in the above numerical example can thus be obtained if $f_\phi$ and $m_{\phi_1}$ 
are smaller. For instance, if they are comparable with the neutrino masses, $f_\phi\sim m_{\phi_1}\sim {\mathcal O}(\text{eV})$, the effective scale would be $\Lambda \sim {\mathcal O}(\text{keV})$. 

Assuming $m_{\phi_1}\ll m_{\phi_2}$ the above model matches onto the simplified model singlet mediator models for enhanced neutrino polarizability, Section \ref{subsec:toymodel}, with $\phi_1$ playing the role of the light mediator, $\phi$,\footnote{Note that $\phi$ from now on denotes the single mediator from Section \ref{subsec:toymodel} and not the field from the beginning of this subsection, Eqs. \eqref{eq:inv:Majoron:nus:prime}-\eqref{eq:Lm:break}.} and 
\beq
\label{eq:cnu:gphigamma}
c_\nu^{ij}=i\frac{(m_\nu)_{ij}}{f_\phi} c_\theta,  \qquad g_{\phi\gamma} = 
\frac{9 \alpha}{16\pi} \frac{n_\Psi  s_w^2}{f_\phi'} s_\theta,
\eeq
the nonzero coefficients in the Lagrangian  \eqref{eq:lagr:phi} (we use the notation in Eq. \eqref{eq:gphigamma}). The contributions from heavier $\phi_2$ state are suppressed.  Note that $\phi=\phi_1$ has both flavor diagonal and off-diagonal couplings to neutrinos, and thus only approximately matches onto the constraints shown in Fig. \ref{fig:Summary:AllBounds} in which flavor universal neutrino couplings were assumed, cf. Eq. \eqref{eq:cnu}. Nevertheless, the constraints on $c_\nu$ shown in Fig. \ref{fig:Summary:AllBounds} should approximate well the constraints on $c_\nu^{ij}$ from Eq. \eqref{eq:cnu:gphigamma}.  Numerically, 
\beq
\big|c_\nu^{ij}|=10^{-4} c_\theta \biggr(\frac{(m_\nu)_{ij}}{0.1\,\text{eV}}\biggr)\biggr(\frac{1\,\text{keV}}{f_\phi}\biggr),  \qquad g_{\phi\gamma} = 3\times 10^{-9} \text{GeV}^{-1} \biggr(\frac{100\,\text{GeV}}{f_\phi'}\biggr)  \biggr(\frac{ s_\theta n_\Psi  }{10^{-3}}\biggr),
\eeq
and thus for $m_\phi=1\,\text{keV}, 1\,\text{MeV} $ mass benchmarks the $U(1)_L\times U(1)'$ model can cover the whole experimentally  still  available parameter space in Fig. \ref{fig:Summary:AllBounds}, even when imposing $f_\phi> m_{\phi}$. For the $m_\phi=1\,\text{GeV}$ mass benchmark the main constraint on couplings to photons would be from searches for on-shell production of electroweak triplets, limiting $f_\phi'$ to be above several 100 GeV, and thus $g_{\phi\gamma}\lesssim 10^{-6}\,\text{GeV}^{-1}$ even for large mixing angles, $\sin\theta\sim {\mathcal O}(1)$. For couplings to the neutrinos the requirement $f_{\phi}>m_\phi$ leads to $\big|c_\nu^{ij}|\lesssim 10^{-7}$.

\section{Conclusions}
\label{sec:Conclusions}

In this paper, we examined the theory and phenomenology of New Physics sources of neutrino polarizability, that is, the electromagnetic interaction of neutrinos with two photons. We present the latter in an EFT framework, where these interactions are described by the Rayleigh operators; see Eq.~\eqref{eq:L:EFT}. While naively one would expect these chirality-flipping operators to be suppressed by the neutrino mass, we show that such a suppression can be compensated if the interaction is mediated by a light scalar or pseudo-scalar particle. 

Such models can have a wide variety of phenomenological consequences, depending on the new particle's mass. We fix four mass benchmarks, $m_\phi = 1$ eV, keV, MeV, GeV, and explore the constraints in the parameter space defined by the coupling to photons, $g_{\phi\g}$, and neutrinos, $c_\nu$. To provide results easily comparable with existing literature, we limit ourselves to the case of a pseudoscalar mediator, with flavor-universal couplings to neutrinos. The main results are summarized in Tables~\ref{table:bounds_summary_EFT}, \ref{table:bounds_summary_PHI:ev,kev} and \ref{table:bounds_summary_PHI:mev,gev}, and in Fig.~\ref{fig:Summary:AllBounds}. 

The first three mass benchmarks, $m_\phi = 1$ eV, keV, MeV, are largely excluded by cosmological and astrophysical observables. This is to be expected since CMB spectrum measurements strongly constrain the number of relativistic degrees of freedom in the Universe, $N_{{\rm eff}}$, up to scales of order $T\sim100$ eV. Similarly, $N_{{\rm eff}}$ affects the abundance of primordial elements produced during BBN, which takes place when the Universe temperature is $T\sim1-2$ MeV. Finally, exotic emission of neutrino and light new particles can be excluded by measurements of Horizontal Branch star cooling rates and neutrino fluxes from SN1987a. The former has a typical temperature of $T_{HB}\sim8$ keV, while the latter can reach $T_{SN}\sim30$ MeV in the inner core, largely setting the $\phi$ mass reach of the corresponding bounds.

We performed comprehensive analysis for coupling values in rather large ranges, $c_\nu \in [10^{-12},1]$ and $g_{\phi\g} \in [10^{-12} \, \text{GeV}^{-1},1\, \text{GeV}^{-1}]$. In this parameter space, the eV benchmark is completely excluded. 
The heavier benchmarks, keV and MeV, allow for very small coupling values, mainly due to the disappearance of CMB bounds. The former benchmark is not excluded for $c_\nu\lesssim10^{-6}$ and $g_{\phi\g}\cdot{\rm GeV}\lesssim10^{-11}$, while the latter is not excluded for $c_\nu\lesssim10^{-9}$ and $g_{\phi\g}\cdot{\rm GeV}\lesssim10^{-9}$.

When the mediator mass is heavier than the typical scales of cosmological and astrophysical processes, we expect these bounds to disappear or become negligible. This is evident for the last mass benchmark, $m_\phi = 1$ GeV, where only terrestrial experiments are able to probe parts of the parameter space. Rare $\tau$ lepton decays bound $c_\nu\lesssim0.3$, while monophoton search in $e^+e^-$ collision leads to $g_{\phi\g}\cdot{\rm GeV}\lesssim10^{-4}$. 

We also discussed  UV complete models that lead to enhanced neutrino polarizability, all of which are based on the appearance of a pseudo-Nambu-Goldstone boson associated with the spontaneous breaking of the global lepton number symmetry, $U(1)_L$. Such a pNGB, the majoron, automatically couples to neutrinos since it participates in neutrino mass generation. In cases where majoron has enhanced couplings to photons, the tree-level exchanges of the majoron result in a parametrically enhanced neutrino polarizability, large enough to saturate the present experimental bounds. Such enhanced couplings to photons are, for instance, generated if majoron couples to a separate sector of heavy charged states. A concrete example of a model in which all such parametric enhancements are present is the non-minimal inverse see-saw model discussed in detail in Section~\ref{sec:enhanced:polariz:model}, which has two global symmetries, the lepton number $U(1)_L$ and the anomalous $U'(1)$. The neutrino polarizability is then generated with very low effective scale suppression, despite being suppressed by the neutrino masses. 

There are several directions in which the study performed in the present manuscript could be extended in future works. For one, the region of parameter space where light scalar couplings to photons and neutrinos are equally important (the hashed bands in Fig. \ref{fig:Summary:AllBounds}) should be explored in more detail. The interplay between couplings to photons and neutrinos would be particularly interesting to investigate for cosmological and SN constraints, where simple scaling with branching ratios, which one can use for the collider constraints, does not apply. In this paper we have also limited the discussion to the current constraints from various experiments, and left projections from planned experiments for future studies. In particular, it would be interesting if dedicated searches at dark matter and neutrino facilities for neutrino polarizability signatures, such as a nuclear recoil accompanied by a single photon (from $\nu A\to \nu A\gamma$), or by two resolved photons (from $\nu A\to \nu A\phi$, $\phi \to \gamma\gamma$), could lead to improved experimental reach. One may furthermore want to attempt an extension of our work where in addition to the light scalar couplings to photons and neutrinos, couplings to gluons (or light quarks) are also taken into account. The motivation for this extension is provided by the model of Section \ref{sec:Majoron:QCD:axion}, where Majoron  acts as a QCD axion and thus couples to neutrinos, photons and gluons. Finally, it would be interesting to investigate if inelastic scattering in the neutrino sector, $\nu A\to N A$, followed by a decay of the sterile neutrino, $N\to \phi \nu, \phi\to \gamma\gamma$, could explain the MiniBoone excess.

{\bf Acknowledgements:} We thank R. Budnik, R. Harnik, J. Kopp, P. Machado and E. Vitagliano for useful discussions. MT acknowledges the financial support from the Slovenian Research Agency (research core funding No. P1-0035). JZ acknowledges support in part by the DOE grant DE-SC0011784. GP and AAP were supported in part by the DOE grant DE-SC0007983.

\begin{appendix}

\section{Notations and conventions}
\label{sec:app:notations}
Throughout the manuscript we use the four-component notation following the conventions of Ref. \cite{Dreiner:2008tw}. For Majorana neutrinos we thus have, 
\beq
\nu=\nu^c=\begin{pmatrix} 
\xi_\alpha
\\
\xi^{\dagger \dot \alpha}  
\end{pmatrix},
\qquad 
\bar \nu=\bar \nu^c=\big( \xi_\alpha, 
\xi^{\dagger \dot \alpha} 
\big), 
\eeq
where $\xi_\alpha$ is a two-component Weyl spinor, so that, for instance, 
\beq
\bar \nu_i P_L\nu_j=\xi_i \xi_j.
\eeq

The normalization we use for the dimension 5 dipole operators and the dimension 7 operators in the EFT Lagrangian, Eq. \eqref{eq:L:EFT}, is straightforwardly related to other notations commonly used in the literature. 
The neutrino dipole moments are conventionally  defined as  
\beq
{\cal L}_{\rm eff}\supset \sum_{i>j}\frac{1}{2}(\lambda_\nu)_{ij}  (\bar \nu_i  \sigma^{\mu \nu} P_L \nu_j) F_{\mu\nu}+{\rm h.c.},
\eeq
 where the antisymmetric $3\times 3$ matrix, $\lambda=\mu-id$, decomposes into the magnetic ($\mu$) and electric ($d$) dipole moments, see, e.g., Ref.~\cite{Miranda:2019wdy}. In terms of the Wilson coefficients in \eqref{eq:L:EFT} we have
\beq
(\lambda_\nu)_{ij} \mu_B \equiv  \frac{\C_{1,ij}^{(5)}}{\Lambda} \frac{e}{4\pi^2},
\eeq
 where  $\mu_B$ is the Bohr magneton.  
 
For polarizabilities, we can follow the conventions used for nucleon polarizabilities, see, e.g., Ref.~\cite{Hagelstein:2015egb}, and define for non-relativistic neutrinos
 \beq
{\cal L}_{\rm NR} = 2\pi \lp \alpha_{E1,i} \vec E^2 + \beta_{M1,i} \vec B^2 \rp \otimes 1_{\nu_i}. 
 \eeq
Here, $\alpha_{E1}$ is the electric,  and $\beta_{M1}$ the magnetic scalar polarizability, with $E_i=F_{0i}$ the electric, and $B_i=\epsilon_{ijk}F_{jk}/2$ the magnetic field, and $1_{\nu_i}$ the neutrino number operator for neutrinos of flavor $i$ (the non-relativistic version of the $\bar \nu_i \nu_i$ operator). At dimension 7 in the EFT expansion in $1/\Lambda$, Eq. \eqref{eq:L:EFT}, we have 
\beq
\label{eq:alpha:beta:rel}
\alpha_{E1,i}=\beta_{M1,i}=\frac{\alpha}{24\pi^2\Lambda^3} \C_{1,ii}^{(7)}.
\eeq
 The relation \eqref{eq:alpha:beta:rel} is broken by $m_\nu^2/\Lambda^2$ suppressed contributions from higher-order operators, for instance from $ (\partial_\mu\bar  \nu_i  P_L \partial_{\tau}\nu_j) F^{\mu\nu} F^{\tau}_\nu$. These can become important only if the effective scale $\Lambda$ is low, i.e., if there are light mediators with mass $m_\phi$ comparable to the neutrino mass that get integrated out in the construction of \eqref{eq:alpha:beta:rel}. In our numerical examples, however, we always have $m_\phi\gg m_\nu$.

\section{Further details on stellar cooling rate calculations}
\label{sec:app:cooling:rates}
In this appendix we collect the $\phi$ production rates relevant for the stellar cooling bounds discussed in Section~\ref{sec:StarCooling}.

\subsection{Primakoff conversion} 
The rate for the Primakoff conversion of a photon $\gamma$ (more precisely, the transverse plasmon, $\gamma_T$) to a pseudoscalar $\phi$ in the field of the nucleus in a plasma, is given by~\cite{Lucente_2020,Carenza:2020zil}, 
\beq
\begin{split}\label{eq:Primakoff}
\Gamma_{\g\to\phi} &= \lp \frac{c_\g' \alpha }{8 \pi f_\phi} \rp^2 \frac{T \kappa^2}{2\pi}\frac{p}{E_\phi} \Big[ \frac{\lp (k + p)^2 + \kappa^2\rp\lp (k - p)^2 + \kappa^2\rp}{4 k\cdot p\,\kappa^2}  \ln\lp \frac{ (k + p)^2 + \kappa^2}{ (k - p)^2 + \kappa^2} \rp - \\
&- \lp \frac{(k^2 - p^2)^2}{4k\cdot p\,\kappa^2} \rp \ln\lp \frac{ (k+p)^2}{(k-p)^2} \rp - 1 \Big]\,,
\end{split}
\eeq
where  $k$ and $p$ are the incoming photon and outgoing $\phi$ momenta, respectively, while $\kappa$ is the Debye screening length, 
\beq\label{eq:Debye}
\kappa^2 = \frac{4\pi\alpha}{T}\biggr( n_e^{\rm eff} + \sum_j Z_j^2 n_j^{\rm eff} \biggr)\,.
\eeq
Here $T$ is the temperature of the star at the radius where the Primakoff conversion occurs, while $n_e^{\rm eff}$ and $n_j^{\rm eff}$ are the effective number densities of electrons and ions, the latter with charge $Z_je$.

The core of a HB star is a non-relativistic, non-degenerate gas of electrons and helium ions, thus $n_X^{\rm eff} = n_X$ for $X=e, \text{He}$. The electrons forms an ideal Fermi gas, so that the number density is given by $n_e = p_F^3/(3\pi^2)$, where the Fermi momentum is $p_F = 88$ keV~\cite{Raffelt:1996wa}. The number density of Helium ions is $n_{\rm He} = \rho/(m_u A_{He})$, where $\rho_{HB} \sim 10^4 {\rm g/cm^3}$ is the HB star density, $m_u = 0.932$ GeV the atomic mass unit, and $A_{\rm He} = 4$ the  atomic number of Helium. Combining the two terms in Eq.~\eqref{eq:Debye}, we get $\kappa_{\rm HB}\simeq27$ keV.  

The electrons in the SN core form a highly degenerate relativistic gas and, as such, do not contribute to the screening. The plasma is composed of a degenerate gas of protons; the degeneracy reduces the effective number of proton targets in the Primakoff process, thus, more care is needed in calculating $n_p^{\rm eff}$. The number density of degenerate non-relativistic protons at a distance $r$ from the SN core center is given by 
\beq
n_p(r) = 2 \int\frac{{\rm d}^3p}{(2\pi)^3} \exp \left[ \lp \frac{p^2}{2 m_p^{\rm eff}(r) T(r)} - \eta_p(r) \rp + 1 \right]^{-1}\,,
\eeq
where $m_p^{\rm eff}$ is the effective proton mass and $\eta_p$ is the proton degeneracy parameter~\cite{Payez:2014xsa}: protons are (non) degenerate for $\eta_p >1 ~ (\eta_p < 1)$. 
Neglecting the proton recoil, the effect of degeneracy on the number of targets can be calculated as~\cite{Payez:2014xsa} (see also Eq.~(D.26) in \cite{Raffelt:1996wa})
\beq
\frac{n_p^{\rm eff}}{n_p} = \frac{2}{n_p} \int \frac{{\rm d}^3p}{(2\pi)^3} \hat f_p \lp 1 - \hat f_p \rp\,,
\eeq
where $\hat f_p$ is the Fermi-Dirac distribution for protons.

The numerical inputs are functions of the SN core profile, i.e., how the temperature and density change with the radial distance $r$ from the center. For numerical results in Sec.~\ref{sec:SN} we use the SN core profile from~\cite{Fischer:2018kdt} at the benchmark time $t=1\,$s after the start of the explosion. For rough numerical estimates we can take $r\sim10\,$km,  $T\sim30\,$MeV, $\eta_p\sim1$, $m_p^{\rm eff}\sim800\,$MeV, which gives $n_p^{\rm eff}\sim 0.6~n_p$ and $\kappa_{\rm SN} \sim {\mathcal O}(40\,\text{MeV})$. 

Inside dense stellar cores the dispersion relation for the transverse plasmon of energy $\omega$ is well approximated by introducing an effective photon thermal mass, $\omega_P$, Eq. \eqref{eq:omegaP}, so that $k = \big(\omega^2 - \omega_P^2\big)^{1/2}$.
The rate for the inverse Primakoff conversion, $\phi\to\g$, is then  given by
\beq\label{eq:InversePrim}
\Gamma_{\phi\to\g} = \frac{2\beta_\g}{\beta_\phi} \Gamma_{\g\to\phi}\,,
\eeq
where $\beta_i$ is the velocity of particle $i$, that is, $\beta_\gamma=(1-\omega_P^2/\omega^2)^{1/2}$ for plasmon of energy $\omega$ and $\beta_\phi=(1-m_\phi^2/E_\phi^2)^{1/2}$ for $\phi$ with energy $E_\phi$. The factor of two in \eqref{eq:InversePrim} is due to the two possible polarizations of a transverse plasmon.

\subsection{Photon and neutrino coalescence}
The rates for photon coalescence, $\g\g\to\phi$, and for neutrino coalescence, $\nu\nu\to\phi$, can be calculated by first considering a generic $\phi$ production process $a_1 + \dots + a_n \to \phi$, where $\{a_1,\dots,a_n\}$ is a set of initial states, e.g., the two photons in photon coalescence. The phase space production rate for the (pseudo)scalars $\phi$ can be extracted from the corresponding Boltzmann equation for the phase space density $\hat f_\phi$. In the limit $\hat f_\phi \ll 1$ this gives, see, e.g., Ref. \cite{Carenza:2020zil},
\beq
\label{eq:fphi:time}
\frac{\partial \hat f_\phi}{\partial t} = \frac{1}{2 E_\phi} \int  \prod_{i\to\{a\}} \frac{{d}^3 k_i}{(2\pi)^3 2 E_i}  (2\pi)^4 \delta^{(4)}\lp p - k_i \rp |\overline {\cal M}|^2  \prod_{i\to\{a\}} \hat f_i (E_i) \,,
\eeq
where $p$ is the $\phi$ four-momentum and $k_i$ the momenta of the initial states, $\overline {\cal M}$ is the spin averaged matrix element for the process, while the integration is performed over  phase space of the initial states. 
  In writing \eqref{eq:fphi:time} we assumed that, once produced, the $\phi$ escapes the SN, and thus we can take $\hat f_\phi \simeq 0$ in the possible collision terms on the right and ignore them. The number of $\phi$ produced is then given by
\beq\label{eq:ProductionNumberRate}
{d}N_\phi = \hat f_\phi \frac{{d}^3p}{(2\pi)^3} \quad\Rightarrow\quad \frac{{d}^2N_\phi}{{d}t\,{d}E_\phi} = \frac{\partial \hat f_\phi}{\partial t} |\vec p\,| E_\phi \frac{{d}\Omega}{(2\pi)^3}\,.
\eeq
\paragraph{Photon coalescence.} In the production of on-shell $\phi$ via annihilation of two photons, $\gamma(k_1) + \gamma(k_2) \to \phi(p)$, the two photon momenta need to satisfy $(k_1+k_2)^2 = m_\phi^2$. The emissivity for the case when $\phi$ is a pseudoscalar, i.e.,  an ALP, is well known in the literature, see for example Ref.~\cite{Lucente_2020}. If $\phi$ is a scalar the amplitude squared changes to 
\beq
|\overline{\cal M}|^2 = \lp \frac{\alpha}{3\sqrt{2}\pi} \frac{c_\g}{f_\phi} \rp^2 m_\phi^4 \biggr[\biggr( 1 -  \frac{4 \omega_P^2}{m_\phi^2} \biggr)^2 + \frac{6\omega_P^4}{m_\phi^4} \biggr]\,.
\eeq
The last term in the square brackets does not appear in the pseudoscalar case; the size of it is, however, relevant only in proximity of the kinematical threshold, $m_\phi = 2 \omega_P$, and becomes quickly negligible for heavier $m_\phi$ masses.
Performing the integration over the initial momenta in  \eqref{eq:fphi:time} gives 
\beq
\frac{\partial \hat f_\phi}{\partial t} = \lp \frac{\alpha}{3\sqrt{2}\pi} \frac{c_\g}{f_\phi} \rp^2 \frac{m_\phi^4}{32\pi E_\phi } \sqrt{1 - \frac{4\omega_P^2}{m_\phi^2}}\biggr[ \biggr( 1 - \frac{4\omega_P^2}{m_\phi^2}\biggr)^2 + \frac{6\omega_P^4}{m_\phi^4}  \biggr] \exp\lp-\frac{E_\phi}{T}\rp\,,
\eeq
where we assumed that the initial photons follow the Maxwell-Boltzmann distribution, where, from conservation of energy, the sum of the two photon energies satisfies $E_1 + E_2 = E_\phi$.
The emissivity due to the photon coalescence is then given by
\beq
\begin{split}
Q_{\phi,\gamma\gamma} &= \lp \frac{\alpha}{3\sqrt{2}\pi} \frac{c_\g}{f_\phi} \rp^2 \frac{1}{32\pi^3} \int_{m_\phi}^\infty {\rm d}E_\phi E_\phi |\vec p| m_\phi^4  \\ 
&\times \sqrt{1 - \frac{4\omega_P^2}{m_\phi^2}}\biggr[ \biggr( 1 - \frac{4\omega_P^2}{m_\phi^2}\biggr)^2 + \frac{6\omega_P^4}{m_\phi^4}  \biggr] \exp\lp-\frac{E_\phi}{T}\rp\,,
\end{split}
\eeq
where $|\vec p| = \sqrt{E_\phi^2 - m_\phi^2}$. The result for pseudoscalar $\phi$ is obtained by neglecting the second term in the square bracket and by replacing $c_\g\to 3/2 ~c_\g'$.

\paragraph{Neutrino coalescence.} The amplitude squared for the $\phi$ production via scattering of two neutrinos, $\nu(k_1) + \nu(k_2) \to \phi(p)$, is given by 
\beq
|\bar{\cal M}|^2 = \frac{c_\nu^2}{2} \lp k_1\cdot k_2 \rp =\frac{c_\nu^2 m_\phi^2}{4}\,,
\eeq
where we used the momentum conservation and assumed massless neutrinos. The neutrinos follow the Fermi-Dirac thermal distribution, $\hat f_\nu(E) = (\exp[(E-\mu)/T] + 1)$, where $\mu$ is the neutrino chemical potential. 

 The Boltzmann equation can be written as
\beq
\begin{split}
    \frac{\partial \hat f_\phi}{\partial t} &= \frac{1}{2 E_\phi}\int \frac{{\rm d}^3k_1}{2 E_1 (2\pi)^3}\frac{{\rm d}^3k_2}{2 E_2 (2\pi)^3} (2\pi)^4 \delta^{(4)}\lp p - k_1 - k_2 \rp \frac{c_\nu^2 m_\phi^2}{4} \hat f_{\nu}(E_1) \hat f_{\nu}(E_2) \\
    &= \frac{c_\nu^2 m_\phi^2}{16 E_\phi (2\pi)^2}\int \frac{{\rm d}^3k_1}{E_1} \delta\lp (p - k_1)^2 \rp \hat f_{\nu}(E_1)\hat f_{\nu}(E_2)\,,
\end{split}
\eeq
where in the second line we integrated over ${\rm d}^4k_2$ using the on-shell condition. The delta funtion can be written as
\beq
\delta\lp (p - k_1)^2 \rp = \delta\lp m_\phi^2 - 2 p\cdot k_1 \rp = \frac{1}{2|\vec p|E_1}\delta\biggr( \cos\theta -  \frac{(m_\phi^2 - 2 E_\phi E_1)}{2|\vec p|E_1}  \biggr)\,,
\eeq
where $\theta$ is the angle between $\vec p$ and $\vec k_1$. Requiring that $-1<\cos\theta<1$ gives the integration limits
\beq
E_{1,{\rm min}} = \frac{E_\phi - \sqrt{E_\phi^2 - m_\phi^2}}{2}\,,\qquad E_{1,{\rm max}} = \frac{E_\phi + \sqrt{E_\phi^2 - m_\phi^2}}{2}\,.
\eeq
Changing to spherical coordinates for the integration over ${\rm d}^3k_1$ then gives, 
\beq
\frac{\partial \hat f_\phi}{\partial t} = \frac{c_\nu^2 m_\phi^2}{64\pi E_\phi} \int_{E_{1,{\rm min}}}^{E_{1,{\rm max}}} \frac{{\rm d}E_1}{|\vec p|} \hat f_{\nu}(E_1)\hat f_{\nu}(E_\phi - E_1)\,.
\eeq
Using the above expression  in  $\phi$ production rate, Eq.~\eqref{eq:ProductionNumberRate}, and then in the general expression for emissivity, Eq.~\eqref{eq:emissivity:general}, gives the emissivity due to the neutrino coalescence,
\beq
Q_{\phi,\nu\nu} = \frac{c_\nu^2 m_\phi^2}{16 (2\pi)^3} \int_{m_\phi}^\infty{\rm d}E_\phi \int_{E_{1,min}}^{E_{1,max}} {\rm d}E_1 E_\phi \hat f_{\nu}(E_1)\hat f_{\nu}(E_\phi - E_1)\,.
\eeq

\section{Rare decays of heavy (pseudo)scalars}
\label{sec:RareDecays}

In this appendix we provide further details on the derivation of the constraints on the $\nu\nu \gamma\gamma$ effective interactions from $S\to \gamma\gamma\to \nu\nu$ decays that were given in Section \ref{sec:ColliderConstraints}. Throughout, we assume that we can use EFT to describe the neutrino-photon interactions, Eq.~\eqref{eq:L:EFT}.

\subsection{Constraints from invisible decay widths} 
To construct the bound, one needs first to compute the discontinuity of the $S \to \nu\nu$ amplitude. While many possible intermediate states can contribute to $S \to \nu\nu$, only on-shell intermediate states generate the absorptive part. The total amplitude can then be obtained from a dispersion relation \cite{Donoghue:1996fv}.
Selecting one intermediate state, $\gamma\gamma$, one obtains a bound on the parameters of the $\nu\nu \gamma\gamma$ interaction from the bounds on the invisible widths of the $\pi$ and $B$-mesons and the Higgs bosons (see also \cite{Bhattacharya:2018msv}).

To obtain the bound, one needs to parameterize the $S\to\gamma\gamma$ decay amplitude, which can be written as \cite{Bosch:2002bv}
\begin{equation}\label{eq:Aplusminus}
{\cal A}(S\to\gamma\gamma)\equiv{\cal A}_+\langle\gamma(q_1,\epsilon_1)\gamma(q_2,\epsilon_2)|F^{\mu\nu}F_{\mu\nu}|0\rangle+{\cal A}_-\langle\gamma(q_1,\epsilon_1)\gamma(q_2,\epsilon_2)|\tilde{F}^{\mu\nu}F_{\mu\nu}|0\rangle\,,
\end{equation}
where $q_1,q_2$ are the photon momenta and $\epsilon_1, \epsilon_2$ their polarizations.
The matrix elements of the operators $F^{\mu\nu}F_{\mu\nu}$ and $\tilde{F}^{\mu\nu}F_{\mu\nu}$ in Eq.~\eqref{eq:Aplusminus} are given by
\beq
\begin{split}
\langle\gamma(q_1,\epsilon_1)\gamma(q_2,\epsilon_2)|F^{\mu\nu}F_{\mu\nu}|0\rangle &= -4\big(q_1\cdot q_2g^{\alpha\beta}- q_1^\beta q_2^\alpha\big) \epsilon_{1\alpha}\epsilon_{2\beta}\,, \\
\langle\gamma(q_1,\epsilon_1)\gamma(q_2,\epsilon_2)|\tilde{F}^{\mu\nu}F_{\mu\nu}|0\rangle &= 4\epsilon^{\alpha\beta\rho\sigma}\epsilon_{1\alpha}\epsilon_{2\beta}q_{1\rho} q_{2\sigma}\,.
\end{split}
\eeq
Note that the CP-even (CP-odd) matrix element is symmetric (anti-symmetric) in $\epsilon_1\leftrightarrow \epsilon_2$ interchange. 

The discontinuity of the $S\to\gamma\gamma\to\nu\nu$ amplitude reads
\begin{equation}
\label{eq:Disc:gamma:nu}
\mbox{Disc }{\cal A}(S\to\gamma\gamma\to\nu\nu)=i\frac{m_S^4}{2\pi}\left[{\cal A}_+\left({\cal A}^R_+\right)^*+{\cal A}_-\left({\cal A}^R_-\right)^*\right],
\end{equation}
where the contributions from the Rayleigh operators are encoded in (cf. Eq.  \eqref{eq:C12:Re})
\beq\label{eq:ARpmcoeff}
{\cal A}^R_+ = \frac{1}{2} \lp \frac{\alpha}{12\pi} \rp \hat \C_1^{\rm Re},
\qquad 
{\cal A}^R_- = \frac{1}{2} \lp \frac{\alpha}{8\pi} \rp \hat \C_2^{\rm Re},
\eeq
Notice that while ${\cal A}^R_\pm$ are real since they arise from an effective Lagrangian, ${\cal A}^S_\pm$ are not necessarily real, as they might receive contributions from other on-shell intermediate states. In fact,  
please note that the discontinuity computed in Eq.~(\ref{eq:Disc:gamma:nu}) assumes that the on-shell transitions $S \to \gamma\gamma$ and $\gamma\gamma \to \nu\nu$ are dominated by local interactions, i.e. there are no discontinuities generated by on-shell contributions in $A_\pm^{(R)}$. This implies that one cannot obtain a meaningful bound from the invisible decays of the $D^0$, as $D^0\to \gamma\gamma$ is dominated by the non-local contributions \cite{Burdman:2001tf}.  

Since $|\mbox{Im }{\cal A}(S\to\gamma\gamma\to\nu\nu)|\leq |{\cal A}(S\to\gamma\gamma\to\nu\nu)|$, the decay rate calculated using $|\mbox{Im }{\cal A}|$ is smaller or equal to the decay rate calculated using  $|{\cal A}|$ which in turn is smaller than $|{\cal A}(S\to{\rm invisible})|$. 
The rate calculated using $|\mbox{Disc }{\cal A}|$ is 
\beq\label{eq:GammaImBound}
\begin{split}
\Gamma_{{\rm Im }} &= \frac{|\mbox{Im }{\cal A}|^2}{16\pi m_S}= \frac{m_S^9}{256\pi^3} \frac{1}{4} \lp \frac{\alpha}{8\pi} \rp^2 \left| \frac{2}{3}{\cal A}_+  \hat \C_1^{\text{Re}*}  + {\cal A}_- \hat \C_2^{\text{Re}*} \right|^2 \leq\Gamma(S\to{\rm inv}).
\end{split}
\eeq
Note that summing over the neutrino spins gives a factor of $m_S^2$ in the rate. If only $\hat \C_1^{\text{Re}}$ or $\hat \C_2^{\text{Re}}$ are nonzero, the above relation can be rewritten in the form of the bound in  \eqref{eq:CRe:bound}, that was used to obtain bounds on Rayleigh operators from the bounds on invisible $\pi^0$ and Higgs decays. 

For the $B^0$ meson decays both ${\cal A}_+$ and ${\cal A}_-$ are nonzero. At leading order in the $1/m_b$ expansion~\cite{Bosch:2002bv} they are given by
\begin{equation}
{\cal A}_+={\cal A}_-=\frac{G_F}{\sqrt{2}}\frac{\alpha}{3\pi}\frac{f_B}{2}V_{td}^*V_{tb}C_{7\gamma}\frac{m_{B^0}}{\lambda_B}\,,
\end{equation}
where $f_B=0.190$ GeV~\cite{FlavourLatticeAveragingGroup:2019iem} is the $B$-meson decay constant, $\lambda_B=0.35$ GeV~\cite{Bosch:2002bv} is the inverse of the first inverse moment of the $B$-meson light-cone distribution amplitude, while $C_{7\gamma}=-0.38$~\cite{Benzke:2010tq} is the Wilson coefficient in the effective weak Hamiltonian. Higher order corrections to the expansion are $\sim10\%$,
and we neglect them in the following. 
Using \eqref{eq:GammaImBound} together with the total decay width, 
$\Gamma_{B^0} = 4.3\times10^{-13}$ GeV and $\text{Br}(B^0\to\text{ inv})\leq2.4\cdot10^{-5}$ gives \eqref{eq:B0:constr}.

\subsection{Recasting the monophoton search}
\label{sec:BaBar:notesMichele}
In this subsection we give further details on recast in Section \ref{sec:ColliderConstraints} of the BaBar monophoton search~\cite{BaBar:2017tiz} in terms of a bound on the couplings of the $\phi$ mediator. The BaBar results in Ref.~\cite{BaBar:2017tiz} are given in terms of bounds on the dark photon mixing parameter, $\varepsilon$, where
the differential cross section for $e^+ e^-\to A^\prime \gamma$ is given by
\beq
\begin{split}
\frac{d\sigma_{A'\gamma}}{d\cos\theta}=\frac{4\pi \alpha^2\varepsilon^2}{s^2(s-m_{A'}^2)}\left[\frac{s^2+m_{A'}^4}{\sin^2\theta}-\frac{(s-m_{A'}^2)^2}{2}\right]
\frac{1}{\left(1+4m_e^2\cot^2\theta\,/s\right)^2}\frac{1}{(\sqrt{1-4m_e^2/s}}
\\
+m_e^2\frac{16\pi \alpha^2\varepsilon^2}{s^2(s-m_{A'}^2)}\frac{s^2\left(s-2m_{A'}^2-4m_e^2\right)\sin^2\theta+(s-m_{A'}^2)^2(s-2m_e^2)\cos^4\theta}{\sin^2\theta\left(1+4m_e^2\cot^2\theta\,/s\right)^2\sqrt{1-4m_e^2/s}}\,
\end{split},
\eeq
where $\theta$ is the angle of photon momentum with respect to the electron axis.
To avoid forward scattering backgrounds BaBar limited the angular acceptance to $| \cos\theta | <0.6$. In this region one can safely neglect the electron mass, giving Eq.~\eqref{eq:BaBar:diffxsec} in the main text. Note also, that for $m_{A'} < 1$ GeV the cross section given above is for all practical purposes independent of the dark photon mass.

The bound on the dark photon mixing parameter $\varepsilon$ can then be recast as the bound on  the $e^+e^-\to \gamma \phi$ cross section, by equating the allowed $e^+ e^-\to A^\prime \gamma$ and $e^+e^-\to \gamma \phi$ cross sections (after integration over $\cos\theta\in[-0.6,0.6]$) and setting $m_{A'}=m_\phi$.  The $e^+e^-\to \gamma \phi$ cross section in \eqref{eq:dsigmaphigamma} follows from the decay amplitude for $e^+ e^-\to\phi \g$, with momenta $p_1,~p_2,~p_3,~k$, respectively
\beq
i{\cal M} = (ie)\lp \frac{i\alpha c_\g'}{8\pi f_\phi} \rp \lp \bar v_1 \g_\mu u_2 \rp \lp \frac{-i}{q^2} \rp \lp 4 \epsilon^{\mu\beta\rho\eta} q_\rho k_\eta \rp \epsilon_\beta^*(k)\,,
\eeq
where $q = p_1 + p_2$ is the momentum exchange, and for simplicity we take $\phi$ to be a pseudoscalar and set $c_\gamma=0$. 
Squaring and averaging over initial spins, and taking the $m_e \to0$ limit gives
\beq\label{eq:Msquared:onshell}
\overline{|{\cal M}|^2} = 4 \lp \frac{e\alpha c_\g'}{2\pi f_\phi} \rp^2 \frac{(k\cdot p_1)(k\cdot q)(p_2\cdot q) + (k\cdot p_2) [ (k\cdot q) (q\cdot p_1) - q^2 (k\cdot p_1) ]}{q^4}\,.
\eeq
Using $q^2 = s$, $p_1\cdot q = p_2 \cdot q = s/2$ and that in the center of mass frame $k\cdot p_1 = E_\g {\sqrt s}(1 - c_\theta)/2$, $k\cdot p_2 = E_\g {\sqrt s}(1 + c_\theta)/2$ and $k\cdot q = E_\g {\sqrt s}$, where $E_\g$ is the photon energy, gives
\beq
\overline{|{\cal M}|^2 }= \lp \frac{e\alpha c_\g'}{2\pi f_\phi} \rp^2 E_\gamma^2 (1+\cos^2\theta)\,.
\eeq
In the center of mass frame $4E_\gamma^2=s(1-m_\phi^2/s)^2$. We can also obtain a similar expression for a scalar $\phi$ by replacing $c_\g'\to 2c_\gamma/3$. There is no interference between the amplitudes for $F_{\mu\nu} F^{\mu\nu}$ (scalar) and $F_{\mu\nu}\tilde  F^{\mu\nu}$ (pseudoscalar).

The differential cross section for the $2\to2$ process is then
\beq
\begin{split}
\frac{d\sigma_{\phi\gamma}}{d\cos\theta}=\frac{\alpha^3}{128\pi^2}\left(\left|\frac23\frac{c_\gamma} {f_\phi}\right|^2+\left|\frac{c^\prime_\gamma} {f_\phi}\right|^2\right)\left(1-\frac{m_\phi^2}{s}\right)^3 (1+\cos^2\theta)=
\\
=\frac{\alpha^3}{256\pi^2}\left(\left|\frac23\frac{c_\gamma} {f_\phi}\right|^2+\left|\frac{c^\prime_\gamma} {f_\phi}\right|^2\right)\left(1-\frac{m_\phi^2}{s}\right)^3 \lp 3 + \cos2\theta\rp\,,
\end{split}
\eeq
in agreement with \eqref{eq:dsigmaphigamma}.

In the EFT limit, $m_\phi\gg \sqrt{s}$, neutrino polarizability leads to a $2\to3$ scattering process, $e^+e^-\to  \nu \nu \gamma$ with momenta $p_1,~p_2,~p_3,~p_4,~k$, respectively. The amplitude is similar to $e^+ e^-\to\phi \g$ and given by
\beq
i{\cal M} = (ie) \lp \bar v_1 \g_\mu u_2 \rp \lp \bar u_3 P_L v_4 \rp \lp \frac{-i}{q^2} \rp \epsilon_\beta^*(k)T^{\alpha\beta}_\pm \C_\pm\,,
\eeq
where $T^{\alpha\beta}_+=4q\cdot k g^{\alpha\beta}-4q^\beta k^\alpha$, $T^{\alpha\beta}_-=-4\epsilon^{\alpha\beta\rho\eta} q_\rho k_\eta $, $\C_+=i\hat \C_{1}^{(7)}\alpha/12\pi$, $\C_-=i\hat \C_{2}^{(7)}\alpha/8\pi$.

The spin-averaged amplitude squared depends on the neutrinos only via the invariant mass of the neutrino-antineutrino pair $m_{\nu\nu}^2=(q-k)^2$. The final state has the usual three-body kinematics familiar from, e.g., muon decay. Using the kinematical relation after Eq. (\ref{eq:Msquared:onshell}) the differential cross section is given by 

\beq
\frac{d\sigma}{d\cos\theta \,dE_\gamma} =\biggl(\left|\frac23\hat\C_1\right|^2+\left|\hat\C_2^2\right|^2\biggr) \frac{n_\nu}{4\pi}\Big(\frac{\alpha}{8\pi}\Big)^3  E_\g^3 \biggr(1 - \frac{2 E_\g}{\sqrt s}\biggr) \big(3 + \cos 2\theta\big)\,,
\eeq
where $n_\nu$ is the number of neutrino flavor. In our numerical analysis we take $n_\nu=3$.  
Note that the photon energy is related to the invariant mass of the neutrino pair through $E_\g = {\sqrt s}/2 - m_{\nu\nu}^2/(2{\sqrt s})$.

\end{appendix}

\bibliographystyle{JHEP}
\bibliography{nubiblio}

\end{document}